\documentclass[manuscript]{aastex}

\slugcomment{Submitted to the Astrophysical Journal}
\shorttitle{\textsl{RXTE} Observations of Cen A}
\shortauthors{Rothschild et al.}
\begin{document}

\title{Twelve and a Half Years of Observations of Centaurus A with \textsl{RXTE}}

\author{R. E. Rothschild, A. Markowitz, L. Rivers, and S. Suchy}
\affil{Center for Astrophysics and Space Sciences, University of California,
    San Diego, CA, USA 92093-0424}
\email{rrothschild, almarkowitz, erivers, ssuchy@ucsd.edu}

\author{K. Pottschmidt}
\affil{CSST, University of Maryland Baltimore County, 1000 Hilltop Circle,
Baltimore, MD, USA 21250\\
CRESST and NASA Goddard Space Flight Center, Astrophysics Science Div., Code 661, Greenbelt, MD 20771}
\email{katja@milkyway.gsfc.nasa.gov}

\and

\author{M. Kadler, C. M\"uller, and J. Wilms}
\affil{Dr. Remeis Sternwarte \& ECAP, Sternwartstr. 7, 96049 Bamberg, Germany}
\email{joern.wilms, matthias.kadler, cornelia.mueller@sternwarte.uni-erlangen.de}

\begin{abstract}

The \textsl{Rossi X-ray Timing Explorer} has observed the nearest radio galaxy, 
Centaurus A, in 13 intervals from 1996 August to 2009 February over the 3 -- 200 keV 
band. Spectra accumulated over the 13 intervals were {\bf well} described with an absorbed power 
law and an iron line. Cut-off power laws and Compton 
reflection from cold matter did not provide a better description.  
{\bf For the 2009 January observation we set} a lower limit on the cutoff energy at over 2 MeV.
The power spectral density function was generated from \textsl{RXTE}/ASM and 
PCA data as well as an \textsl{XMM-Newton} long look, and clear evidence for a 
break at $18^{+18}_{-7}$ days {\bf (68\% conf.)} was seen. Given Cen A's high black hole mass 
and very low value of $L_{\rm X}/L_{\rm Edd}$,  
the break was a factor of 17$\pm$9 times higher than the break frequency predicted by the 
McHardy and coworkers relation, which was empirically derived for {\bf a sample of objects, which are } radio-quiet {\bf and} accreting at relatively high values of $L_{\rm bol}/L_{\rm Edd}$. 
We have interpreted our observations in the context of a clumpy molecular torus. The variability characteristics and the broadband spectral energy distribution, when compared to Seyferts, 
imply that  the bright hard X-ray continuum emission may originate at the base of 
the jet, yet from behind the absorbing line of sight material, in contrast to what is commonly 
observed from blazars. 
\end{abstract}


\section{Introduction}

The prototype {\bf Fanaroff-Riley Class I \citep[FR I;][]{Fanaroff74}} low luminosity radio 
source Centaurus~A (hereafter Cen~A) played a
significant role in the early understanding of the nature of active
galaxies and jet emission. Cen~A was found to ``correspond within
limits of experimental error'' to the ``extra-galactic nebula'' NGC
5128 by \citet{Bolton49}. Early radio interferometry by
\citet{Mills52} established that Cen~A was an extended source and
therefore not a stellar object. In addition, Mills found it to be
nearly 2$\degr$ in diameter in radio with ``a strong concentration near its
centre''. A year later \citet{Mills53} refined the sizes to
1.5$^\circ$ for the extended flux and 5$\arcmin$ for the compact
core. Arguments by \citet{Baade54} demonstrated that Cen~A was
extragalactic, and \citet{Humason56} revealed an uncorrected redshift
close to the present day value ($z=0.001825$). The giant radio lobes were
resolved by \citet{Wade59} after removal of the bright compact
source coincident with NGC 5128. The synchrotron nature of the radio
emission was put forth by Shklovsky and collaborators \citep[see, 
e.g.,][for references]{Burbidge57}.

The Centaurus region of the southern sky was scanned in the late 1960s
by rocket and balloon X-ray payloads without detecting Cen~A
\citep{cooke67,harries67,lewin68}. The 1969 rocket flight of
\cite{Bowyer70} was the first claimed detection, and was followed by many
rocket, balloon and satellite observations \citep[see][for a list of
early results]{Stark76}. A power law continuum with photon index of 
$\Gamma \sim$ 1.8 extending beyond 200 keV
was soon established with low energy absorption due to intervening
matter having an equivalent hydrogen column density, $N_\mathrm{H}$, along the line of 
sight near 10$^{23}$ atoms cm$^{-2}$
\citep[e.g.,][]{Baity81}. \textsl{CGRO} observations of Cen~A extended the 
flux detections to 1 MeV \citep{Kinzer95}, and to greater than 
100 MeV \citep{Steinle98}. Recently, emission from the core of Cen~A \citep{Abdo10b}
and its extended emission \citep{Abdo10a} have been detected by the \textsl{Fermi}/Large Area Telescope (LAT) at GeV 
energies, and by \textsl{H.E.S.S.} at TeV energies \citep{Aharonian09}, further extending 
the detected emissions from this remarkable object over the entire electromagnetic spectrum. 

By virtue of being the nearest radio galaxy and one of the few bright active
galaxies in the 2--10 keV X-ray band, detailed X-ray observations of Cen~A revealed
much about the nuclei of low power, radio-loud active galaxies in general 
\citep[see][for a review]{Israel98}. With present day imaging instruments, we have a 
remarkably detailed view of Cen~A. \textsl{Chandra}'s 
spectacular images have revealed details for a wealth of features: jet structure, 
knots in the jets, discrete sources, and shocks in the interstellar medium 
\citep[e.g.,][]{Kraft00,Kraft01,Kraft03}.   \citet{Evans04} used \textsl{Chandra} and \textsl{XMM-Newton}
observations to conclude that the iron line at 6.4 keV originated in cool material that 
either surrounds the black hole or forms part of a torus, most of which is out of the line of 
sight. In either case the iron emitting region was distant from the central source.

While comparison of the blazar Cen~A to Seyfert 
 galaxies is not strictly proper due to jet dominated blazar spectra versus disk dominated 
 Seyfert spectra, Cen~A displays {\bf at least one} aspect of disk dominated spectra, i.e. iron line 
 emission from cold material. The jet-like emission comes from a jet whose primary beam is 
 oriented far from the observer's line of sight and thus some Seyfert-like aspects of the spectra 
 become detectable.  One must keep this in mind when comparing Cen~A results 
 to those of Seyfert AGN and to those of blazars.

Infrared observations from the ground with adaptive optics and from
\textsl{Hubble Space Telescope} yielded estimates of the mass of the
supermassive black hole from gas kinematics and from stellar dynamics \citep[e.g., ][]{Silge05,Marconi06}. 
Integral-field observations in the near infrared with the SINFONI 
spectrograph with adaptive optics on the VLT gave 
$M_{\rm BH}$=(5.5$\pm$3.0)$\times$10$^7$ M$_\odot$ based upon
stellar kinematics and (4.5$^{+1.7}_{-1.0}$)$\times$10$^7$ M$_\odot$ from 
H$_2$ kinematics \citep{Cappellari09,Neumayer10}. 
\cite{Krajnovic07}, using the CIRPASS spectrograph on Gemini 
South, found $M_{\rm BH}$=(8.25$^{+2.25}_{-4.25}$)$\times$10$^7$ M$_\odot$
 based on gas dynamics. For a distance of 3.42 Mpc \citep{Ferrarese07}, 3.5 Mpc 
 \citep{Hui93}, or 3.8 Mpc \citep{Rejkuba04,Harris10}, this mass must be contained 
 within $\simeq$0.6 pc \citep{Marconi06}. {\bf We adopt the values of 3.8 Mpc and $6\times10^7$ M$_\odot$ 
 for the following analyses and discussions.}
 
 Radio observations of the inner tens of parsecs reveal an unresolved core with a jet and counter 
 jet \citep{Horiuchi06} with the inner $\pm$20 mas being resolved into multiple components. 
 TANAMI radio observations \citep{Ojha10} find that the core region has an inverted spectrum indicative of on-going 
 synchrotron and possible synchrotron self-Compton (SSC) or free--free absorption 
 processes \citep{Mueller10}. While not unexpected, this would support a model of 
 synchrotron self-Compton for producing the hard X-ray flux observed \citep{Chiaberge01}.

With its broad band X-ray coverage, highly manueverable spacecraft, and flexible 
scheduling, the \textsl{Rossi X-ray Timing Explorer} (\textsl{RXTE}) is the premier 
mission for monitoring and multiple observations of X-ray sources on timescales from milliseconds to years. 
Multi-timescale monitoring campaigns to probe X-ray variability in active galactic nuclei (AGN) on time 
scales from hours to years began with the \textsl{EXOSAT} era in the 1980s 
\citep[e.g., ][]{Lawrence87,Lawrence93,Green93}, and has continued to present day with \textsl{RXTE} 
\citep[e.g., ][]{Edelson99,Markowitz03,McHardy06}. 
The resulting broadband Power Spectral Density functions (PSDs), 
derived mainly for radio quiet ``normal'' broad line Seyfert 1s and
narrow-line Seyfert 1s, have yielded evidence for breaks at temporal 
frequencies $f_{\rm b}$ in the range $\sim 10^{-6}$ to 10$^{-3}$ Hz with PSD power-law slopes breaking from 
about --2 to about --1 above and below $f_{\rm b}$, respectively.
\citet{McHardy06} and references therein have shown 
that the temporal variability resulting from accretion onto black holes, 
as characterized by $f_{\rm b}$, scales inversely with black hole mass and directly with bolometric 
luminosity. Comparatively less is known about the X-ray variability properties of 
non-Seyferts, including jet-dominated blazars and radio galaxies 
such as Cen~A. Preliminary X-ray structure functions and PSDs for blazars, though 
frequently based on long-term light curves which are gap-dominated, 
have revealed breaks which, like Seyferts, correspond to timescales 
of a few days to a couple of weeks \citep{Kataoka01,Kataoka02,Marscher04}.

In this paper we present spectral and timing results on Cen~A  derived from 
13 observations spanning 1996 to 2009 made with \textsl{RXTE}. Our goals include
characterizing broadband X-ray spectral variability,
constraining the geometry of the circumnuclear accreting material by studying both the
line of sight absorbing gas and the Fe K$\alpha$ line-emitting gas,
and quantifying X-ray variability across a range of time scales in order
to better understand Cen~A for comparison with other AGN.
The remainder of this paper is structured as follows: Section 2 gives the details of 
the observations, Section 3 describes data selection and 
analysis techniques, Section 4 gives the results of spectral and temporal analyses, 
including a first-ever broadband PSD (covering below $\sim 10^{-5}$ Hz) for Cen~A, 
while Section 5 discusses the results of these 
observations of Cen~A, and finishes with our conclusions.

\section{Observations}

\begin{figure}
\includegraphics[width=7.in]{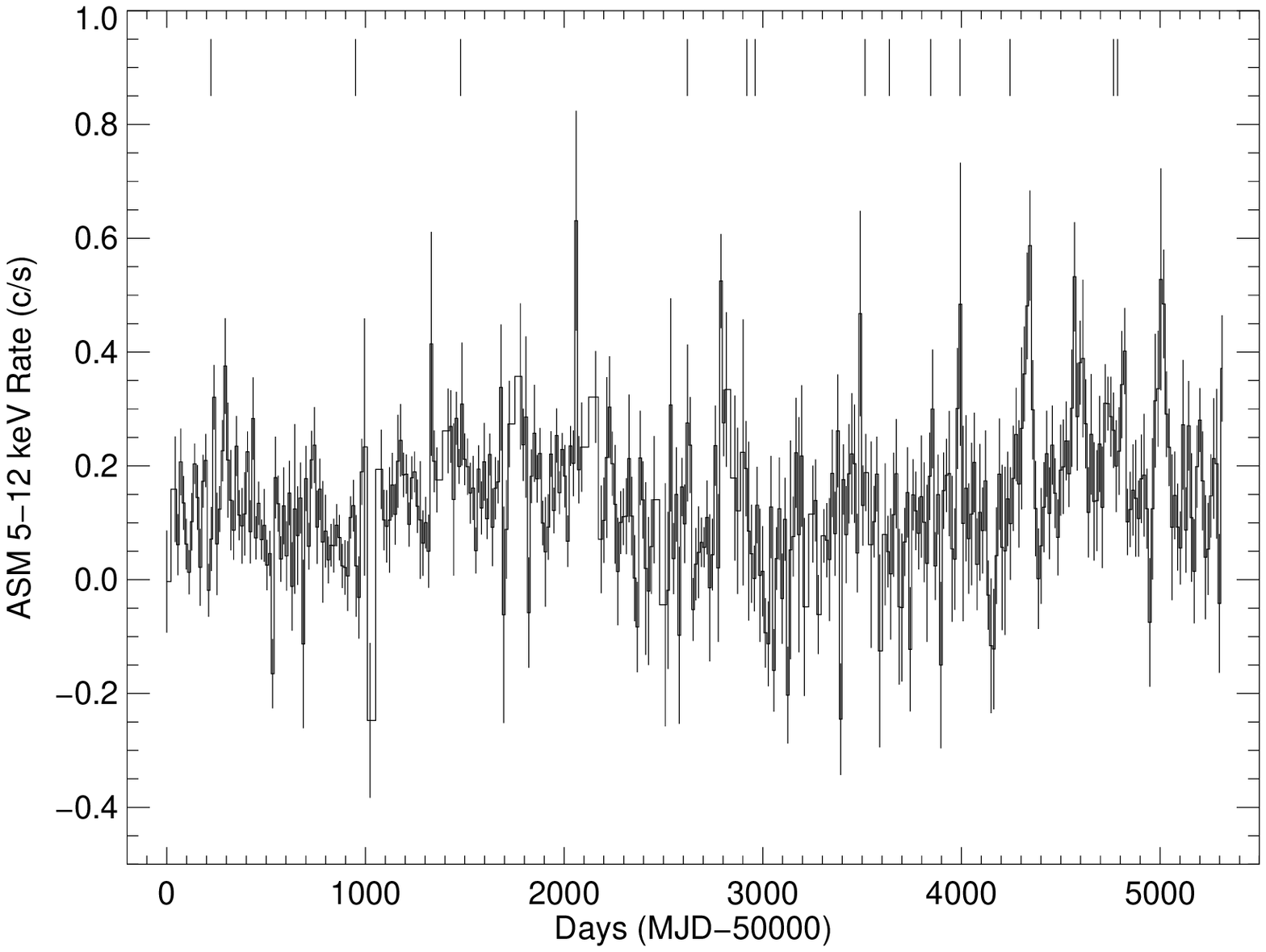}
\caption{The flux history of Cen~A over the \textsl{RXTE} mission as measured by the All-Sky Monitor 
5--12.1 keV channel 3, with 2 week binning. The verticle bars at the top of the plot denote when the 
\textsl{RXTE} pointed observations were made.\label{fig:asm}}
\end{figure}

The flux history of Cen~A over the duration of the  \textsl{RXTE} mission from 1996 through mid 2010, 
as measured in the 5--12.1 keV band by the \textsl{RXTE}/All-Sky Monitor (Fig.~\ref{fig:asm}), reveals 
variations of a factor of 3 or more. Thirteen observing campaigns totaling 129 separate pointings were executed 
using \textsl{RXTE}'s pointed-mode instruments, the Proportional Counter Array (PCA) and the
High Energy X-ray Timing Experiment (HEXTE), as indicated in Fig.~\ref{fig:asm}.
Each campaign consisted of 1 to 22 individual
pointings, and spanned durations of less than a day to several days.
The 13 campaigns could be divided into observations on 51 separate days (from which 47 daily spectra were produced; see \S\ref{sec:daily}),
and adjacent campaigns were separated by less than a month to years. 
We thus have the ability to probe spectral and temporal variability in Cen A 
on a wide range of time scales, including hours, individual days, and
weeks to years.

Table~\ref{tab:obstime} gives the livetimes for the
13 separate observing intervals of Cen~A from 1996 August to 2009
February. The counting rates are for 3--60 keV from the top layer of PCA detector PCU2 and 15--200 keV 
for HEXTE. Since the on-source/off-source rocking of HEXTE cluster A was terminated on 2006
July 13, the livetimes and rates are given for the combined clusters up to that
date as well as just  cluster B throughout the mission. See the next Section for a description of the PCA and HEXTE.

\section{Data Analysis}

Version v6.7 of the HEASOFT software package
release\footnote{\tt http://heasarc.gsfc.nasa.gov/ftools/} with 
updated PCA response v11.7 (2009 May 11) and background estimation files Faint/L7 and 
Sky\_VLE dated 2005 Nov. 28 were used
throughout, as was \texttt{XSPEC} 12.5.1n \citep{Arnaud96}. All errors presented are 
90\% confidence intervals, with the exceptions of iron line equivalent widths, 
which have 99\% uncertainties, and the counting rates in Table~\ref{tab:obstime} 
are 68\% errors.

\subsection{Proportional Counter Array}

For the PCA \citep{Jahoda06}, only data from the top layer 
of the second Proportional Counter Unit (PCU) --- PCU2 --- were used in the present analysis, since
its calibration was the most refined of the 5 PCU counters \citep{Jahoda06}, and since
it was the one PCU in common for all of the Cen~A observations. The
PCU2 data were accumulated under the conditions that the source be greater than
10$\degr$ above the Earth's limb, the pointing direction be within
0$\degr$.01 of the source position, the satellite was more than 30
minutes from the last onset of a South Atlantic Anomaly (SAA) passage,
and the veto rate indicative of precipitating electrons was less than
0.1\footnote{\bf {\tt http://heasarc.gsfc.nasa.gov/docs/xte/recipes/pca\_event\_spectra.html\#reduction}}. The PCU2 Standard Data counts histograms were rebinned into wider energy bins
above channel 60 by 2, above channel 80 by 4, and above channel 100 by
29 in all analyses. PCU2 data from the full 3--60 keV
energy range were subject to fitting for this paper, and no systematic
errors were added to the PCU2 data. 

The Faint background model was used for all but the 2009
January/February observations, for which the SkyVLE background model
was used. The choice of background models was determined by whether or
not the PCU2 toplayer rate exceeded 50 counts s$^{-1}$, and only the 
2009 January and February rates exceeded this value. The PCU2
background models were calculated from particle rates from the
detector based upon numerous blank sky pointings throughout the
\textsl{RXTE} mission \citep{Jahoda06}, and as such, individual
observations were susceptible to minor differences in normalization
between the model background and reality. Similarly, the deadtime
correction was based upon detector rates and modeled appropriately. In
order to adjust spectral analyses for small deviations of the
background and deadtime models from reality, the background model counts histogram
was included in the \texttt{XSPEC} fitting as a correction file (see 
\S~\ref{sec:fitting} for a description of this procedure). 

\subsection{High Energy X-ray Timing Experiment}

The HEXTE \citep{Rothschild98}
data from on-source and two off-source positions
were collected from each cluster separately until the cluster A
rocking was discontinued on 2006 July 13. After that the cluster B data were still
collected from both off-source positions, but no off-source data were
collected from cluster A. The same elevation, pointing accuracy, and
time since the SAA criteria were used in the selection of good data
for HEXTE. The HEASARC tool HEXTErock was used to determine if a
known source would be in any of the HEXTE off-source
positions. None were. Consequently, the two
background spectral accumulations, when available for cluster A 
and always for cluster B, were combined to
form a single background spectrum for each cluster. The on-source
accumulations for both clusters were then combined as well as the
appropriate background files. Combining these data sets directly was possible 
due to the HEXTE automatic and continuous gain control system that 
allowed for the setting of all detector gains to be the same.
This resulted in a single on-source and
single background file for each Cen A observation for which both
clusters were rocking. In those instances where cluster A was not
rocking, only cluster B data were used. No systematic errors were added to the HEXTE data.

\begin{figure}
\includegraphics[width=3.5in]{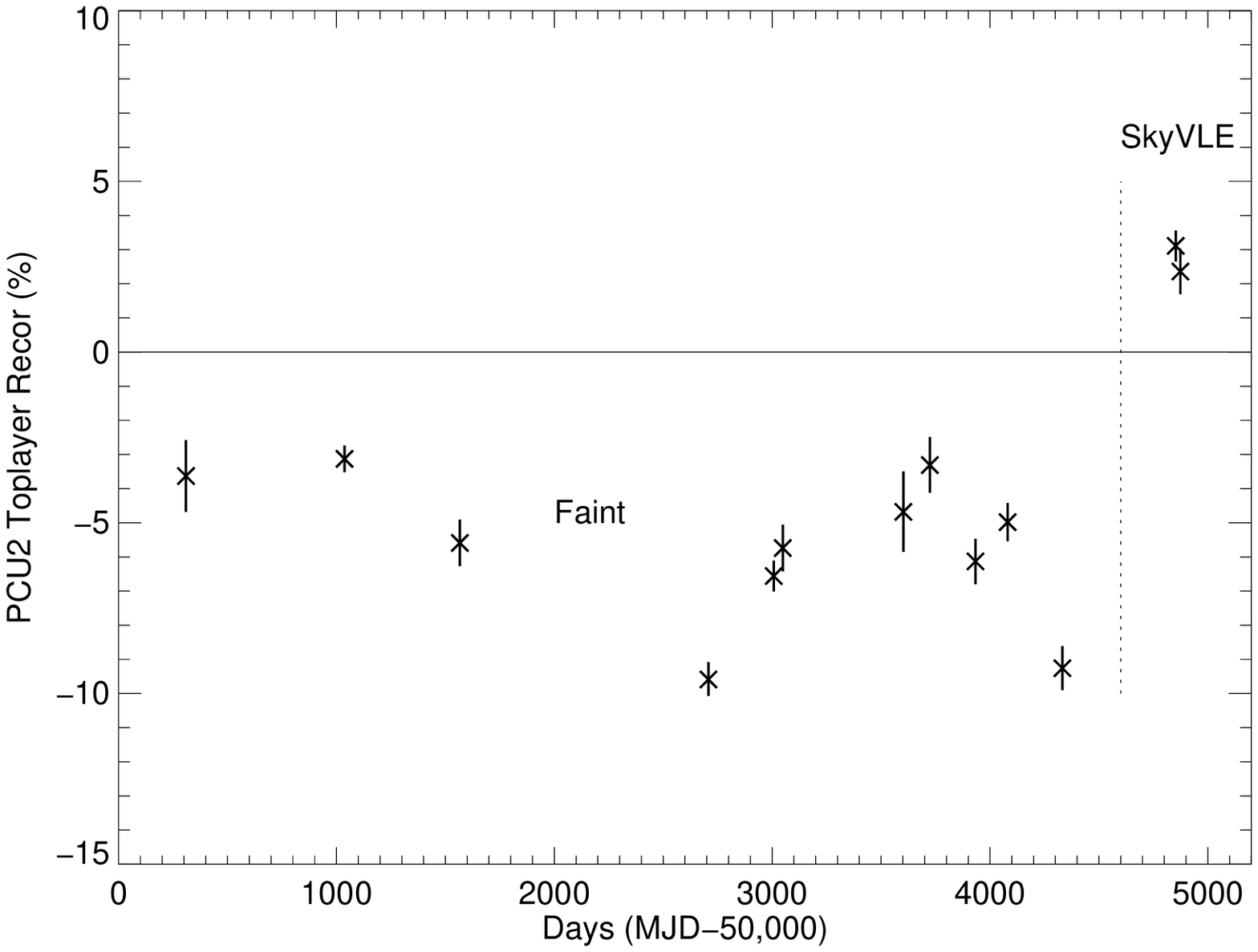}
\includegraphics[width=3.5in]{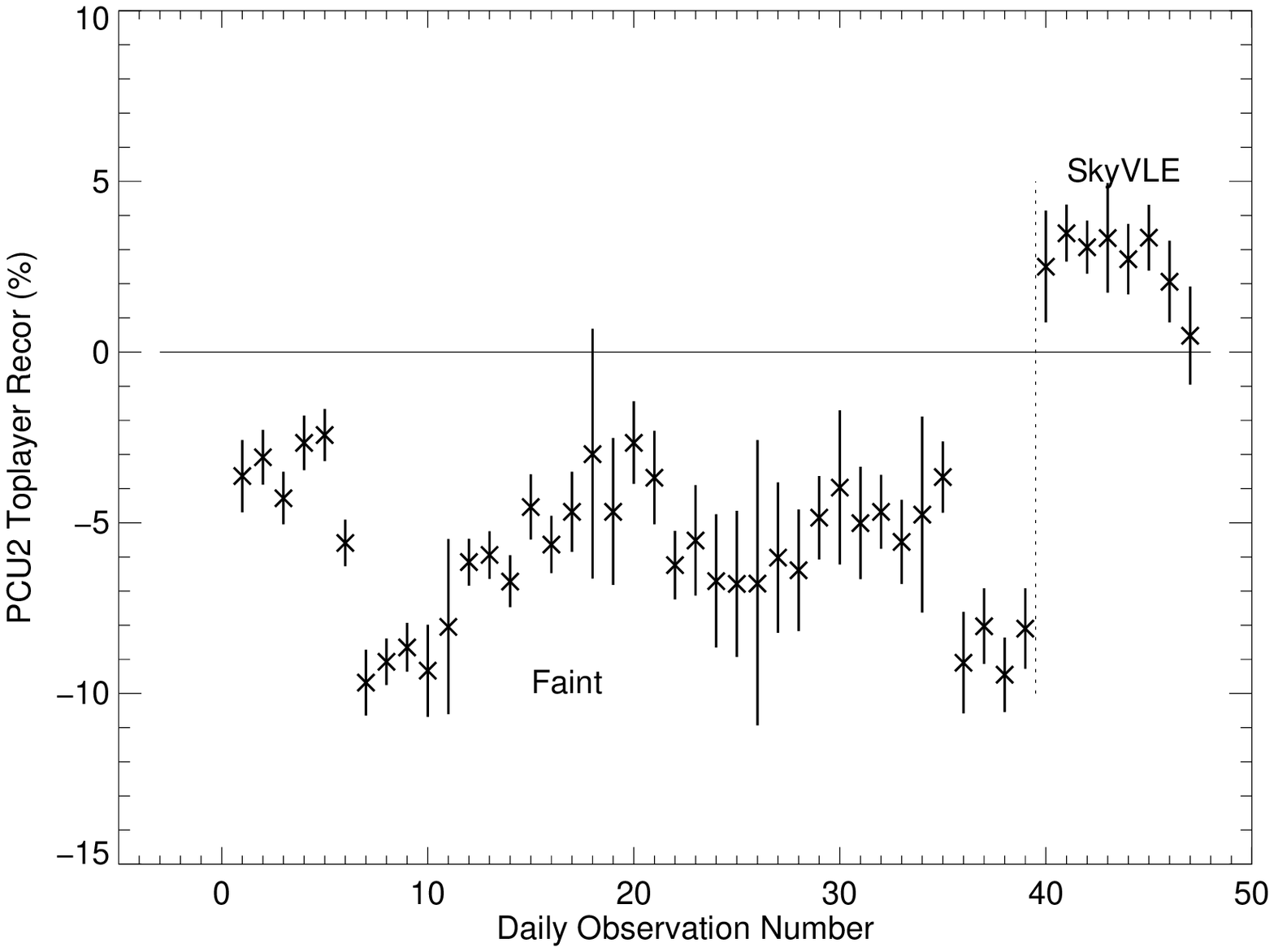}\\
\includegraphics[width=3.5in]{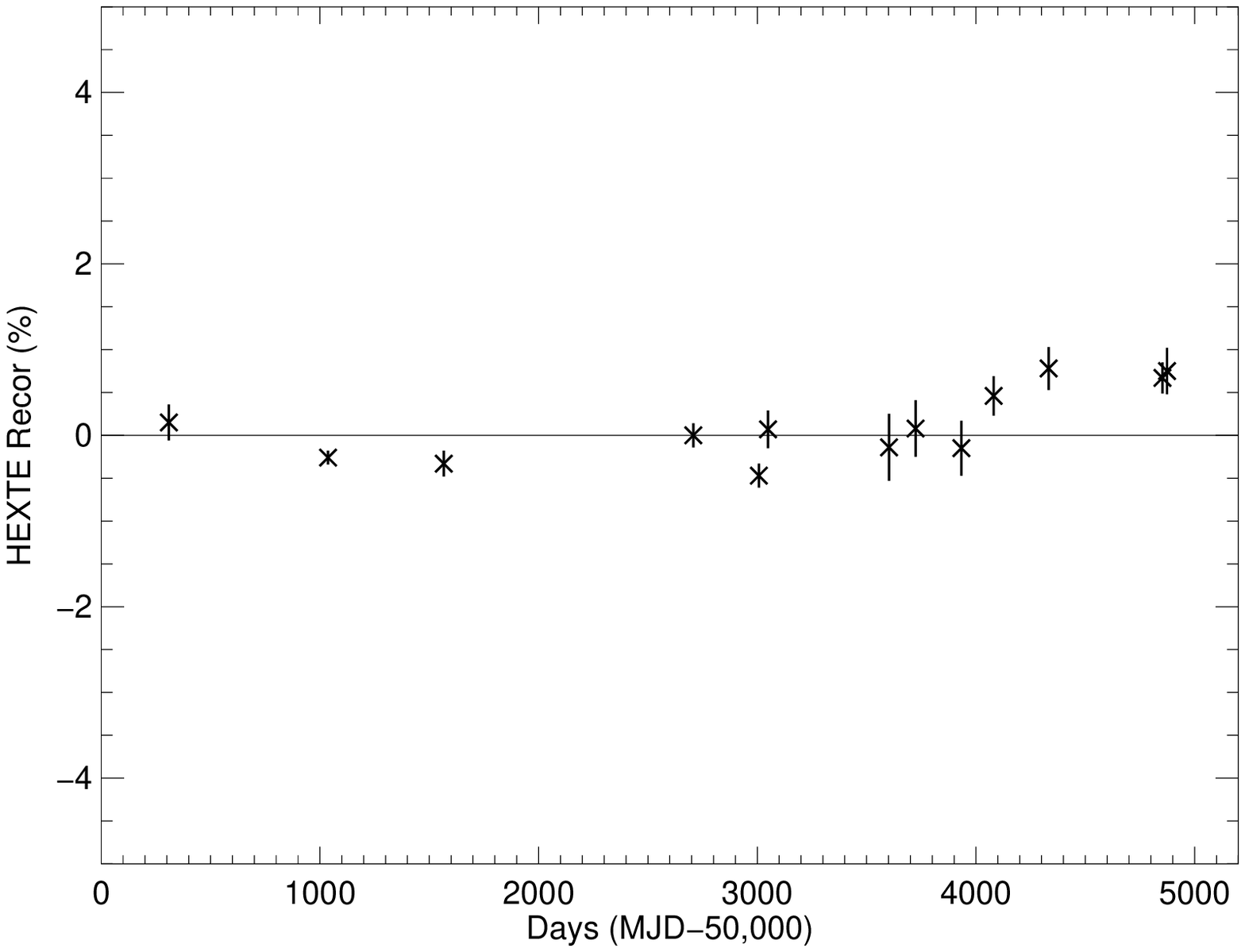}
\includegraphics[width=3.5in]{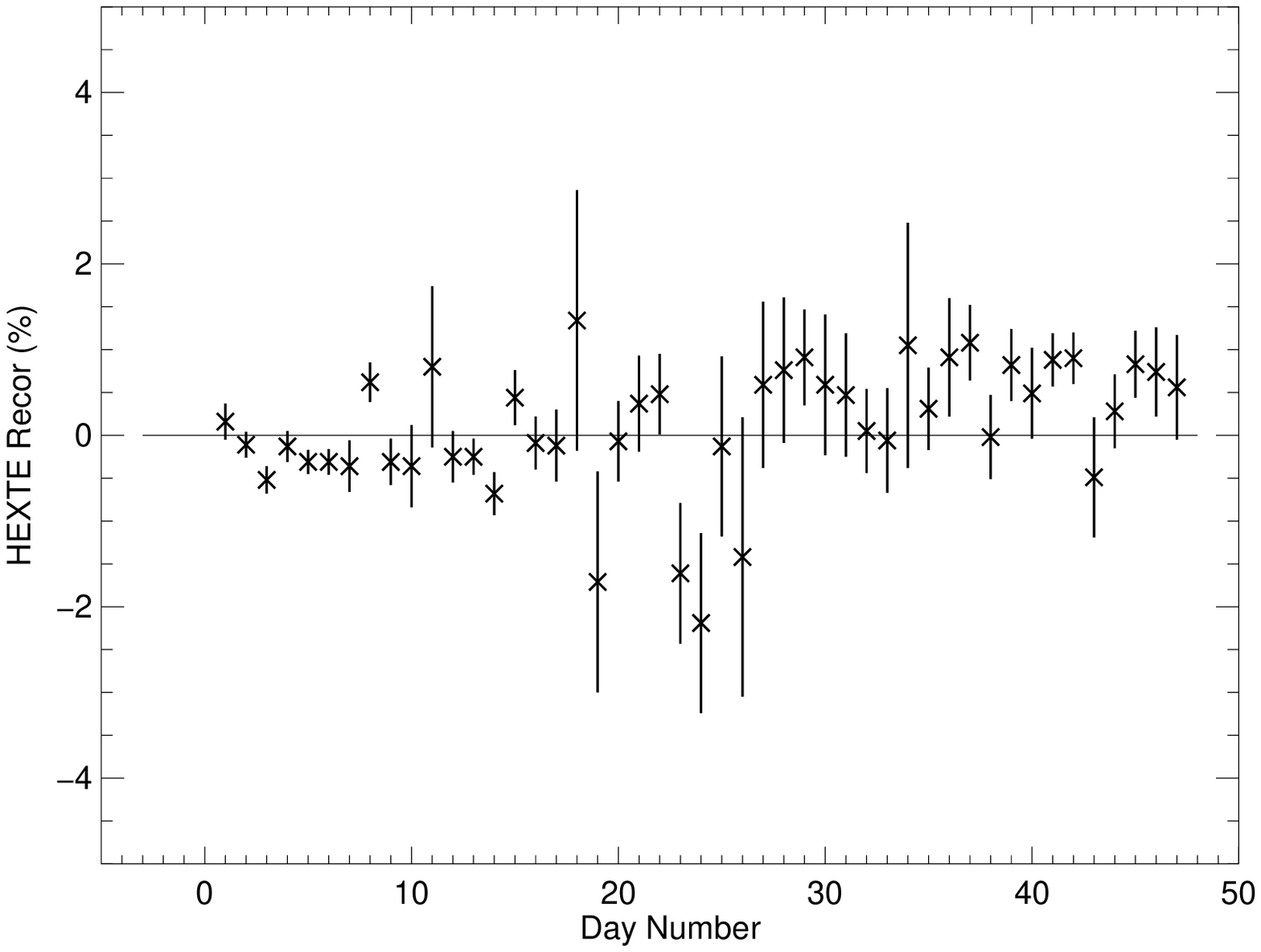}\\
\caption{The best-fit correction factors applied to background/deadtime estimates for the 13 observation intervals and the 47 daily 
spectra histograms for PCU2 (Top) and HEXTE {Bottom} derived from fitting of Cen~A spectra. 
The vertical dotted line in the PCU2 figures delineates the separation of Faint and SkyVLE 
background estimate use. Horizontal lines at a value of zero are included.\label{fig:recor}}
\end{figure}

HEXTE background files were accumulated in real-time during an
observation as a result of the HEXTE cluster rocking program, and were
not a product of a modeling program. The deadtime calculated for each
data segment was, however, subject to a model based upon particle event
rates, and as such could produce imperfect background subtraction. The
same technique for dealing with this effect as for PCU2 was used, and
the correction amounted to less than 1\% in all
cases (Figure~\ref{fig:recor}). The HEXTE data were rebinned by 5 above channel 60, by 10 above
channel 160, and by 20 above channel 210. HEXTE data from 18--200 keV
were used in this paper.

\subsection{Simultaneous PCA and HEXTE Spectral Fitting}\label{sec:fitting}

Simultaneous fitting of the PCU2 and HEXTE data was
performed on both daily and the 13 observation interval data where the
HEXTE data had sufficient statistical quality to affect the result.
This provided the best-fit \textsl{RXTE} broad band 3--200 keV spectral
parameters for Cen A with overlapping PCU2 and HEXTE data in the 18--60 keV band. 
A constant multiplying the HEXTE model was fitted to
correct for the relative normalization of HEXTE with respect to PCU2. The
fitted value of this constant was about 80\% for the observations. 
This reflected the differences in assumed Crab normalization between 
PCA and HEXTE, where the PCA normalization is assumed in the reported 
fluxes.

When fitting Cen A data using \texttt{XSPEC 12.5.1n} (which allows for use of background
correction files for more than one detector), the correction factor for imperfect 
knowledge of the PCU2 background and PCU2 and HEXTE dead time estimates 
was an integral part of the fitting procedure, and in this manner uncertainties in the correction 
were included in the calculation of parameter uncertainties. The use of this 
procedure required that the \texttt{XSPEC} parameter \texttt{delta}, which defines the step size 
used in numerical determination of the derivatives used in the fitting process,
be set to 1$\times$10$^{-4}$ {\bf (K. Arnaud, private communication)} and that numerical differentiation be used (its
value in the \texttt{.xspec/Xspec.init} file must be set to true\footnote{\bf {\tt http://heasarc.gsfc.nasa.gov/docs/xanadu/xspec/manual/XSmodelRecorn.html}}). The resulting
negative values of the PCU2 corrections for the first 11 observation
intervals ranged from a few to less than 10\%, which is consistent
with the values found by \citet{Roth06} for the first 6
\textsl{RXTE}/PCU2 observations of Cen A. The 2009 January and February 
observations using the \texttt{SkyVLE} background required positive corrections of a 
few percent. Figure~\ref{fig:recor} shows the values of the background correction 
factor for the 13 observation intervals and the 47 daily data sets for both PCU2 and HEXTE. The corrections for 
the two different PCU2 background estimations are quite different.

\section{Results}

\subsection{The Energy Spectrum of Cen~A}

All data were fitted with a model representing red-shifted line of sight
absorption \texttt{(XSPEC} model \texttt{ZPHABS)} with abundances set to those of
\citet{Wilms00} and utilizing the \citet{Verner96} cross sections, 
a power law (\texttt{XSPEC} model \texttt{PEGPWRLW}) with a normalization equal to the unabsorbed flux in the
2--10 keV range, and a red-shifted Gaussian component representing iron K$\alpha$
emission with width set to the \textsl{Suzaku} value of 30 eV
\citep{Markowitz07}. Instrumental line-like residuals near 4.5, 8.0,  and 29 keV, when present,
were fit as Gaussians as part of the model. Including one or more of these Gaussians did not 
significantly affect the best-fit parameters of the rest of the model, but did reduce 
$\chi ^2$ nearer to 1 in most cases.

The best-fit spectral 
parameters for the power law fit to the 13 observation intervals
are given in Table~\ref{tab:13obs} and the 2--10 keV and 20--100 keV 
fluxes are given in Table~\ref{tab:13_other_obs}. In order for comparisons to spectra modeled 
with the \texttt{XSPEC} model \texttt{POWERLAW}, the best-fit flux at 1 keV is also given in Table~\ref{tab:13_other_obs} 
for additional fitting done with the \texttt{POWERLAW} model component replacing the \texttt{PEGPWRLW} description 
of the power law. We achieved good fits with both versions of this model in all cases. 

\begin{figure}
\includegraphics[angle=270,width=6.0in]{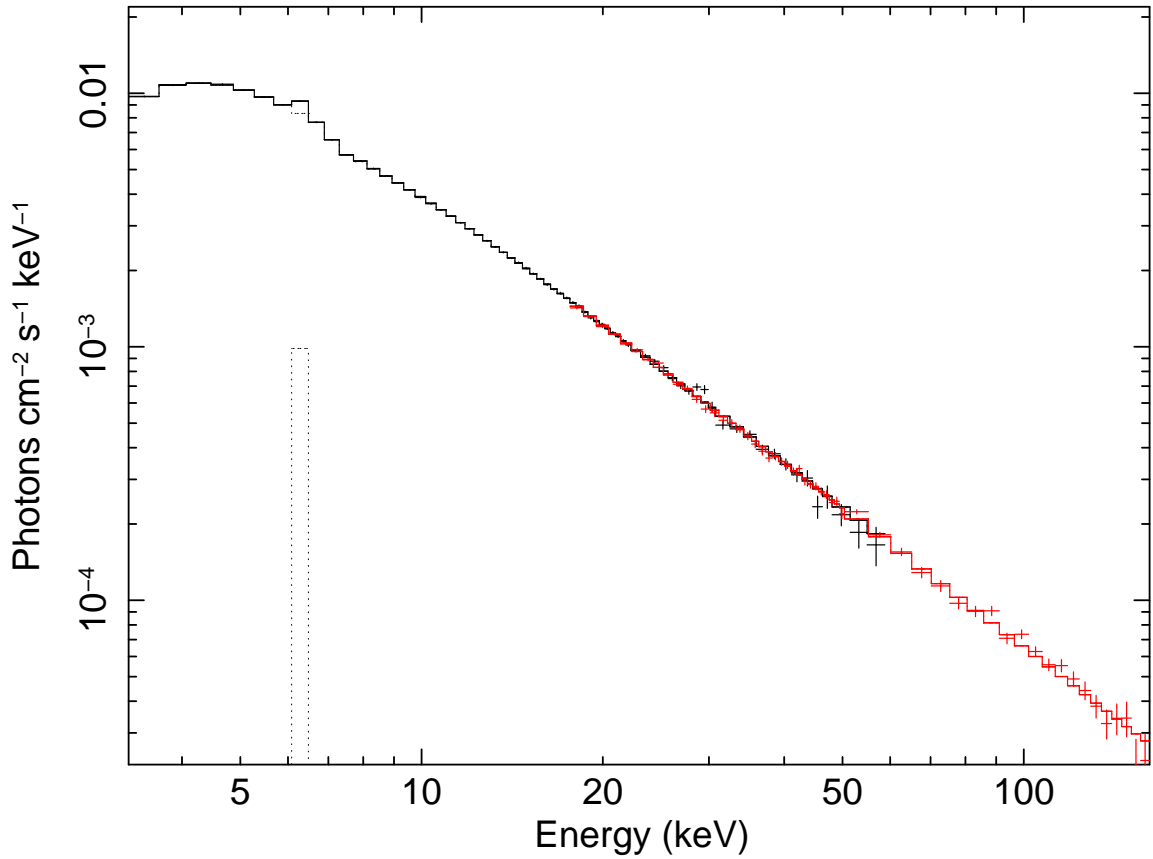}
\caption{The inferred incident spectrum of the 2009 January set of observations. PCU2 toplayer 3--60 keV and HEXTE 18--200 keV data points are shown along with the model of a single absorbed power law with an iron emission line at 6.4 keV. The line component is indicated by the dotted lines. \label{fig:cutoff}}
\end{figure}

A second model was tested with the power law replaced with the \texttt{XSPEC}
model \texttt{CUTOFFPL}, which is a power law times an exponential to
approximate any rollover of the spectrum due to Comptonization
processes.  This second model did not result in significantly lower
$\chi ^2$ values in any case, and since it contained an added parameter (the cutoff
energy), we used the results from the pure power law fits
in the discussions that follow. Lower limits, and in 2 cases best-fit
values, to the rollover energy are given in Table~\ref{tab:13_other_obs}
for comparison to such fits by other authors. The 2 observations yielding best-fit values of the rollover energy
have a common range of 382 keV to 693 keV. This is consistent with the \citet{Rivers10} 
lower limit of 490 keV from fitting the sum of all the Cen~A data. The observations with the strongest lower 
limits to the cutoff energy ($\geq$2796 keV and $\geq$2499 keV), however, occur at the highest 
flux values in 2009. This latter result is strong
evidence for no cutoff being necessary, as measured over the 3--200 keV band (Fig.~\ref{fig:cutoff}).
Future instrumentation with good sensitivity to hundreds of keV
acquired over a realistic amount of observing time will be required to
derive a meaningful value for the rollover energy, if indeed
that is the proper description of the high energy spectrum of Cen~A.

A third model included the \texttt{XSPEC} model \texttt{PEXRAV} component of  Compton reflection
from cold/neutral material in a semi-infinite slab \citep[i.e., from the accretion disk; ][]{Magdziarz95} 
along with the observed power law. The folding energy was set to 1000 keV and the inclination angle 
was fixed at 62$\degr$.6 \citep{Burtscher10} from the dust emission in the mid-infrared.  This
model also did not improve the fit over a single power law, but did
provide upper limits on the reflection component
(Table~\ref{tab:13_other_obs}). All upper limits were consistent with the \citet{Rivers10} 
upper limit of 0.5\% from a similar fit to the entire Cen~A set of observations.

\subsubsection{Time Average Cen~A Spectra}

In order to set the stage for studying spectral variability of Cen~A in the next section, we present the results of a 
separate study of the time averaged spectra of 23 bright active galaxies observed by 
\textsl{RXTE} over the duration of the \textsl{RXTE} mission \citep{Rivers10}. 
In the case of Cen~A, the time average spectrum included all of the observations through 
2009 February. \citet{Rivers10} found a mean Cen~A power law index $\Gamma$=1.83$\pm$0.01, 
a mean $N_\mathrm{H}$=(1.69$\pm$0.03)$\times 10^{23}$ cm$^{-2}$, a mean iron line 
energy of 6.38$\pm$0.09 keV, and a mean iron line flux of (4.9$\pm 0.7) \times 10^{-4}$ photons cm$^{-2}$ s$^{-1}$. 
The upper limit to a Compton reflection component was \textsl{R}$<$0.5\%, and the lower limit to 
the high energy rollover was $>$490 keV. As will be demonstrated below, these values are consistent with the mean values found by averaging the 13 observation intervals, i.e., a mean power law index of 1.822$\pm$0.004, a mean iron line energy of 6.361$\pm$0.014 keV, and a mean iron line flux of (4.55$\pm$0.14) $\times$ 10$^{-4}$ photons cm$^{-2}$ s$^{-1}$.

\subsection{Spectral Variability}

\subsubsection{Months to Years}

\begin{figure}
\includegraphics[width=3.20in]{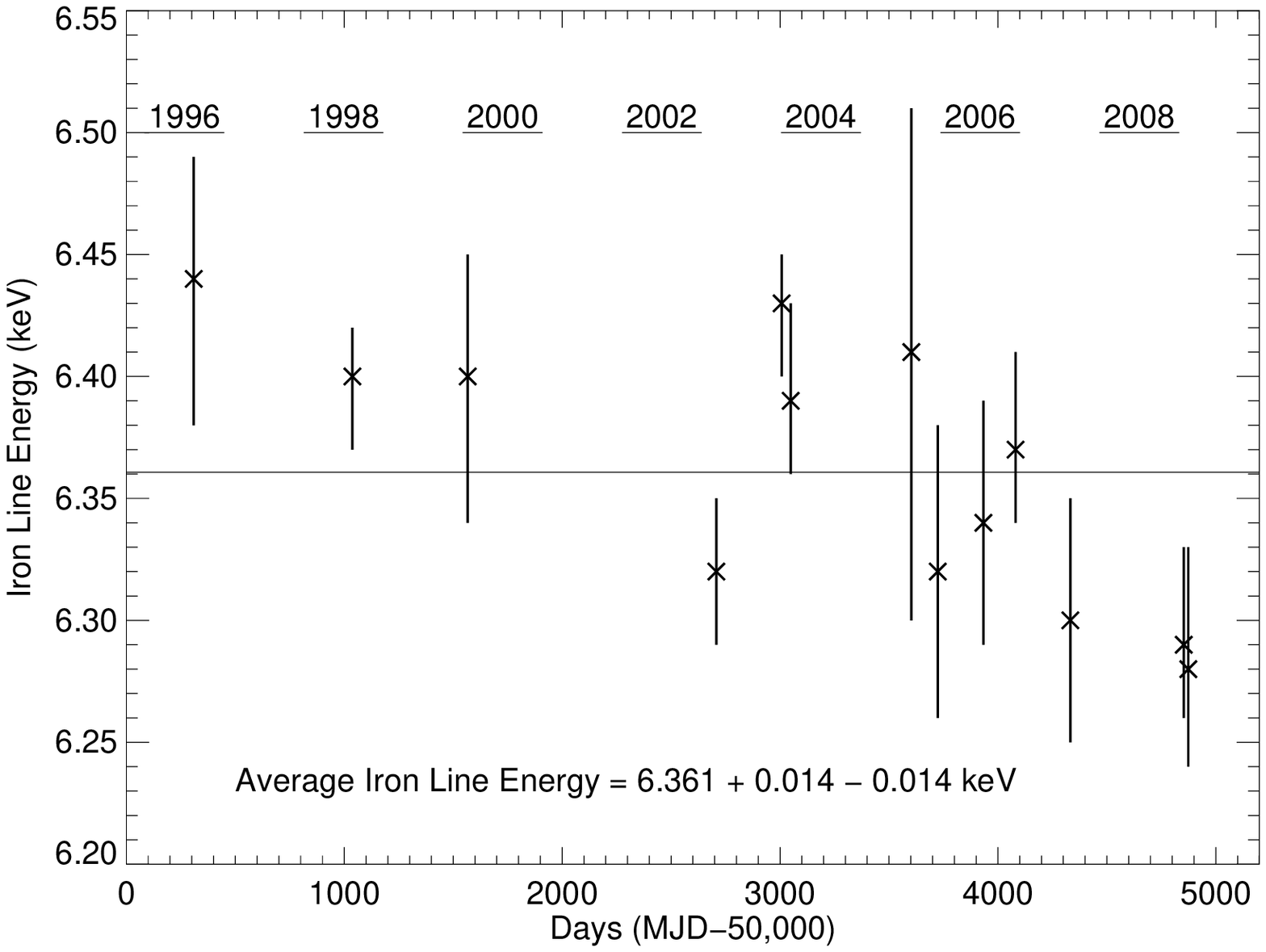}
\includegraphics[width=3.20in]{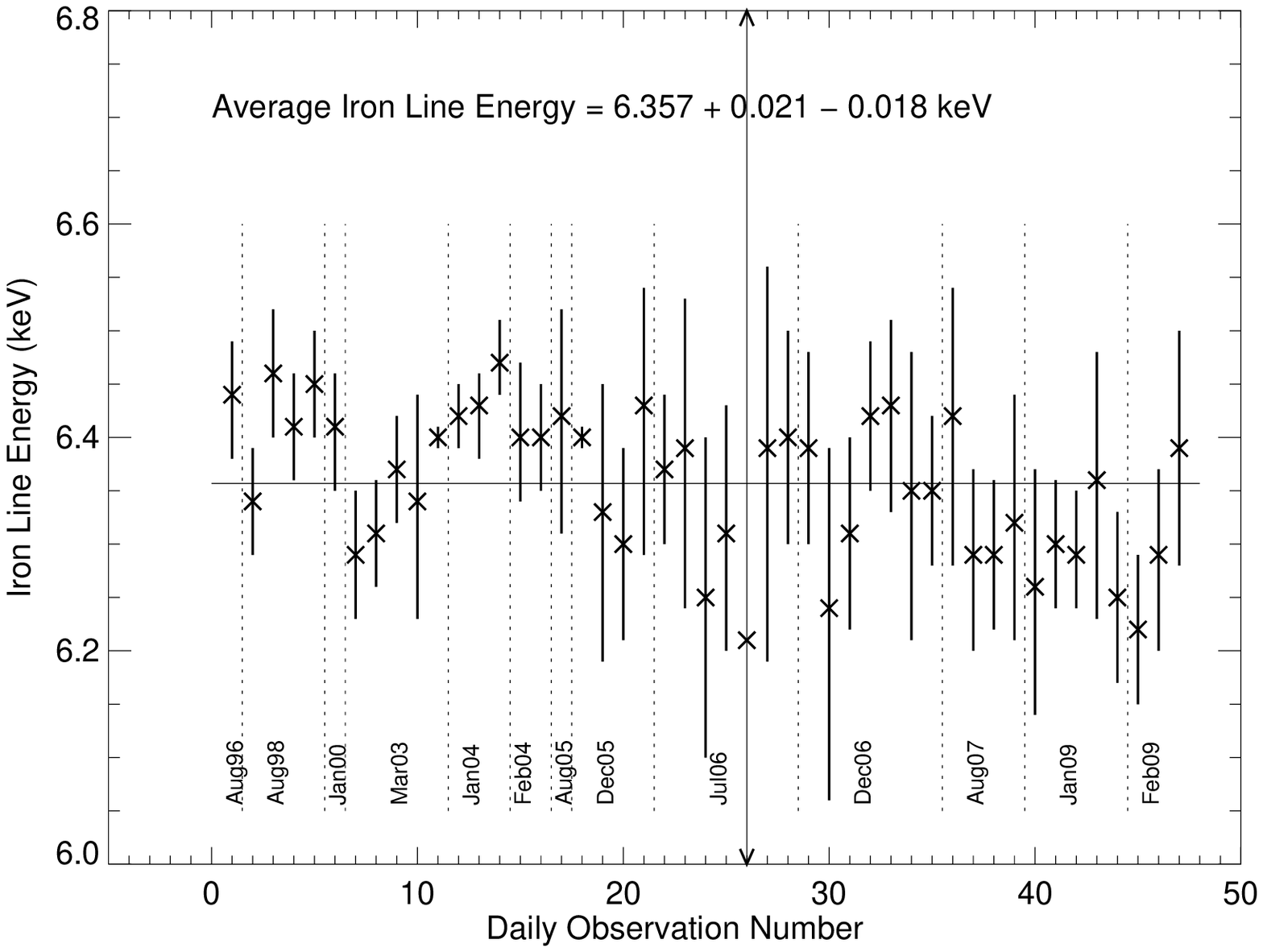}\\
\includegraphics[width=3.20in]{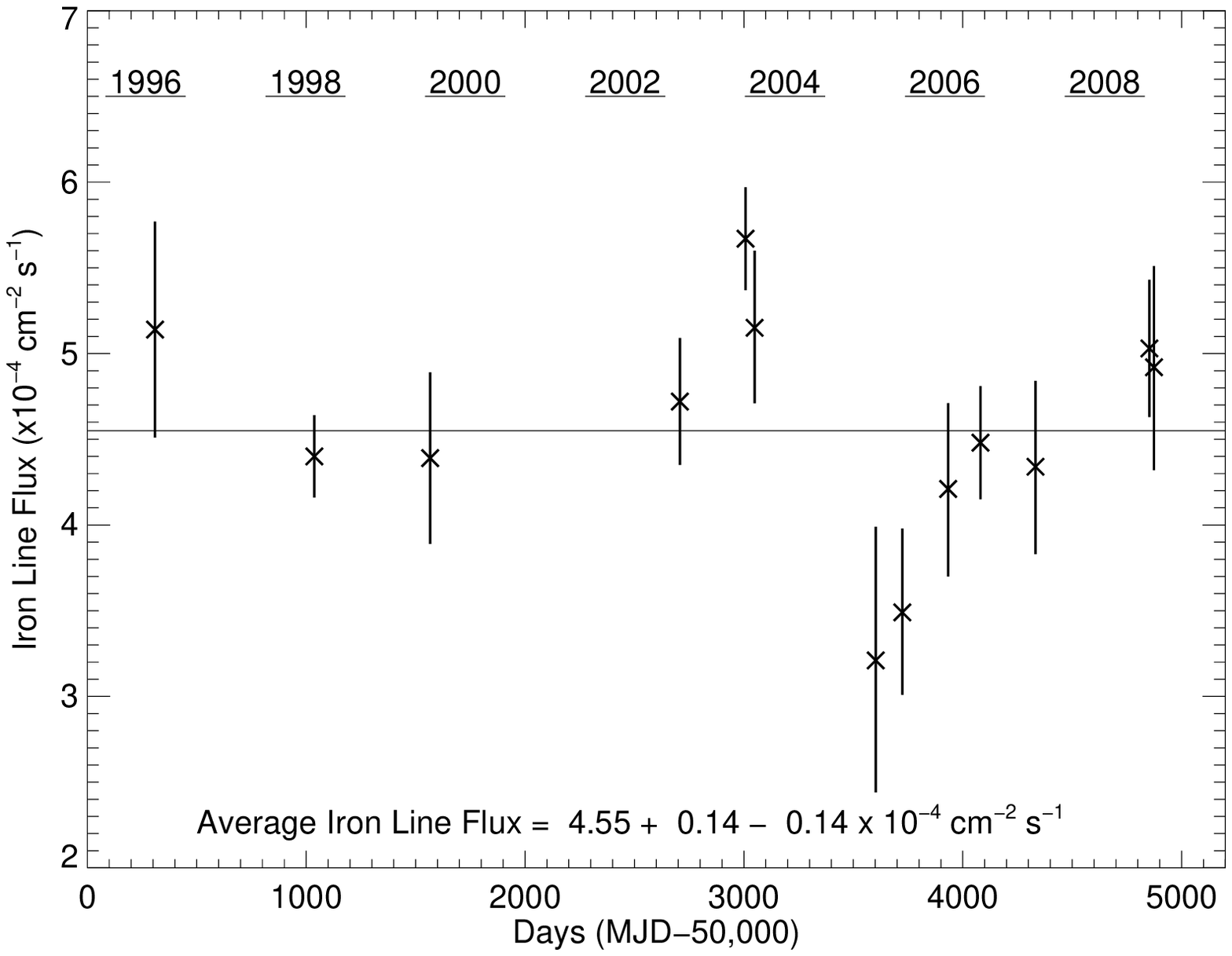}
\includegraphics[width=3.20in]{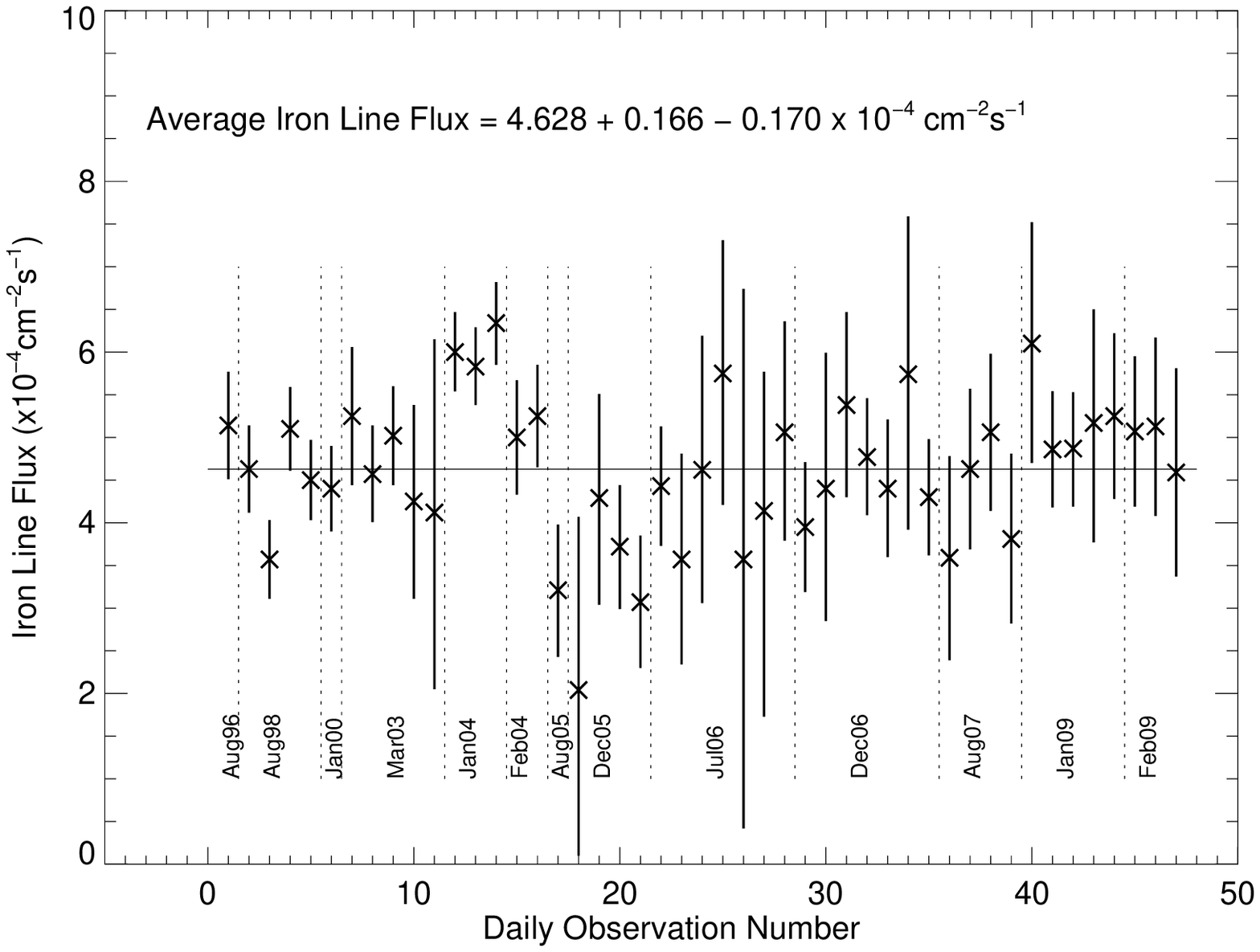}\\
\includegraphics[width=3.20in]{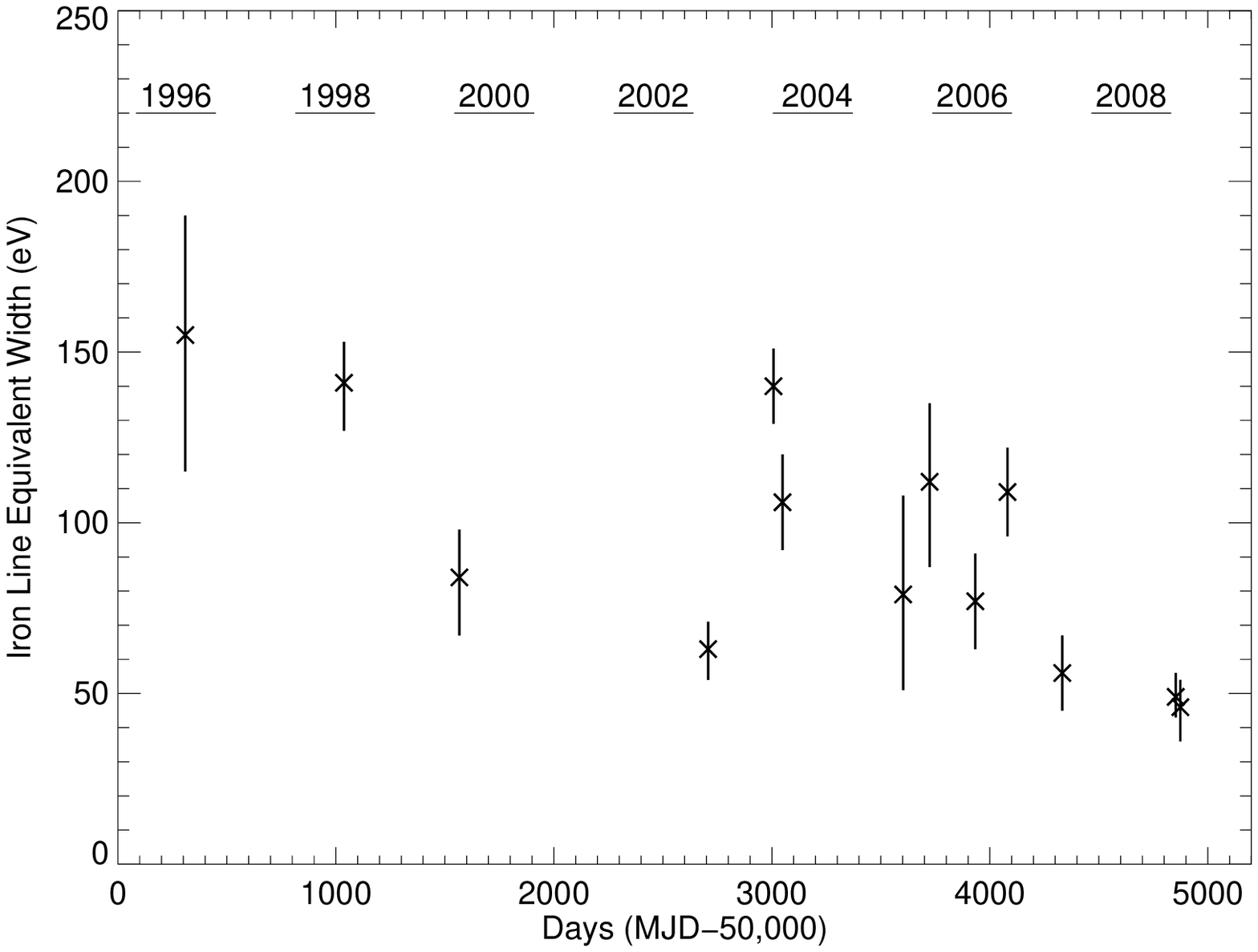}
\includegraphics[width=3.20in]{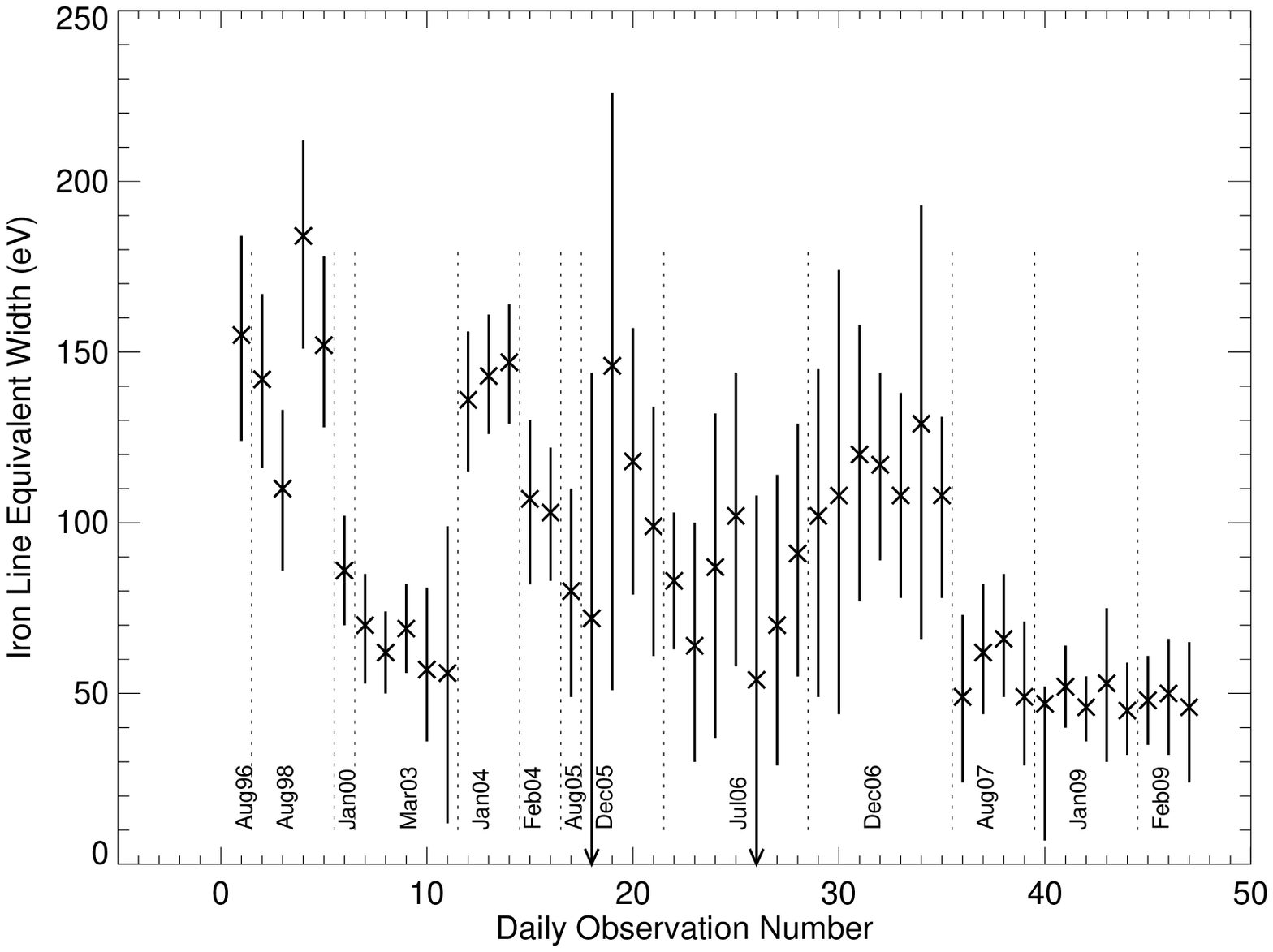}\\
\caption{The best-fit iron line energy (Top), iron line flux (Middle), and equivalent width (Bottom) spectral 
parameters resulting from fitting the 13 separate observation interval spectra and the 47 daily spectra. 
Error bars represent 90\% uncertainties. Average values plus 90\% uncertainties are given for the iron line 
energy and flux. Years and observing intervals are shown.\label{fig:iron}}
\end{figure}

\begin{figure}
\includegraphics[width=3.5in]{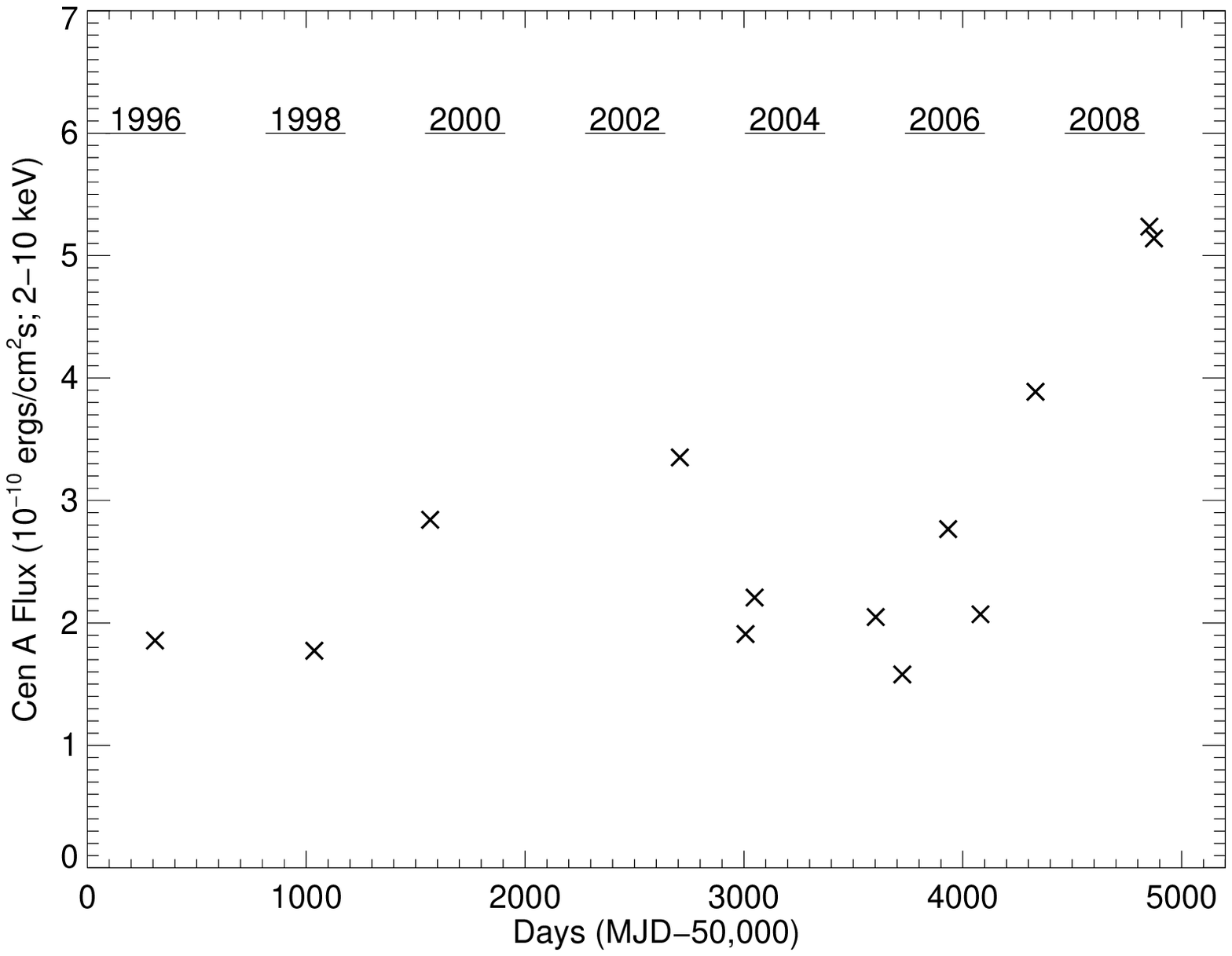}
\includegraphics[width=3.5in]{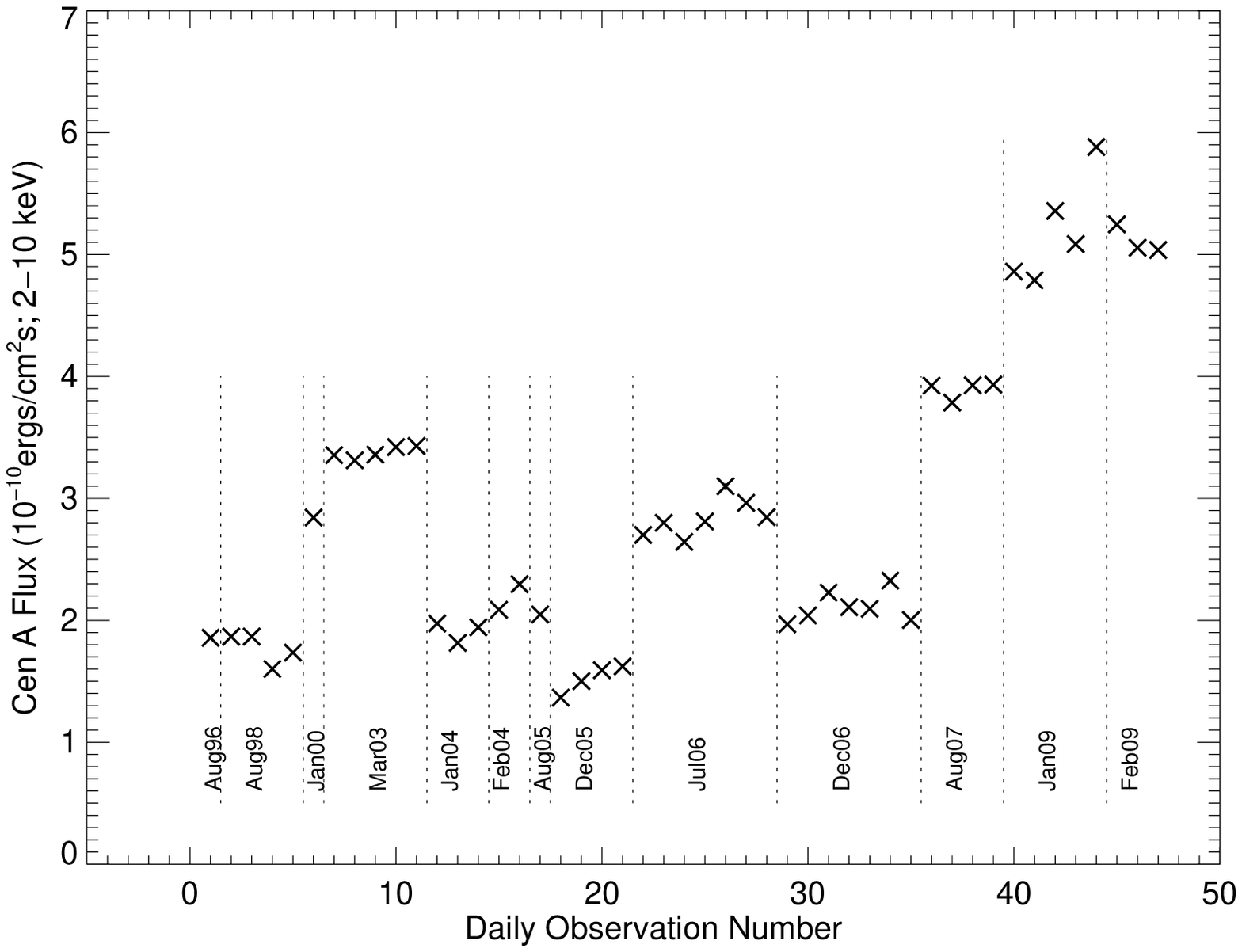}\\
\includegraphics[width=3.5in]{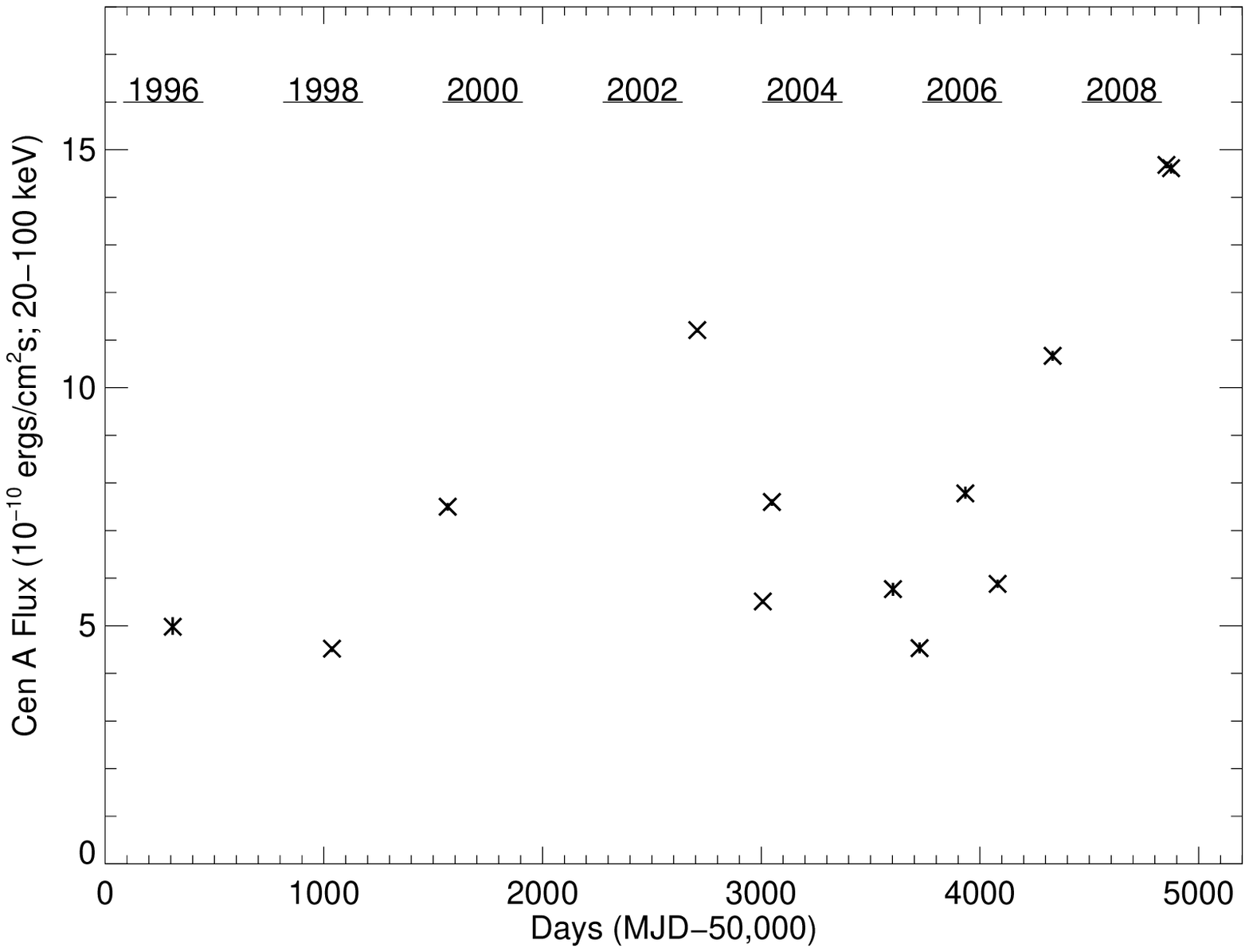}
\includegraphics[width=3.5in]{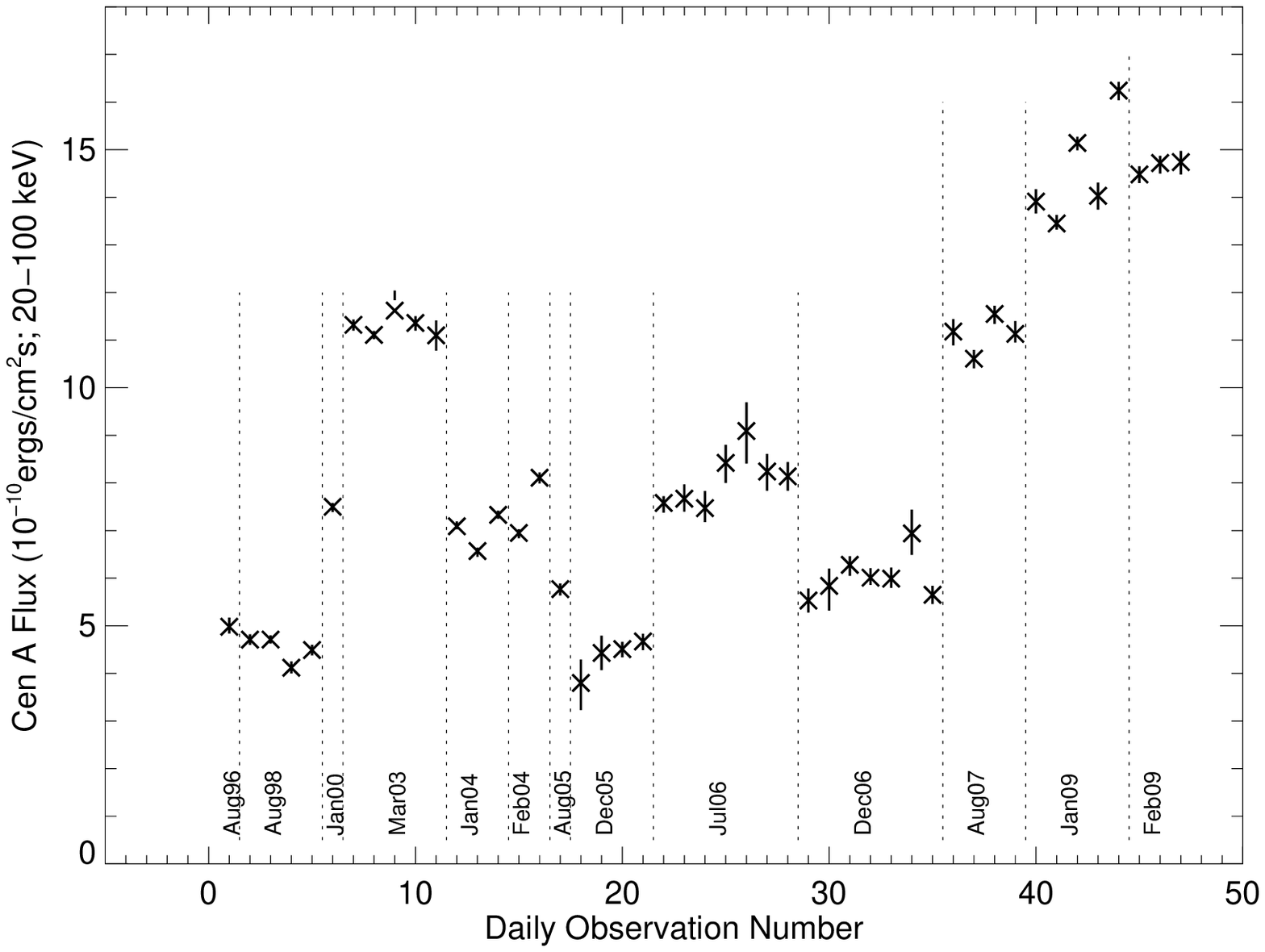}
\caption{The \textsl{RXTE} Cen~A 2--10 keV (Top) and 20--100 keV (Bottom) fluxes for 
the 13 observation intervals and the 47 daily spectra. Years and observing intervals are 
shown. Error bars representing 90\% uncertainties
are generally smaller than the symbols.\label{fig:flux}}
\end{figure}

\begin{figure}
\includegraphics[width=3.25in]{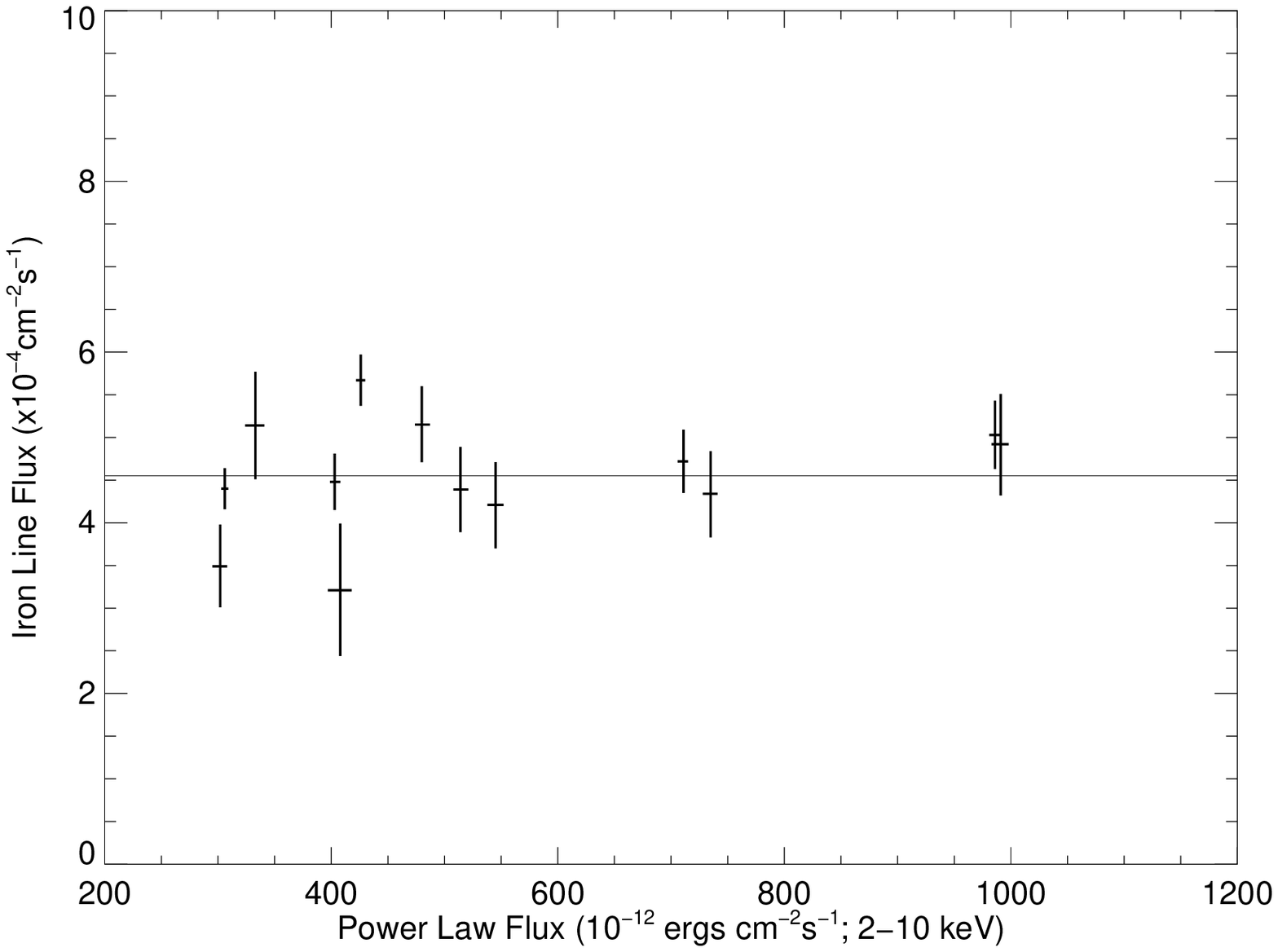}
\includegraphics[width=3.25in]{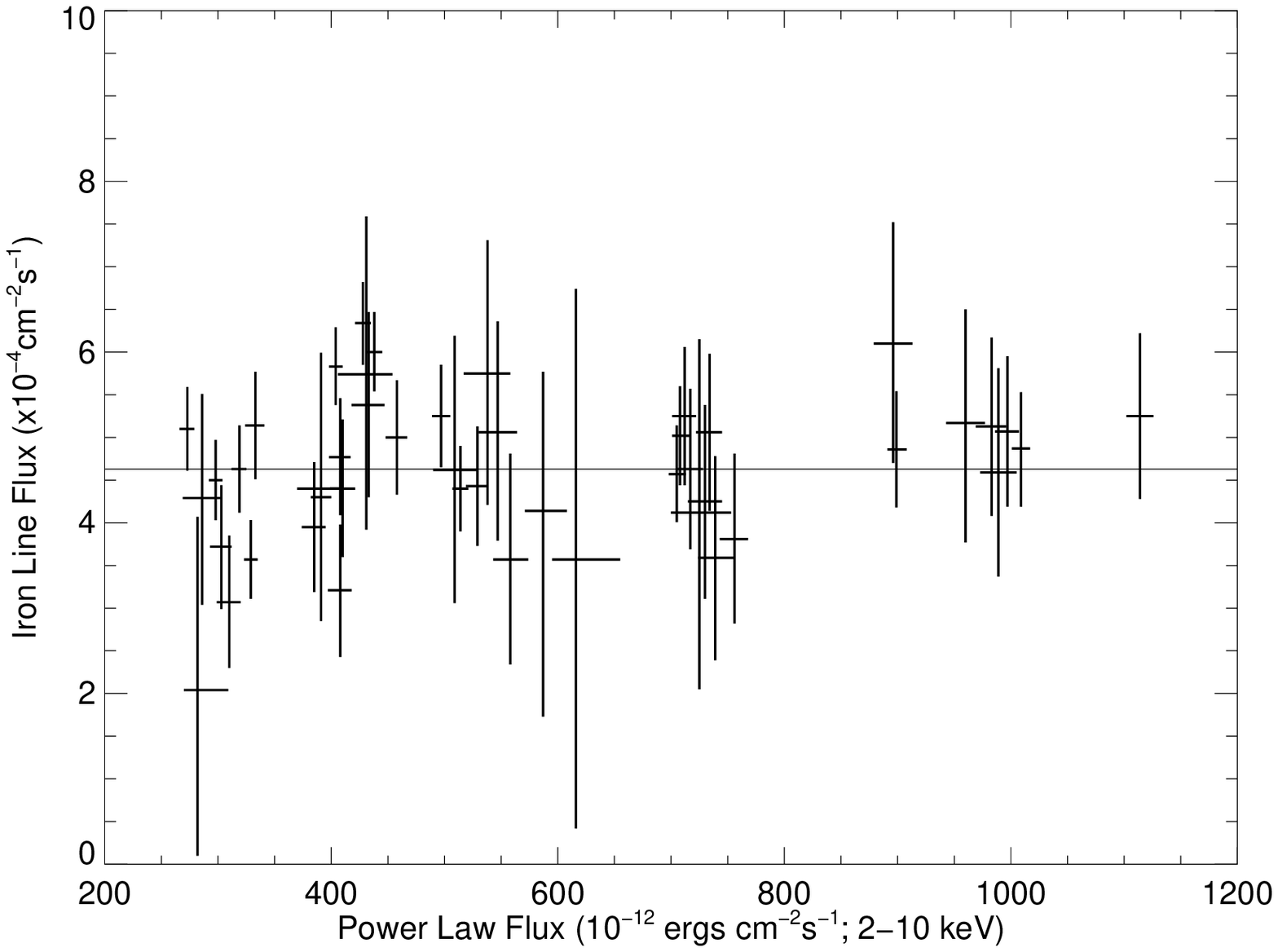}\\
\includegraphics[width=3.25in]{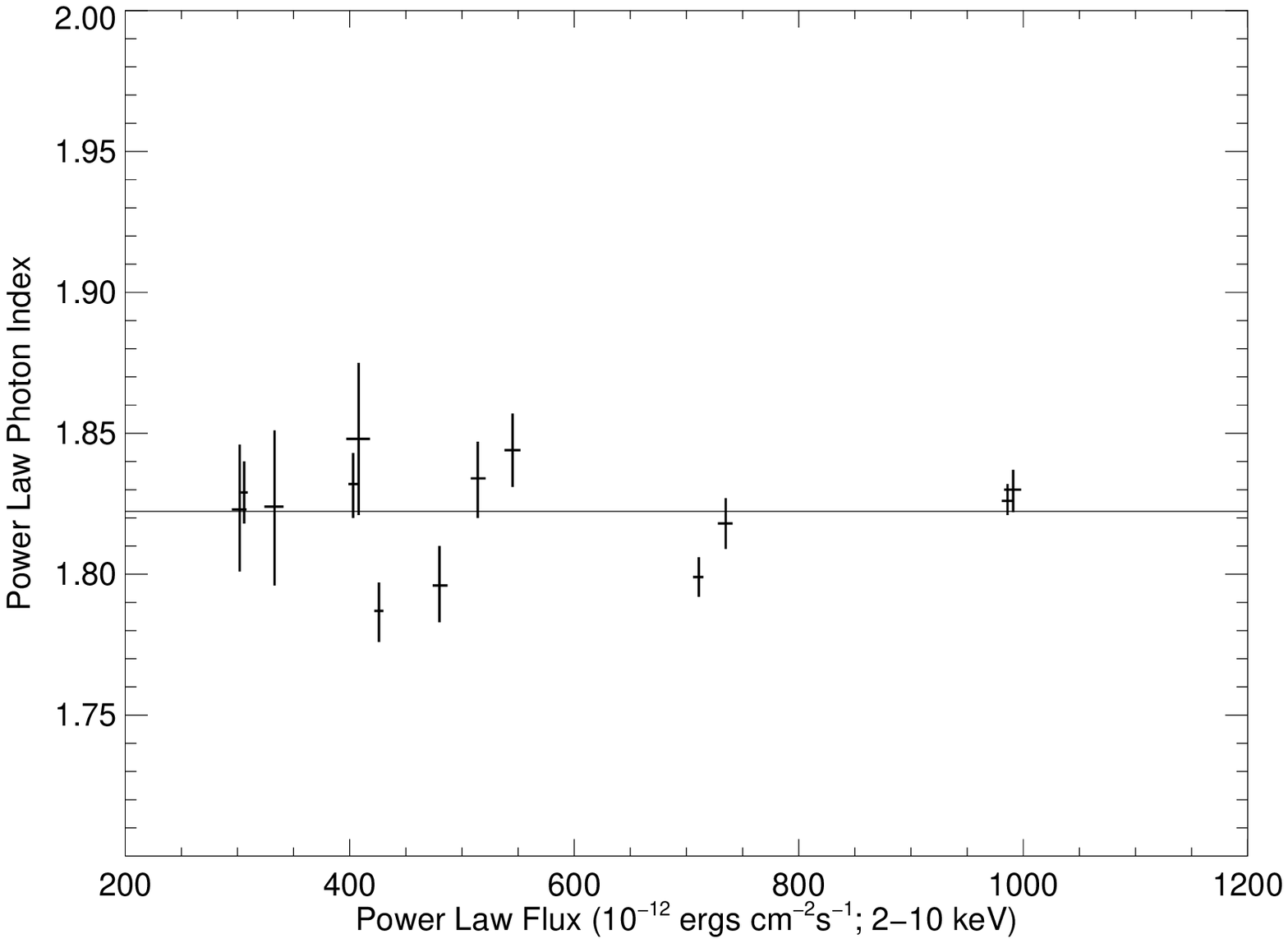}
\includegraphics[width=3.25in]{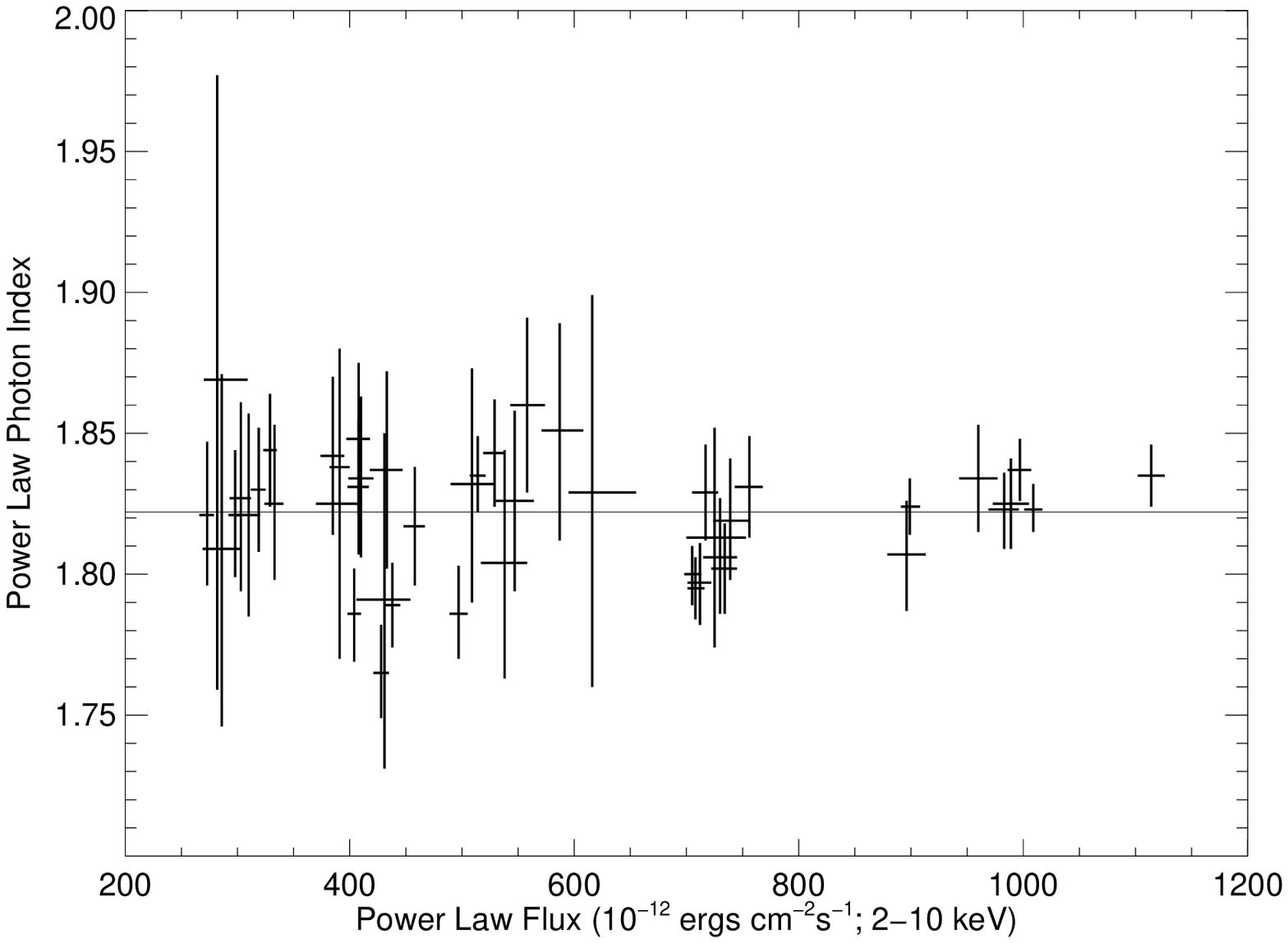}\\
\caption{The iron line flux (Top) and power law index (Bottom) as a function of unabsorbed power law 
flux for the 13 observation intervals (Left) and 45 daily spectra (Right). The average iron line flux (Top) 
and average power law index (Bottom) are shown as a horizontal lines. Error bars represent 90\% uncertainties. 
\label{fig:versus}}
\end{figure}

\begin{figure}
\includegraphics[width=3.25in]{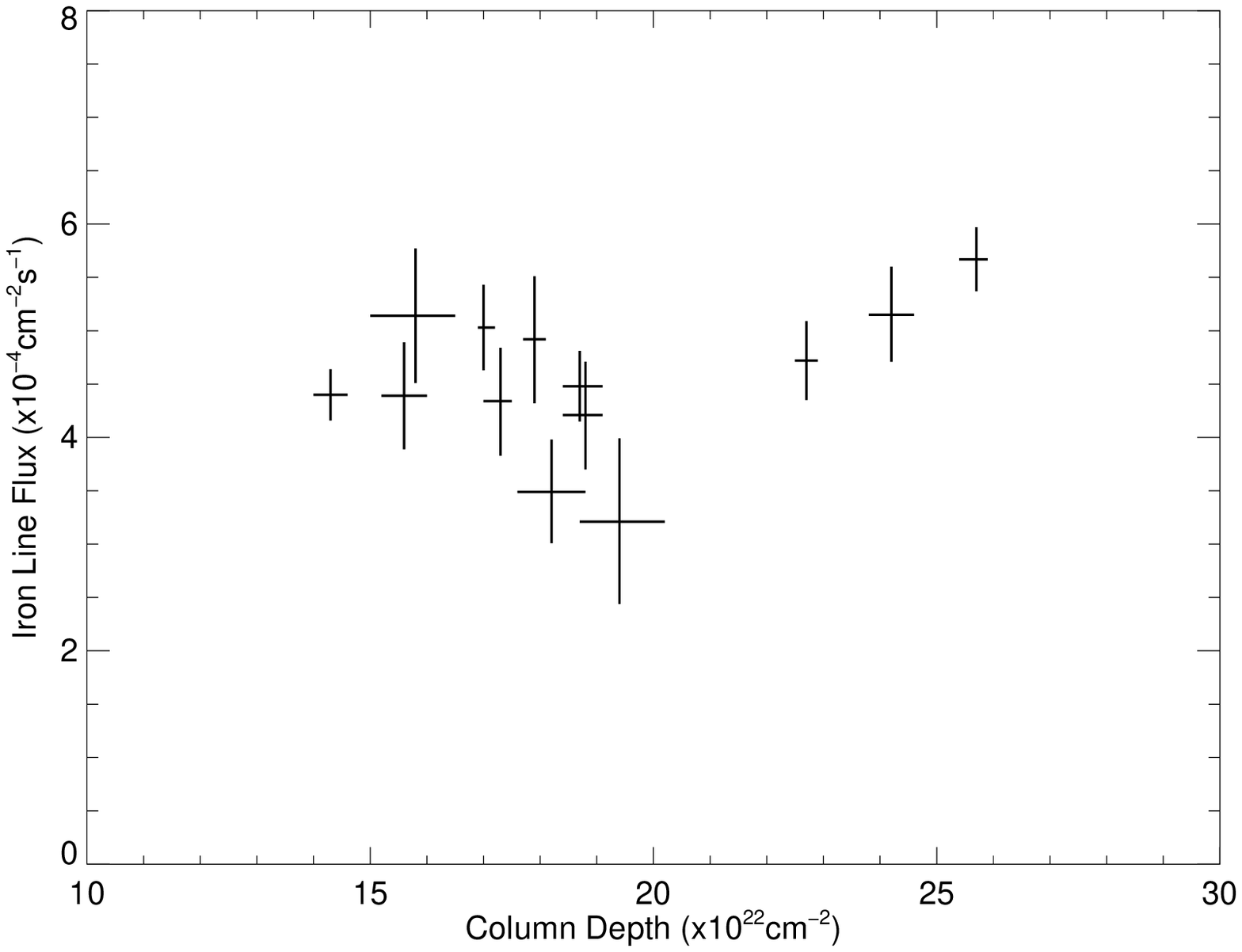}
\includegraphics[width=3.25in]{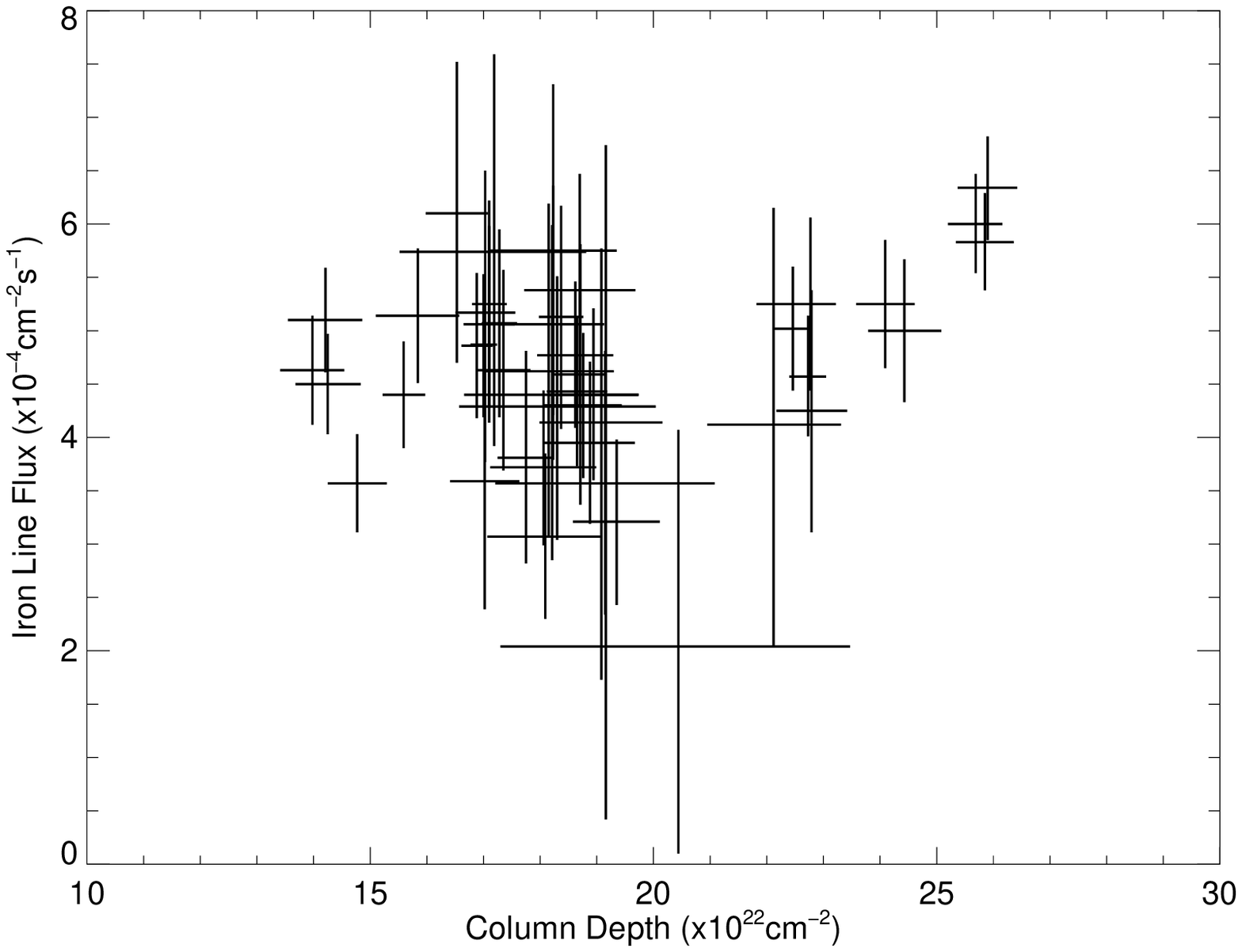}\\
\includegraphics[width=3.25in]{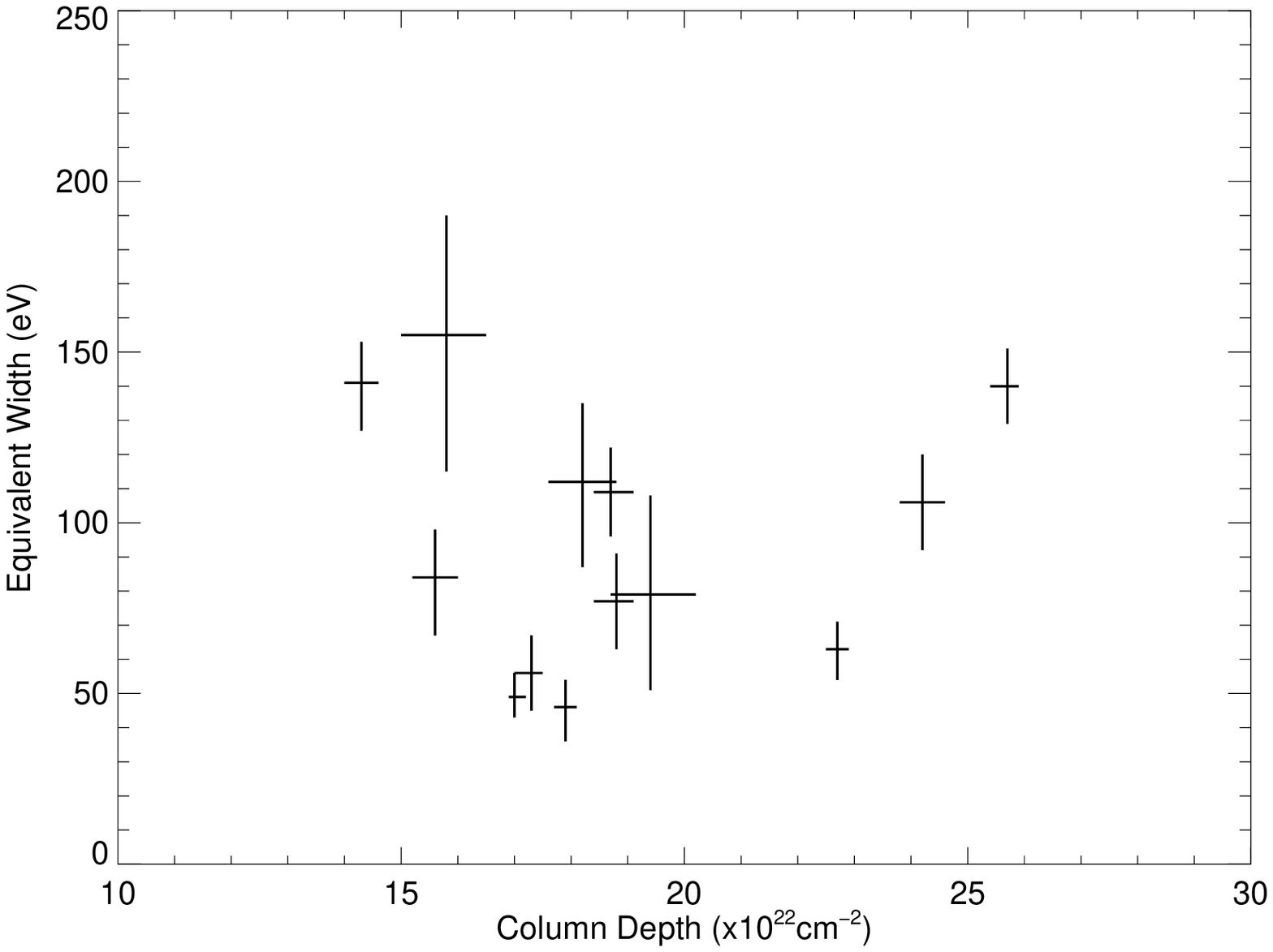}
\includegraphics[width=3.25in]{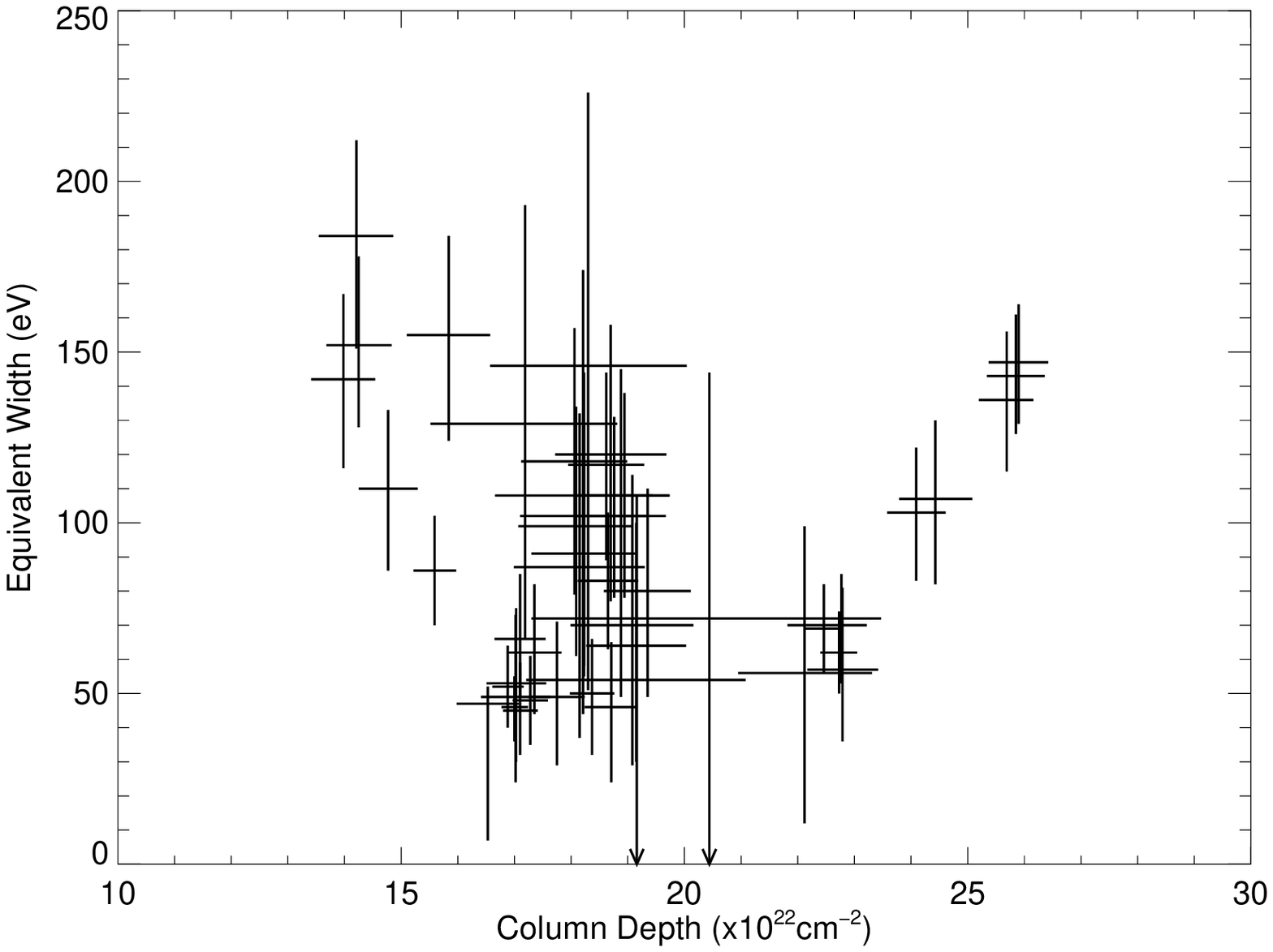}
\caption{The iron line flux as a function of the low energy column depth  (Top) and the iron line 
equivalent width as a function of the column depth (Bottom) for the 13 observation intervals (Left) 
and 47 daily spectra (Right). Error bars represent 90\% uncertainties. \label{fig:eqw_nh}}
\end{figure}

Individual Cen~A observations spanned intervals from less than 1 ks to a few hours, with separations 
between observations of days to 3 years. Consequently the spectral analysis could be performed on a
range of temporal scales. As will be shown, the best fit parameters of the 
day-to-day observations did not show large variability and thus only 
the best-fit parameters from the 13 observational intervals are given in Table~\ref{tab:13obs}. Figures~\ref{fig:power} to 
\ref{fig:eqw_nh} display the 13 sets of best fit spectral parameters for the 
observation intervals on the left and those from fitting the individual days 
on the right. Figure~\ref{fig:power} shows the best-fit unabsorbed power law 
2--10 keV flux, column density, and power law index; Fig.~\ref{fig:iron} shows 
the best-fit iron line energy, flux, and equivalent width, and Fig.~\ref{fig:flux} 
shows the 2--10 keV and 20--100 keV total fluxes from Cen~A. 

The inferred line of sight column density experienced nearly a factor of two in 
variability, with an interval of high density in the 2003 March to
2004 February data and lower density otherwise. The power law index
remained essentially constant ($\Gamma$=1.822$\pm$0.004) over 
12.5 years, with a slight ($\sim$2\%) reduction at the time of the increased 
column density. The best-fit iron line energy values (6.361$\pm$0.014 keV)
were consistent with fluorescence of cold/neutral material at 6.4 keV, and 
the iron line flux appears to vary minimally over the observing decade with a mean value of 
4.55$\pm$0.14 photons cm$^{-2}$ s$^{-1}$. 
The equivalent width of the iron emission ranged from 46$^{+10}_{-8}$ to 155$^{40}_{-35}$ eV, 
and the variation was as expected for a constant flux and varying power 
law continuum responsible for the fluorescing flux. 

Figure~\ref{fig:versus} (Top) shows the values of the
iron line flux plotted versus the power law normalization for the 13 observing 
intervals (Left) and the 47 daily measurements (Right). The average iron line flux is given by 
the horizontal line indicating no overall dependence of the iron line flux with 
the power law normalization. Similarly, the power law index showed no systematic 
variation with instantaneous power law flux (Fig.~\ref{fig:versus}-Bottom). 
The very sparse sampling of the \textsl{RXTE} observations over 12.5 years prevented a meaningful 
conclusion vis-\`a-vis the relation between the power law emission and fluorescent 
iron line fluxes, i.e., any time delay between power law and iron line fluxes. 

\begin{figure}
\includegraphics[width=3.25in]{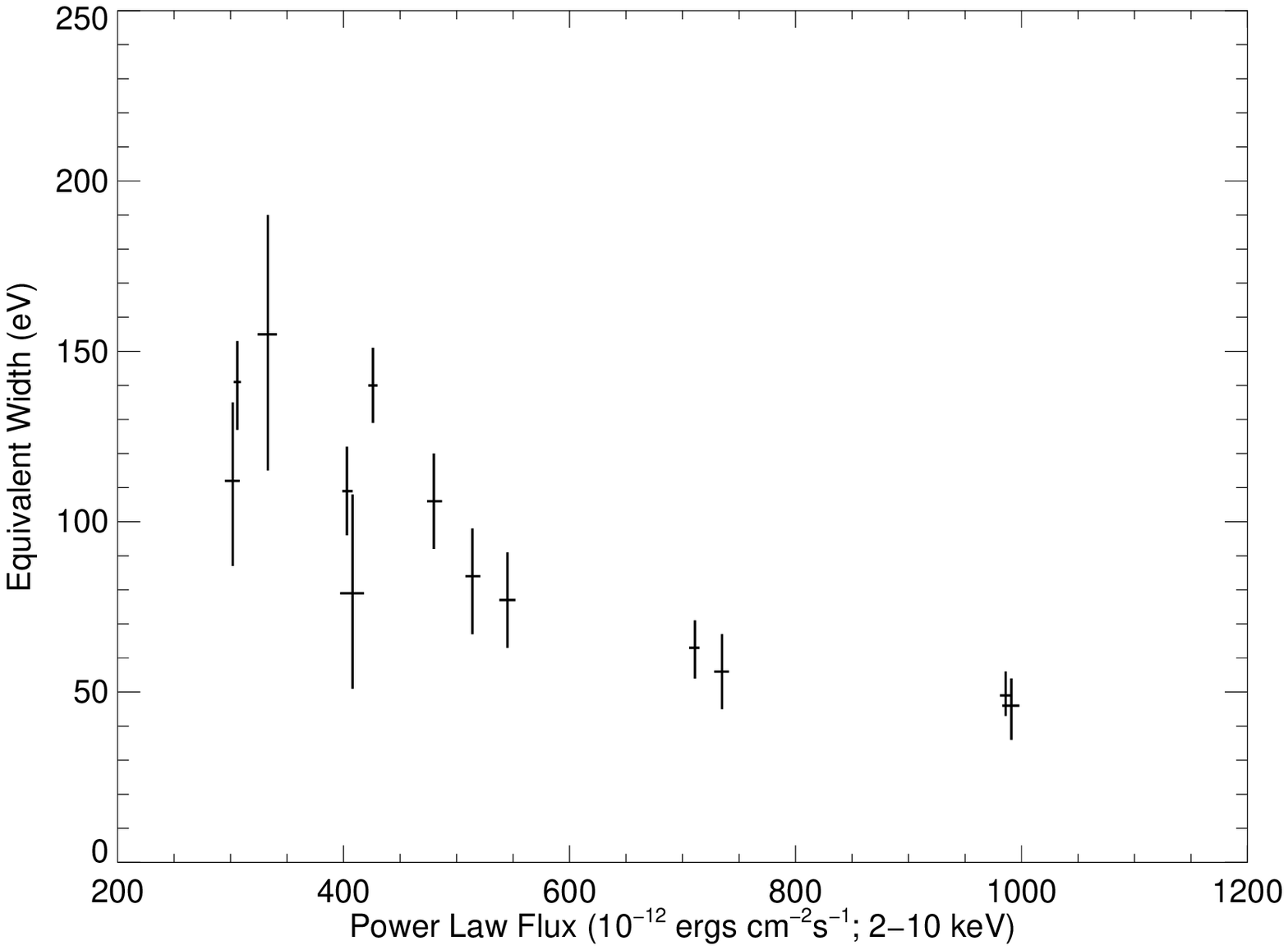}
\includegraphics[width=3.25in]{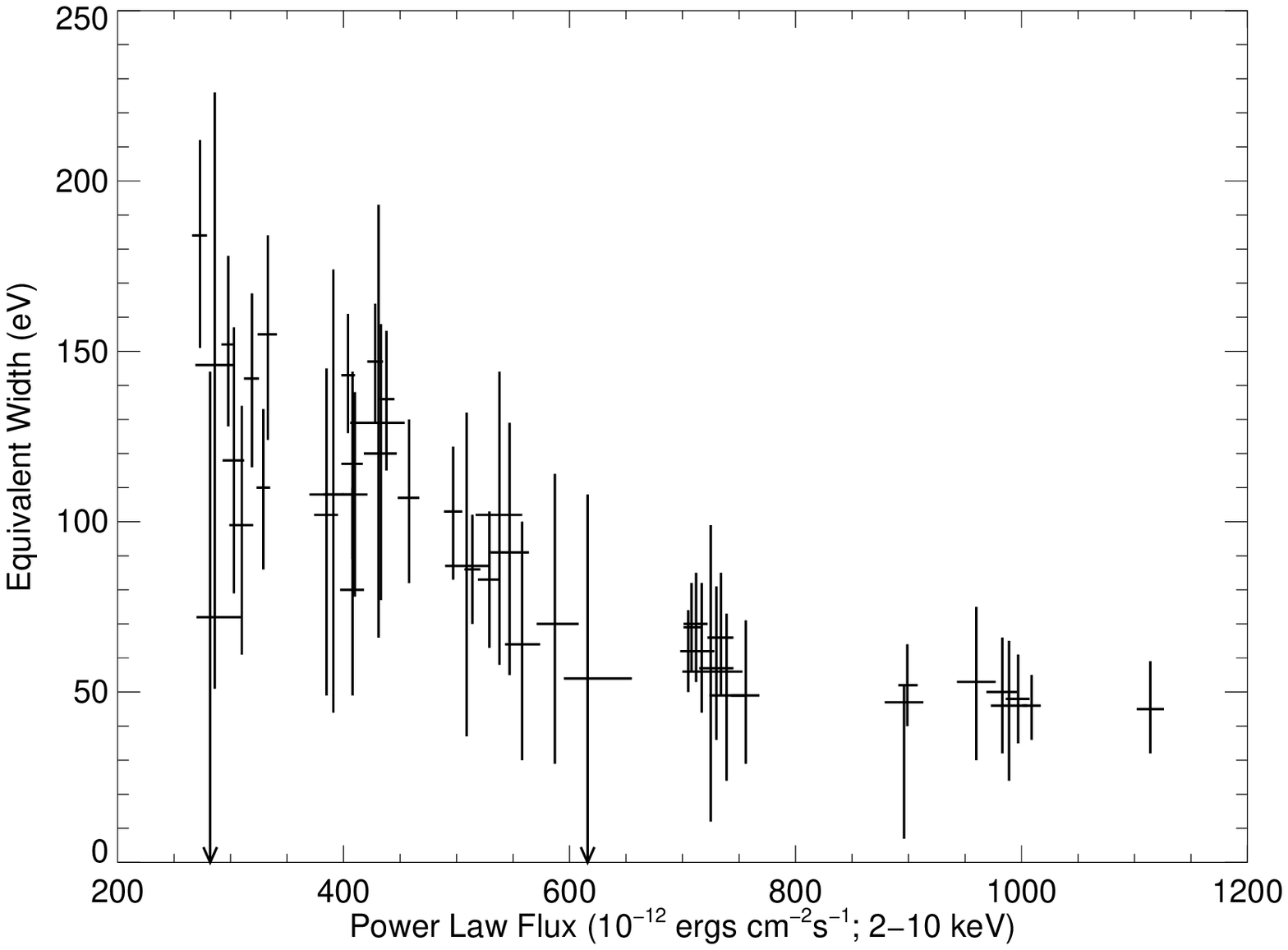}\\
\caption{The equivalent width of the iron line is presented as a function of the unabsorbed power 
law 2-10 keV flux. The anti-correlation between the two parameters is clear, which indicates that 
the iron line was not responding to the power law continuum on the timescales available.\label{fig:eqw_norm}}
\end{figure}

\begin{figure}
\includegraphics[width=3.0in]{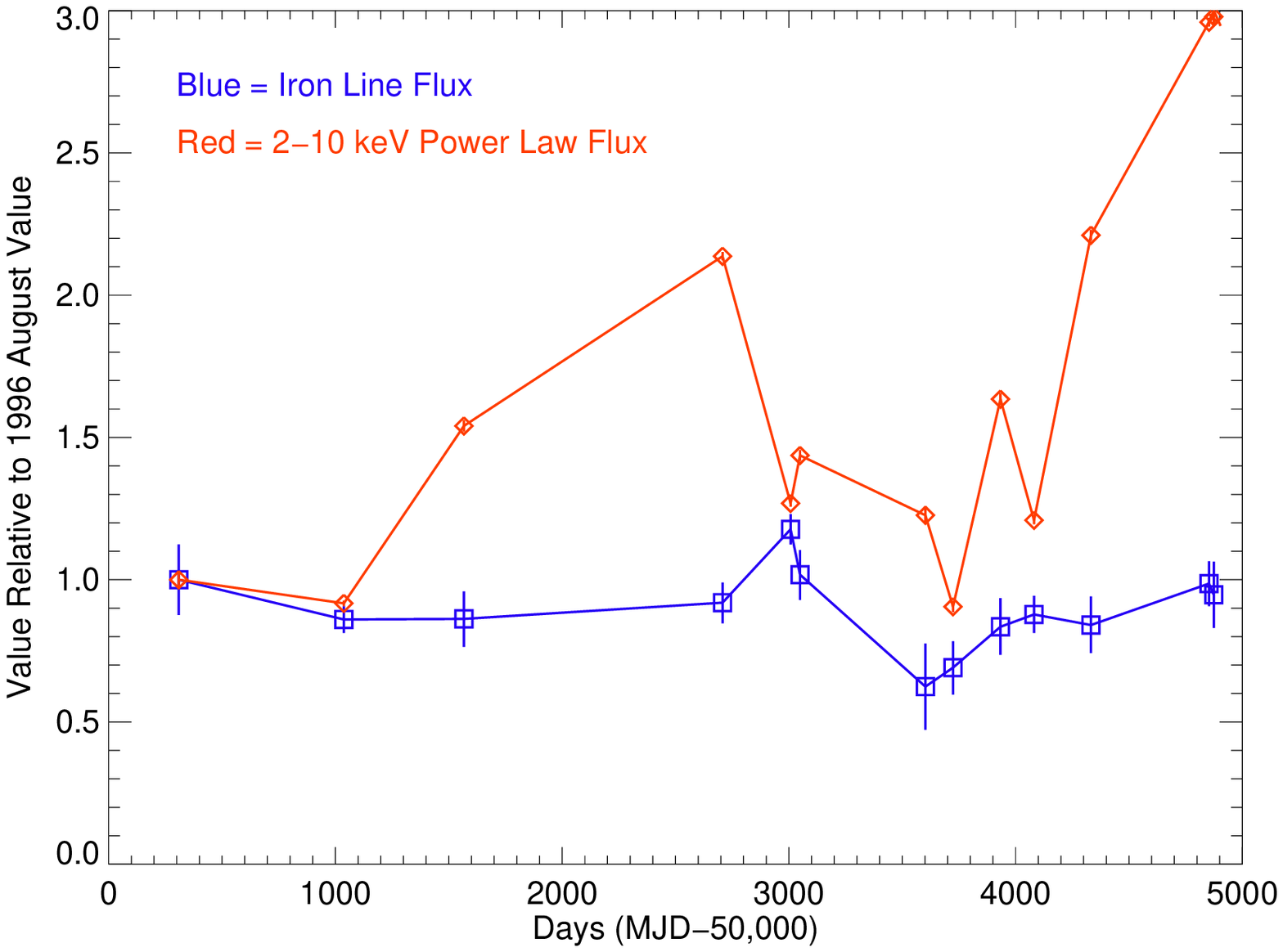}\\
\includegraphics[width=3.0in]{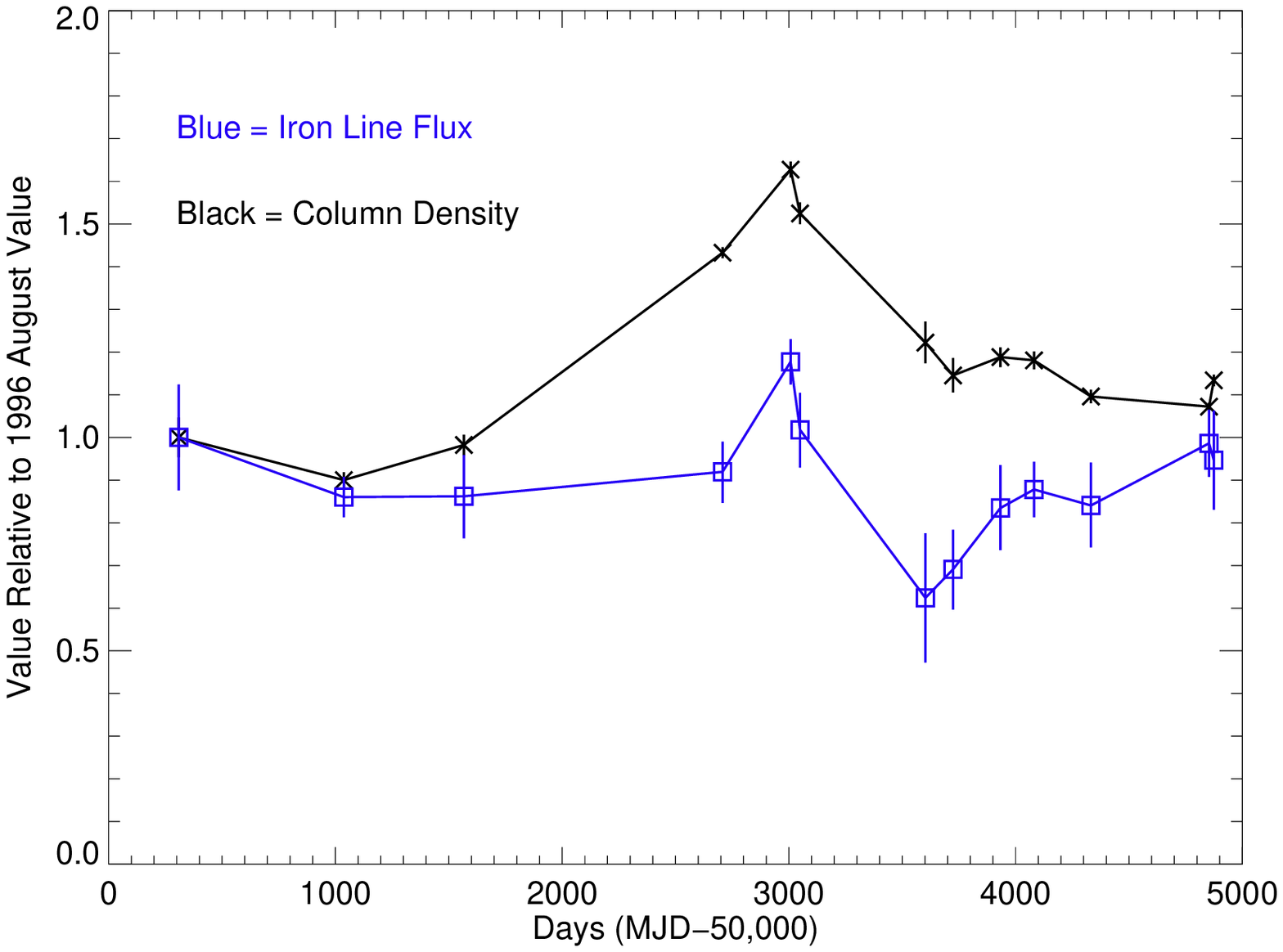}\\
\includegraphics[width=3.0in]{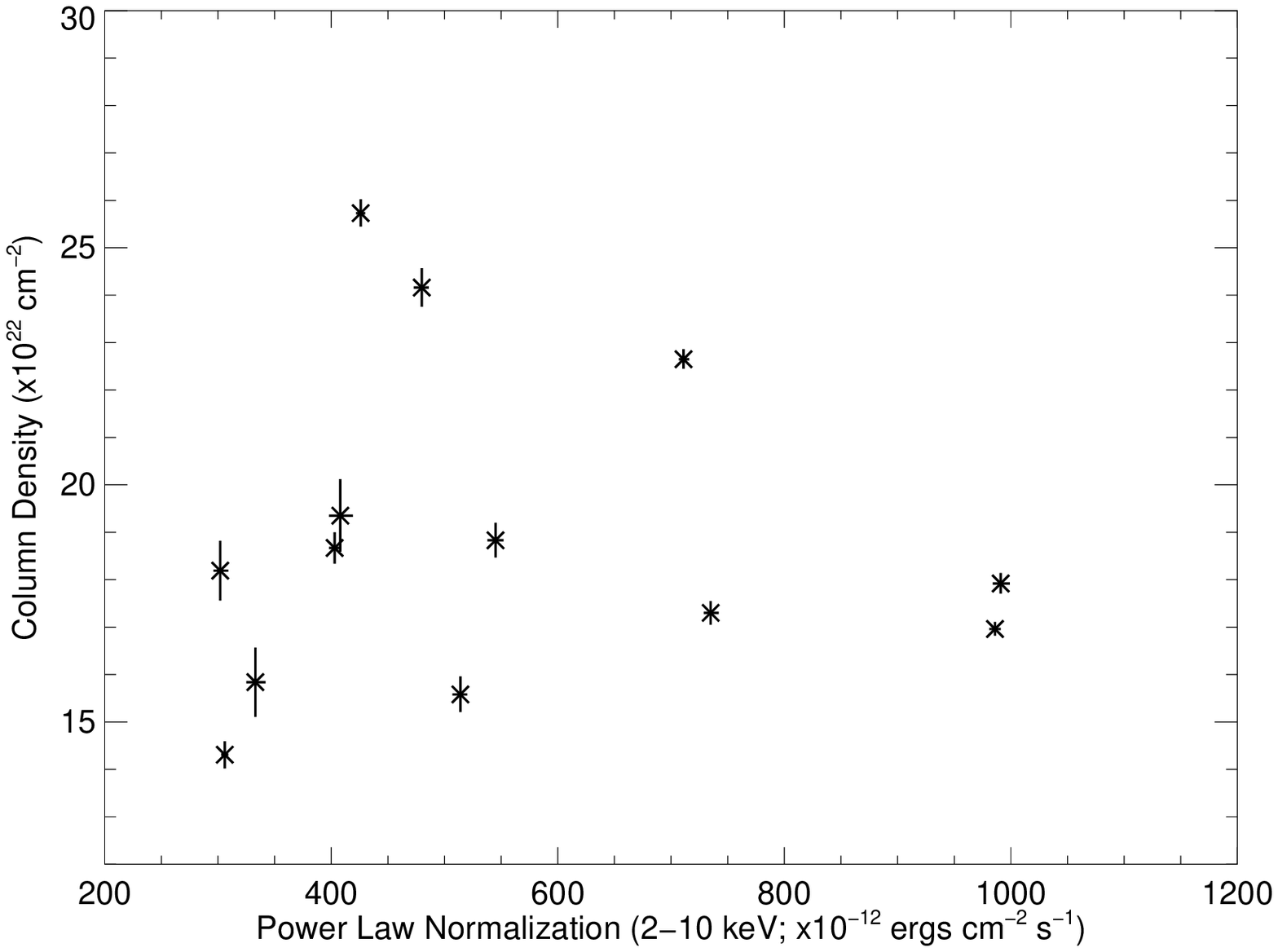}
\caption{The variation in the iron line flux and the 2--10 keV unabsorbed power law flux (Top) and 
the column density and iron line flux (Middle) as a function of time. The values are normalized to 
the first observation in 1996 August. The lack of correlation is clear in both plots. {\bf The column density as a function of the power law flux (Bottom) shows no correlation between the two parameters.}\label{fig:fe_norm}}
\end{figure}

Figure~\ref{fig:eqw_nh} (Top) shows the variations in iron line flux versus column 
depth for the 13 observing intervals and for the daily observations, while Fig.~\ref{fig:eqw_nh} 
(Bottom) displays the equivalent width of the iron line versus column depth. 
Fig.~\ref{fig:eqw_norm} shows that the equivalent width of the iron line is 
inversely correlated with the power law flux, which is to be expected if the iron line flux 
is essentially constant with varying power law flux.
Figure~\ref{fig:fe_norm} reveals that there is no temporal correlation of the iron 
line with either the 2--10 keV power law flux or the column density. {\bf In addition, 
Figure~\ref{fig:fe_norm} (Bottom) plots the measured column density versus the 
measured power law flux, clearly showing that no correlation exists between these 
two parameters also.} Hence, we find no correlations between any of the spectral 
parameters, other than the anti-correlation 
of equivalent width of the iron line and unabsorbed power law flux.

\subsubsection{Day to Day}
\label{sec:daily}

We have combined data from individual ObsIds to produce 47 daily spectra from
1996 August 8 to 2009 February 21. As can be seen in Table~\ref{tab:lctimes},
several sets of one or more contiguous days are present. The data from late 2009 January 15 and early 
January 16 have been combined into a single day since the data is effectively 
contiguous. Similarly the data for 2009 January 18, 19, and 20 have been combined, 
as well as 2009 January 25, 26, 27, 28. From these data sets 
we can derive indications of the variations from day to day exhibited by 
Cen~A. Additionally, one can compare spectral parameters from observations a few days apart.

\begin{figure}
\includegraphics[width=3.25in]{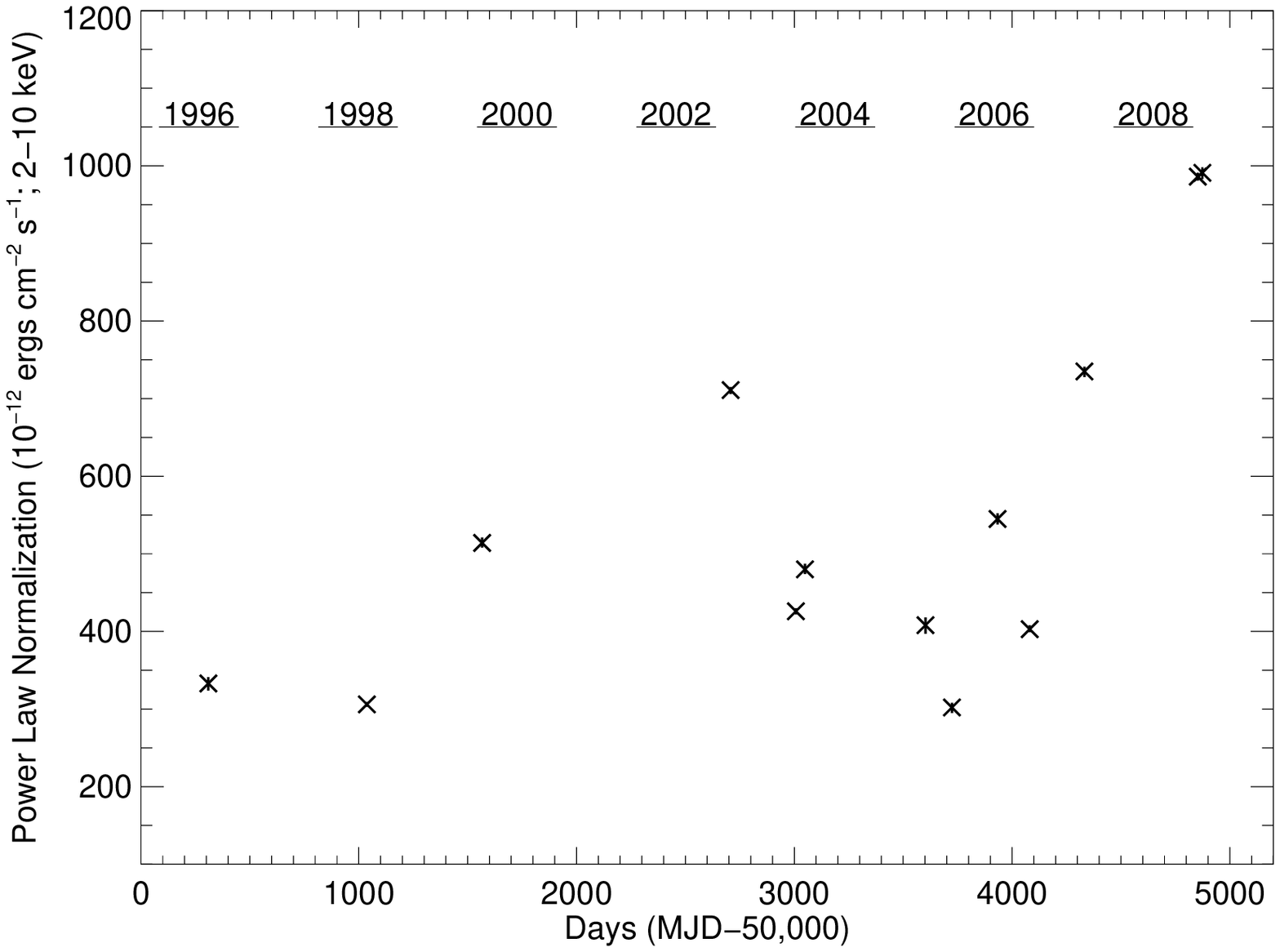}
\includegraphics[width=3.25in]{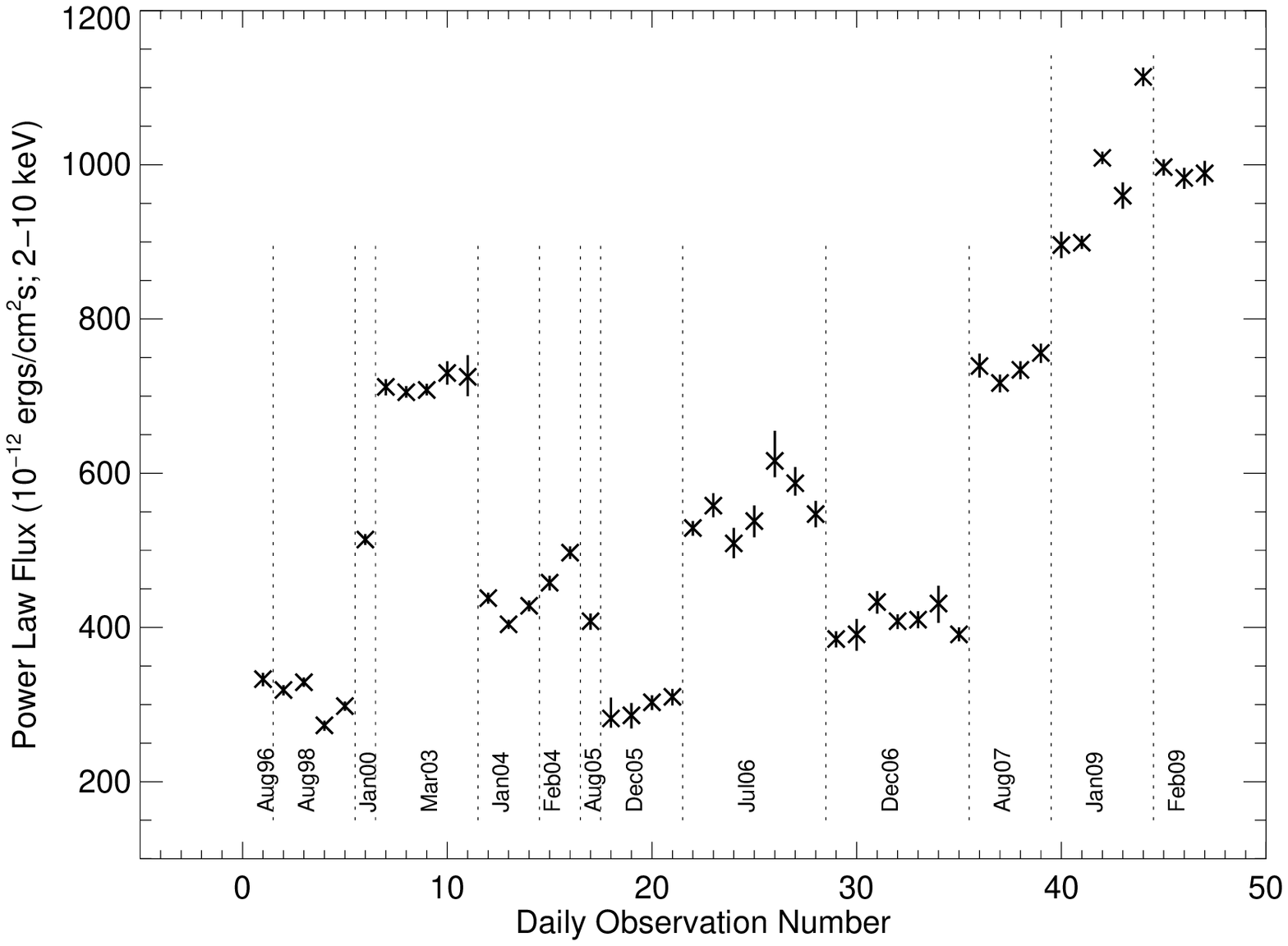}\\
\includegraphics[width=3.25in]{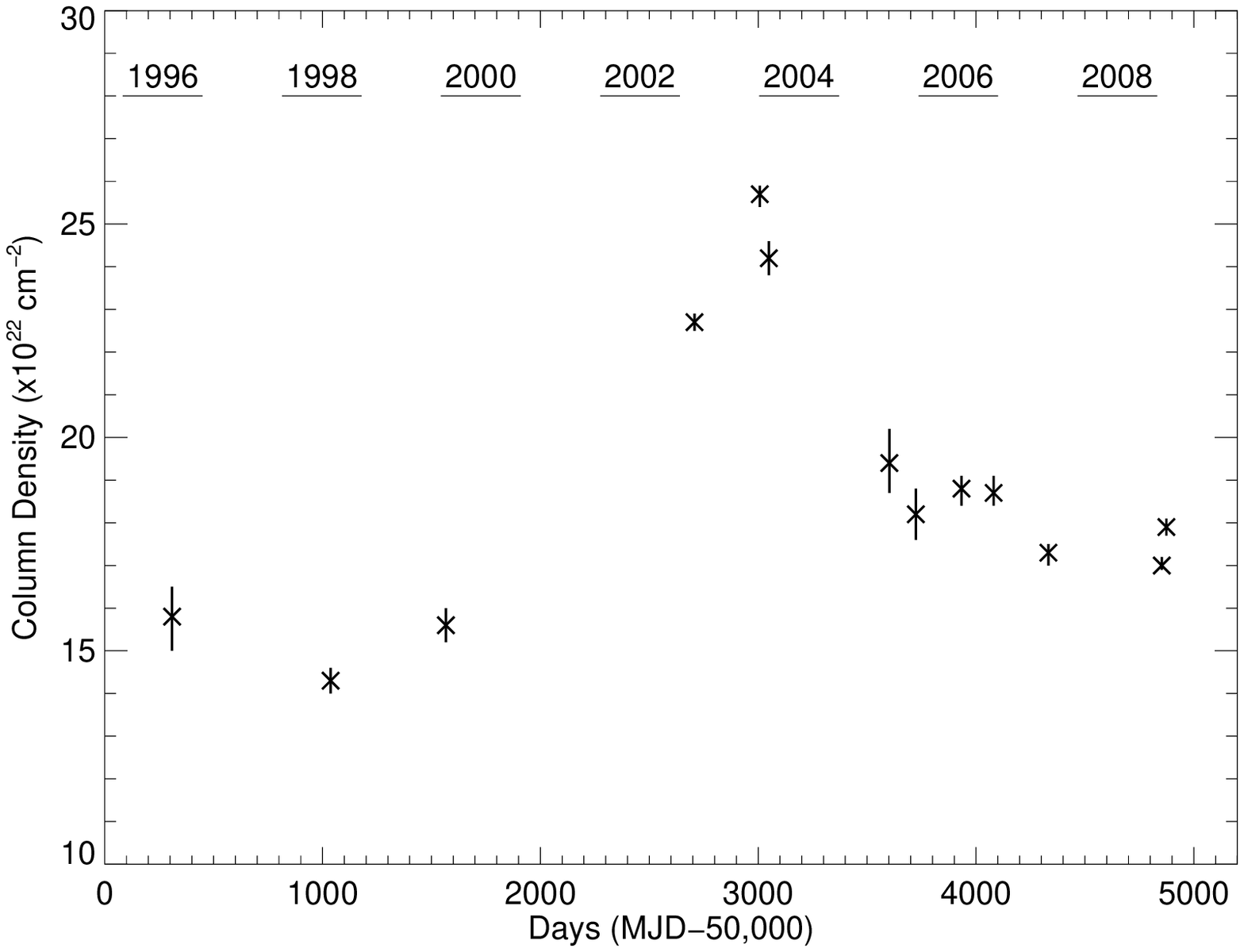}
\includegraphics[width=3.25in]{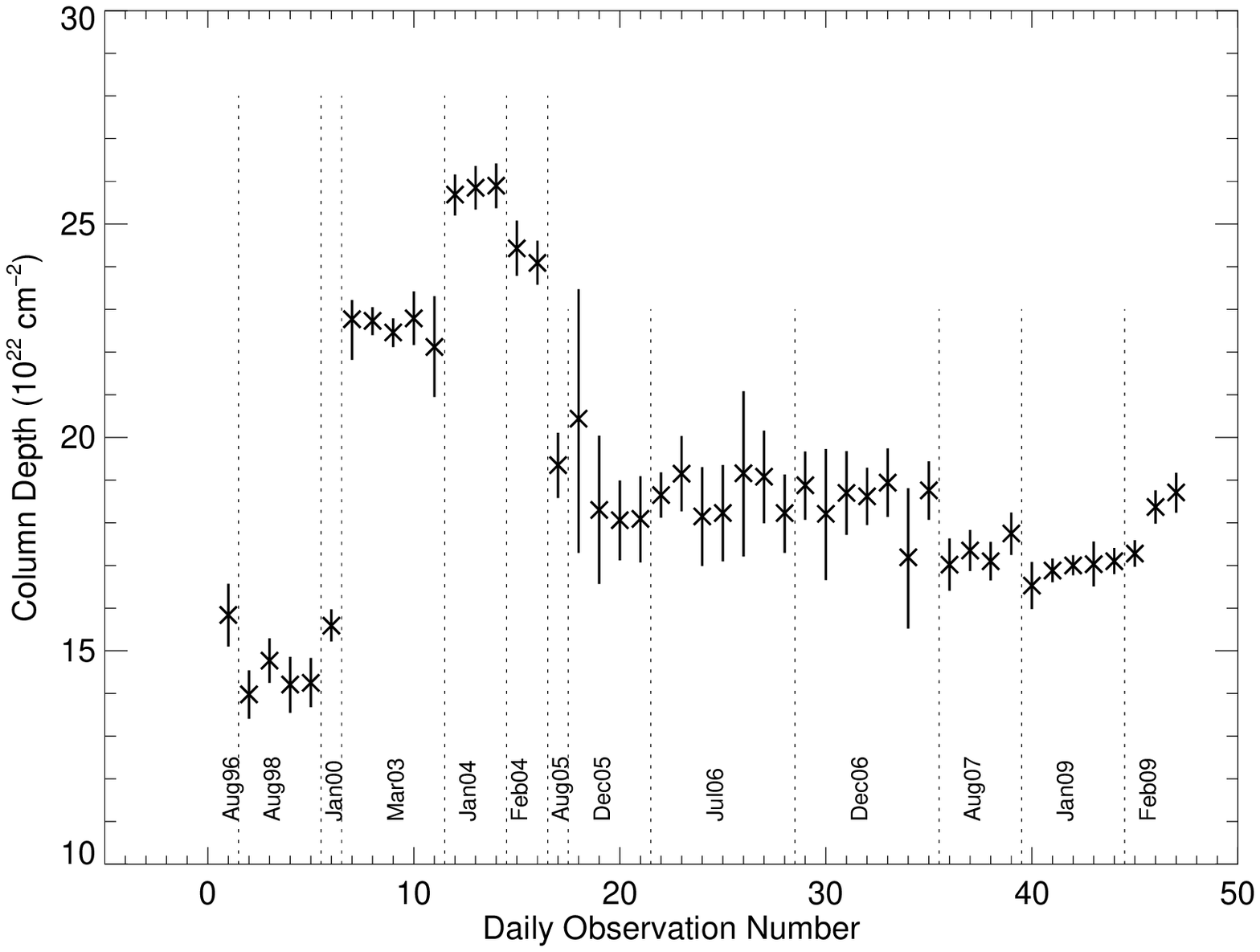}\\
\includegraphics[width=3.25in]{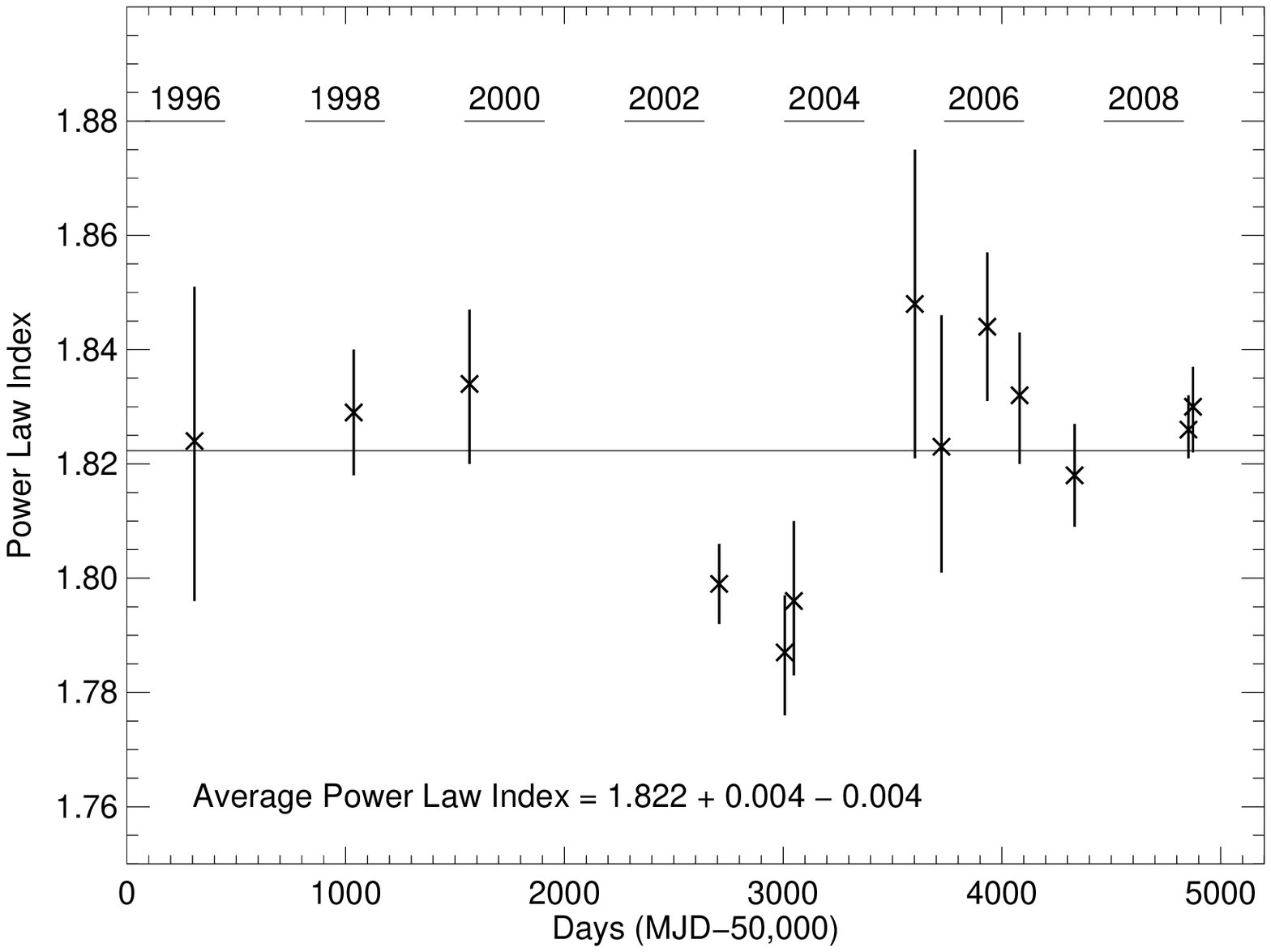}
\includegraphics[width=3.25in]{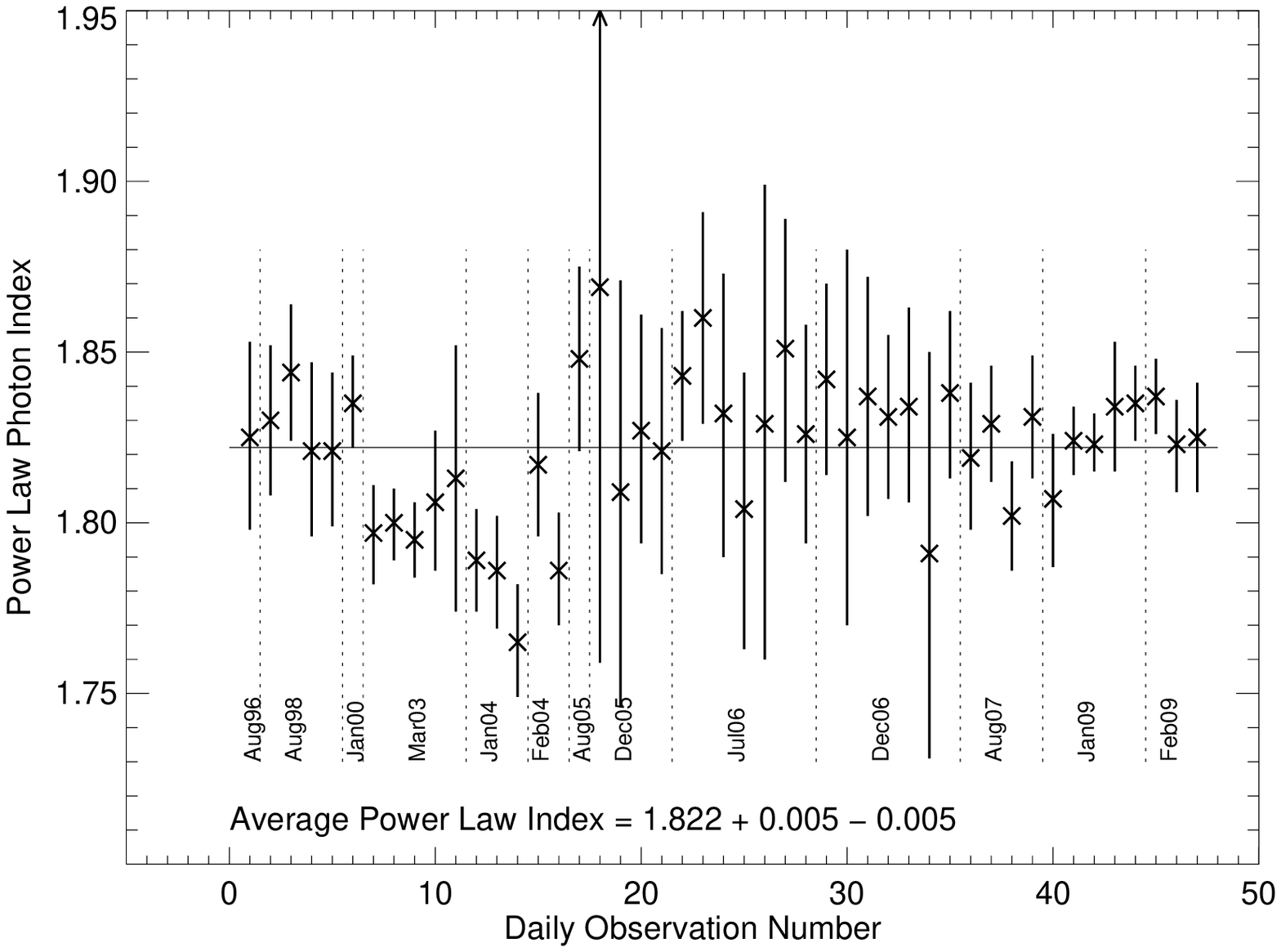}\\
\caption{The best-fit 2--10 keV normalizations of the power law (Top), low energy absorption (Middle), 
and power law photon indices (Bottom) for the 13 observation intervals (Left) and the 47 daily spectra 
(Right). Years and observing intervals are shown. Average values plus 90\% uncertainties are given 
for the power law index. Error bars represent 90\% uncertainties. \label{fig:power}}
\end{figure}

\noindent \textsl{Column Density $N_\mathrm{H}$}: Fig.~\ref{fig:power} (Middle/Right panel) shows 
the best-fit values of $N_\mathrm{H}$ for each of the 47 daily
spectra with the 13 observational intervals indicated. Significant variations 
in $N_\mathrm{H}$ are not seen on a day to day
scale. The change over the 3 weeks between 2009 February 2 and 20 represents 
a 6\% rise, while the interval from 2004 January 4 to February 13 had a 6\% drop in column 
density over 40 days. On the 8 month timescale, a drop of 10\% is seen 
from 2006 December 15 to 2007 August 18.

\noindent \textsl{Power Law Photon Index $\Gamma$}: The value of the power law index was 
clearly below average in 2003 and early 2004 (Fig.~\ref{fig:power} Bottom/Right panel). This 
decrement amounts to only 2\% however.

\noindent \textsl{Power Law Flux}: The unabsorbed power law 2--10 keV flux is seen to vary
with the observational interval fluxes (Fig.~\ref{fig:power} Top/Right panel) in general.
Within each interval, the daily values remain relatively constant, except for the January 2009 
interval where the flux increased by about 20\% with the power law index remaining constant.
This variability is seen in both the 2--10 keV and 20--100 keV fluxes (Fig.~\ref{fig:flux}), thus indicating 
that the power law intensity is the variable and not the power law index or absorbing column. 
{\bf The lack of a correlation between the absorbing column density and the unabsorbed power 
law flux is clearly demonstrated in Fig.~\ref{fig:fe_norm} (Bottom).}

\noindent \textsl{Iron Line}: The iron line centroid, flux, and equivalent width for each 
daily observation are shown in the right-hand panels of Fig.~\ref{fig:iron}. The January 2004 
values of the centroid are barely 1\% above the average, while the flux is $\sim$30\% 
above the mean. While this is simultaneous with the highest value of $N_\mathrm{H}$ and the 2\% drop 
in the power law index, it is not considered significant when the overall variations are 
taken into account (see Fig.~\ref{fig:fe_norm}).

From this we can conclude, as with the 13 observational intervals, the Cen~A spectrum 
only varies in power law normalization and column density, with relatively small variations 
in individual spectral parameters beyond that. Sustained monitoring at regular intervals over several 
months would be necessary for possible reverberation mapping of the iron line emitting material and for
detailed studies of the variation in column density with respect to a clumpy torus model. 
Such monitoring by \textsl{RXTE} began in 2010 January and continues at present.

\subsection{Temporal Variability}

\subsubsection{Lightcurves}

\begin{figure}
\includegraphics[width=6.0in]{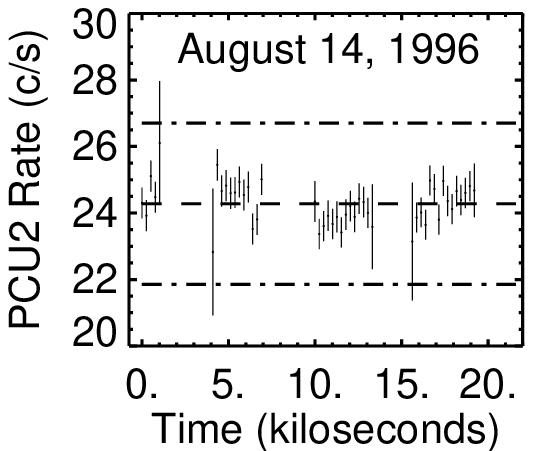}\\
\includegraphics[width=6.0in]{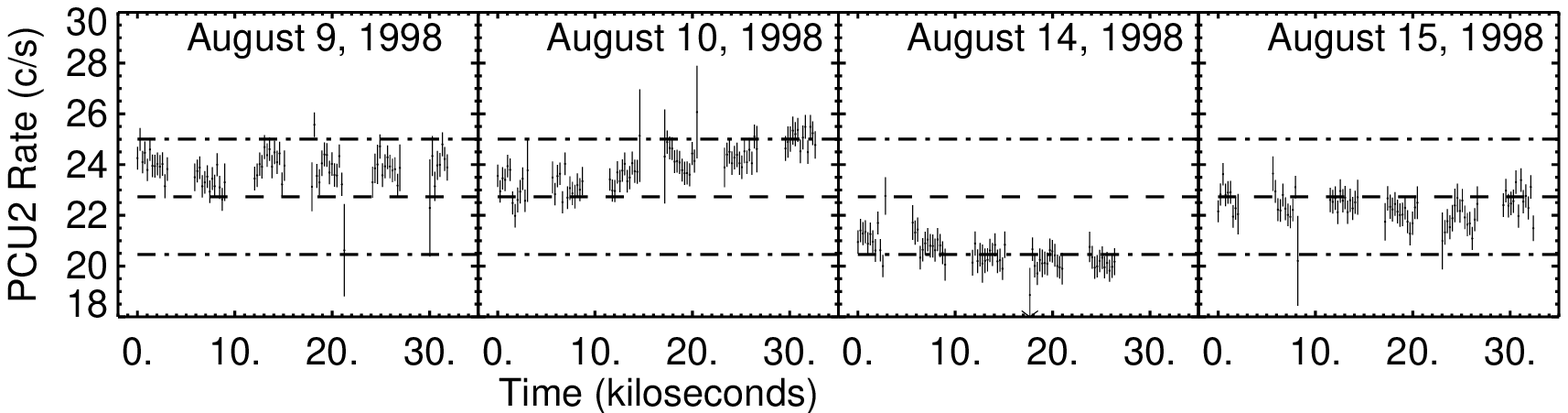}\\
\includegraphics[width=6.0in]{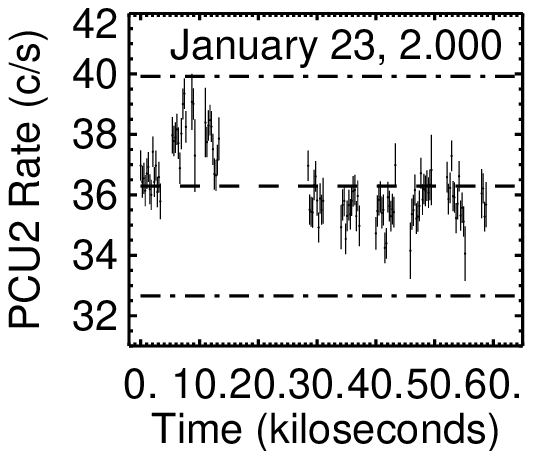}\\
\includegraphics[width=6.0in]{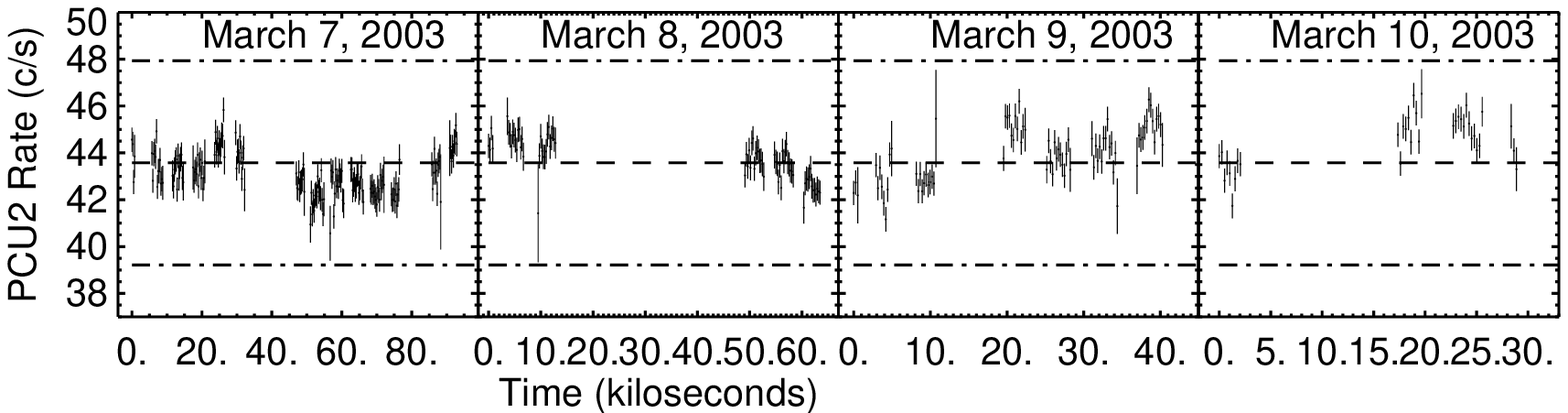}\\
\includegraphics[width=6.0in]{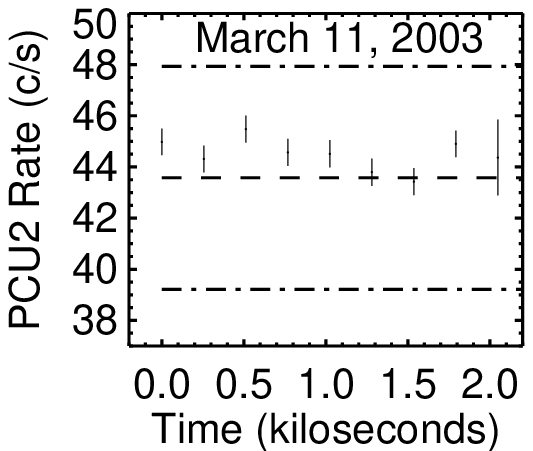}
\caption{Daily light curves of Cen~A with 256 s time bins over the full PCU2 energy range 
for observation intervals August 1996, August 1998, January 2000, and March 2003. 
Dashed line is the average rate for the observational interval and the dash-dot lines 
represent $\pm$10\% in the rate.\label{fig:lcurves}}
\end{figure}

The PCU2 light curves of the first 6 observations (1996 August to 2004 February) 
were originally published in \citet{Roth06}, and the revised light curves for all 
observations are given here, reflecting improvements in the PCA response and 
background estimation since then. Background subtracted light curves with 256 s time bins 
over the complete PCU2 energy band (2-60 keV) were generated for the 53 days 
with 4 or more time bins. In three cases observations overlapped slightly into the 
next day, and this is indicated in Table~\ref{tab:lctimes} (which gives the start and 
stop times of each daily light curve) by stop times greater than 24 hours. The resulting 
51 light curves are plotted versus time since the time of the first temporal bin (Fig.~\ref{fig:lcurves}). 
The short gaps in the light curves are the result of Earth occultations and passages 
through the SAA, while longer gaps and multiple day observations are the result of
\textsl{RXTE} scheduling of other observations with higher priority. Also shown on the daily 
light curve figures are the average rate for the associated observation interval 
(dashed line) and the $\pm$10\% rates (dot-dash lines). 

It is clear from the figures that Cen~A varies by $\pm$10\% quite regularly over a week's 
time, with some variations occurring on a daily timescale. This is especially clear 
in the 2004 January 2--4 and 2004 February 13--14 light curves. A 5\% change over a
few hours was seen between the end of 2006 December 12 and the beginning of December 13. 

\subsubsection{Power Spectral Density Function}

In this section, we present the broadband PSD analysis for Cen A by
combining data from three X-ray instruments. While the resulting PSD
does not have the same temporal frequency coverage as
those for many previously-measured Seyferts, we were still able to 
derive a break in the power spectrum and demonstrate that the PSD shape
was similar to those seen in both Seyferts and X-ray black hole binaries.

To facilitate PSD analysis, and allow use of the Discrete Fourier
Transform (DFT) \citep{Oppenheim75}, we used only light curves which 
are evenly-spaced and relatively continuous (no large gaps). As shown below, for Cen A,
we combined data from \textsl{ RXTE}/All Sky Monitor (ASM), \textsl{RXTE}/PCA, 
and \textsl{XMM-Newton} to probe low, medium, and high temporal frequencies,
respectively.

The sampling obtained by \textsl{RXTE}{\bf /PCA} was relatively inhomogeneous,
and we found only one ``high-quality'' PCA
light curve which, after binning on the satellite orbital timescale,
would make using analysis via a DFT straightforward: {\bf that from the 2004 January 2--4 observation. These data were
obtained from observation IDs
70152-01-01-14, 70152-01-02-000, 70152-01-02-010, and 70152-01-02-[00-16];
the resulting light curve, binned to the satellite orbital timescale of 5.7 ks,
yielded a light curve with only 3/36 points missing.}
With a duration of 2.2 days, we were able to probe variability on temporal 
frequencies near $10^{-5} - 10^{-4}$ Hz.

\textsl{RXTE} is in a low-Earth orbit, and the resulting
Earth occultation {\bf of the PCA} every $\sim$6 ks severely complicates PSD analysis
for temporal frequencies above 1/(2 $\times$ 6 ks).
For the high temporal frequency PSD
($10^{-4} - 10^{-3}$ Hz), we relied on the uninterrupted
light curves provided by \textsl{XMM-Newton} European Photon Imaging
Camera (EPIC) pn. \textsl{XMM-Newton} observed the nucleus of Cen A twice, once in 
2001 February for a duration of 23.4 ks (good exposure time 19.4 ks
after screening), and again in 2002 February for a duration of
15.3 ks (good exposure time 8.9 ks).
We used the longer-duration 2001 data (ObsID 0093650201; data obtained
from the HEASARC public archive), 
which used the medium filter. Using {\tt XSELECT} version 2.4a,  
we extracted 2--10 keV source and background light curves.
{\bf We used only \textsc{PATTERN}=0 events to reduce the impact of pile-up.} 
The source region was a circle of radius
40$\arcsec$ centered on the source; the background was 
extracted from an identical size 
circle $\sim 3\arcmin$ away on the same CCD chip.
We also extracted and inspected the 10--13 keV pn background
light curve for flares, but found none.
The light curves were binned to 300 s; variability at shorter time scales was 
dominated by Poisson noise. 

The decade-long light curve from pointed PCA observations is too
gap-dominated for reliable PSD analysis using a DFT, and so we rely on
the light curve from the ASM.
We downloaded 1-day averaged sum-band (1.5--12 keV) light curves
from the MIT ASM database (http://xte.mit.edu), obtained between MJD
50087--54976. We removed those few 1-day data points which had negative fluxes 
or uncertainties greater than the average flux value, and binned the
light curve to 20 days; power at temporal frequencies higher than $\sim$1/(2$\times$20d)
was dominated by power due to Poisson noise and systematics associated with 
background subtraction, source confusion, etc.
PSD measurements using 1- and 5-d binned light curves
allowed us to empirically determine this level of power, $P_{\rm Psn}$; fitting
a constant to the binned PSD above 10$^{-6.6}$ Hz yielded $P_{\rm Psn} = 33100$ Hz$^{-1}$.

Periodograms were measured for each light curve separately,
binned into a PSD, and then the three PSD segments were combined \citep[e.g., ][]{Edelson99}.
Due to PSD measurement distortion effects (namely, aliasing, which affects only
the PCA light curve, and red-noise leakage),
the model-dependent Monte Carlo method described by \citet{Uttley02}
{\bf was used to assign proper uncertainties to each binned PSD
point and determine the intrinsic, underlying PSD shape.  The reader
is referred to Uttley et al.\ (2002) for the definitions of the
$\chi^2_{\rm dist}$ statistic used to compare observed and modeled
PSDs and the rejection probability used to determine goodness of fit
of model PSD shapes tested.}

Initial PSD construction closely followed $\S$3.1 of \citet{Markowitz03}.
Light curves were linearly interpolated across gaps, though such gaps were rare.
Each light curve's mean was subtracted.
Following \citet{Papadakis93} and \citet{Vaughan05}, 
the periodogram was logarithmically binned by a
factor of $\sim$ 0.20 in the logarithm (roughly {\bf a factor of} 1.6 in $f$)              
to produce the observed PSD, $P(f)$; the two lowest
temporal frequency bins were widened to accommodate three periodogram points.
The constant level of power due to Poisson noise was not subtracted from these
PSDs, but instead modeled in the Monte Carlo analysis. 
The individual long-, medium-, and short-term PSDs were combined to yield the final,
broadband observed PSD, which is shown in Fig.~\ref{fig:psd}(a).
The PSD normalization of \citet{Miyamoto91} and \citet{vdKlis97}
was used to permit combining PSD segments from different missions;
no additional renormalization of the individual measured PSDs was done.

We first tested an unbroken power-law model of the form
$P(f) = A_0 (f/f_0)^{-\alpha}$, where $\alpha$ is the power-law slope
and the normalization $A_0$ is the PSD amplitude at $f_0$, arbitrarily chosen to 
be $10^{-6}$ Hz. We stepped through $\alpha$ from 0.0 to 3.2 in increments of 0.01.
The best-fit model, plotted in Figure~\ref{fig:psd}(a) as the dotted lines,
was obtained for $\alpha$ = 1.73 $\pm$ 0.18.  
The rejection probability $R_{\rm unbr}$ was 0.948.
The residuals are plotted in Figure~\ref{fig:psd}(b).
The errors reported here correspond to a value 1$\sigma$ above 
$R_{\rm unbr}$ for the best-fit value on a Gaussian probability distribution.
For instance, $R_{\rm unbr}$ corresponds to 
1.94$\sigma$; the error on $\alpha$ corresponds to 2.94$\sigma$ or $R_{\rm unbr}$ = 0.997. 
(However, see \citet{Mueller09} for warnings regarding using 
rejection probabilities as confidence regions for PSD model parameters.)

We then tested a singly-broken PSD model shape of the form 
$P(f)=  A_1(f/f_{\rm b})^{- \alpha_{\rm lo}}$ for   $ f \le f_{\rm b}$,
or      $A_1(f/f_{\rm b})^{- \alpha_{\rm hi}}$ for   $ f > f_{\rm b}$.
where the normalization $A_1$ is the PSD 
amplitude at the break frequency $f_{\rm b}$, and
$\alpha_{\rm lo}$ and $\alpha_{\rm hi}$ are the low- and high-frequency power law slopes, respectively,
with the constraint $\alpha_{\rm lo} < \alpha_{\rm hi}$.
PSD slopes were tested in increments of 0.1, as was log($f_{\rm b}$).
The limited amount of PSD data precluded testing more complex PSD shapes.

The best-fit model had a rejection probability $R_{\rm brkn}$ of 0.208
for log($f_{\rm b}$) = -- ($6.2^{+0.3}_{-0.2}$) (corresponding to $T_{\rm b} = 18.3^{+18.3}_{-6.7}$ d),
$\alpha_{\rm hi} =  2.5^{+0.4}_{-0.1}$, and $\alpha_{\rm lo} = 0.9^{+0.3}_{-0.2}$. These errors correspond to values 
1$\sigma$ above the rejection probability 
for the best-fit
value on a Gaussian probability distribution;
the best-fit model's rejection probability corresponds to
0.27$\sigma$; the errors correspond to $R_{\rm brkn} = 0.796$ or
1.27$\sigma$ \citep{Markowitz03}.
The amplitude $A_1$ was $7.9^{+3.1}_{-2.3}  \times 10^4$ Hz$^{-1}$; 
The best-fit value of $A_1 \times f_{\rm b}$ is  0.05, 
broadly consistent with values of $\sim$0.01--0.02  measured for Seyferts.

The best-fit model is shown in Figure~\ref{fig:psd}(a) as a solid line, and as shown in Figure~\ref{fig:psd}(c), the 
residuals are much smaller compared to the unbroken power-law fit, particularly in the ASM
PSD segment. Following \citet{Markowitz03}, 
using the ratio of likelihoods of acceptance $L_{\rm brkn}/L_{\rm unbr}$ ($L_{\rm brkn} \equiv 1 - R_{\rm brkn}$;
$L_{\rm unbr} \equiv 1 - L_{\rm unbr}$), the PSD break was significant at $\sim$15$\sigma$.

Can the flattening in PSD slope seen in the ASM PSD segment be an artifact?
For an energy spectrum with the shape of that for Cen A falling off rapidly below $\sim$3 keV,
the {\bf spectrum-weighted effective area} of the ASM peaks near roughly 4 keV, while that for the PCA peaks near 5--6 keV,
so the difference in average photon energy may play a minor role at best.
PSD power-law slopes above the break have been observed to flatten with increasing photon energy
for an assumed  energy-independent $f_{\rm b}$ \citep[e.g.,][]{Nandra01,Vaughan03}, but the
effect is in the opposite sense to that observed in the Cen A PSD.
Finally, as the ASM light curve is virtually continuous, a constant level
of power due to aliasing is not expected in this PSD segment.
We conclude that the break in the PSD is real. Current \textsl{RXTE} monitoring
to bridge the ``gap'' in the current PSD and cover the temporal frequency range
$10^{-6.6}$ to $10^{-5.2}$ Hz is on-going and will
be useful in further constraining the break frequency and PSD power-law slopes.

{\bf Finally, the effect of the observed evolution in line of sight column density
$N_{\rm H}$ on the ASM PSD can be estimated as follows.  The increase
in $N_{\rm H}$ corresponds to a reduction in observed 2--10 keV flux
by $\sim35\%$. We added a $\sim$2-year long trend
with flux increasing/decreasing by 35\% during 2003--2004 to the binned
ASM light curve as a rough estimate of the intrinsic or unabsorbed
light curve.  The observed 2--10 keV flux of Cen A during 2003--2004
was below the long-term average, and so this action changed the fractional variability amplitude
$F_{\rm var}$ (Vaughan et al.\ 2003) only slightly: a decrease of
3$\%$ from $F_{\rm var}$ = 29$\%$ to 26$\%$.  That is, the unabsorbed
light curve is slightly less variable than the observed light curve;
the intrinsic PSD, compared to the observed PSD, may be lower by up to
17$\%$ in linear space, or 0.07 in log space, on temporal frequencies
$\lesssim 1.0 \times 10^{-8}$ Hz. In other words, the intrinsic
broadband PSD may bend even more strongly than the observed PSD,
albeit by a very small amount.}

\section{Discussion and Conclusions}

A summary of the results from the \textsl{RXTE} observations includes: (1) the power law 
index was constant (to within a few percent) and independent of {\bf 2--10 keV power law flux, which varied by a factor of 3}, 
(2) the column density experienced a factor of 2 variation within a 1--4 year time span, (3) the iron 
line flux was independent of the power law flux, (4) the equivalent width was not proportional 
to the iron line flux, but varied inversely with the power law flux, (5) the iron line flux was not 
correlated with the column density variations, (6) a Compton reflection component was not detected, and
(7) any high energy spectral roll over was greater than $\sim$500 keV and could be as high as 2 MeV.

In order to establish a picture of the geometry of the X-ray continuum emitter and the 
circumnuclear accreting material, these results must 
be combined with results from sub-parsec scale mid-infrared observations that reveal a 
nuclear, molecular torus with 0.6 pc diameter and a 62$\degr$.6 inclination dusty disk
\citep{Burtscher10}, theoretical work describing the torus as a 
clumpy medium \citep{Nenkova02,Nenkova08a,Nenkova08b}, and when combined with 
the steeper spectra seen at GeV and TeV energies, the combined X-ray/gamma ray data 
define the general shape and location of the blazar inverse Compton peak. 

In the following sections we discuss the implications of the iron line diagnostics of the 
circumnuclear gas, inferences about a clumpy torus, the X-ray variability, the implications 
of not observing a correlation between the power law index and the power law flux, and 
the possible origin of the continuum emission. Our overall conclusions are then presented.

\subsection{Fe K$\alpha$ Line Diagnostics of the Circumnuclear Gas}

The lack of a strong observed Compton reflection component
suggests that the bulk of the circumnuclear material is
Compton-thin, or, if there does exist Compton-thick
circumnuclear gas, it does not contribute significantly to the observed 
spectrum and/or is poorly illuminated by the central
X-ray source. In either case, the bulk of the observed Fe K$\alpha$ 
emission line originates in Compton-thin gas, with at most
a negligible contribution from Compton-thick gas.

The Fe emission line's observed equivalent width $EW_{\rm obs}$ can be used as a
diagnostic of the geometry of the Fe-line emitting gas under certain
assumptions. For the moment, let us assume a simplified geometry in which
the Fe-line emitting gas is distributed in an optically-thin, uniform spherical shell
with line of sight column density $N_{\rm H}$ and a
covering fraction as seen from the central X-ray continuum source $f_{\rm cov}$. 
We use $EW_{\rm obs}$ = 90 $\pm$ 10 eV, the long-term average obtained by 
\citet{Rivers10} from fits to the total summed PCA + HEXTE
spectrum. We can then use Eq.\ 5 from \citet{Murphy09}
to relate $EW_{\rm obs}$ to $N_{\rm H}$ and $f_{\rm cov}$.
Assuming solar abundances ($Z_{\rm Fe} = 1$), the abundances of \citet{Wilms00}, and an 
illuminating continuum with $\Gamma = 1.83$ , we find
$EW_{\rm obs} = 40\, \mathrm{eV} \times (N_\mathrm{H}/10^{23}\mathrm{cm}^{-2}) \times f_\mathrm{cov}$.
A value of $N_\mathrm{H}$ of $2.1\times 10^{23}$ cm$^{-2}$ along with $f_\mathrm{cov} = 1$
will yield  $EW_{\rm obs} = 90\, \mathrm{eV}$.
Similarly, if the long-term average observed line of sight $N_{\rm H}$ from
\citet{Rivers10}, $1.5\times10^{23}$ cm$^{-2}$, is used,
a covering fraction of unity and $Z_{\rm Fe} = 1.5$ will yield the same value of $EW_{\rm obs}$.
That is, for all values of $N_\mathrm{H}$ less than a few times $10^{23}$ cm$^{-2}$,   
covering fractions near unity are required. This may indicate that the Fe line emitting gas
is highly spatially extended as seen from the central source, consistent with the 
observed lack of strong variability in the intensity of the Fe line between 1996 and 2009,
with response to the most rapid continuum variations getting smeared out.

Alternatively, if the Fe line emitting gas is not distributed uniformly, the observed equivalent width 
can still be attained with gas having a column density near $10^{24}$ cm$^{-2}$ and 
$f_\mathrm{cov}$ near one-fifth; a torus-like structure lying mostly out of our line of sight
would be consistent with this scenario. Following Fig.\ 8 of \citet{Murphy09},
assuming a torus-like structure with $N_\mathrm{H}$ of a few $\times 10^{23}$ cm$^{-2}$ 
inclined so that it is not intersecting our line of sight, 
$EW_{\rm obs} \sim 90\, \mathrm{eV}$ can be obtained if $Z_{\rm Fe} \sim 2$.
If this torus does intersect our line of sight (consistent with full-covering absorption
observed in hard X-ray spectra of Cen A), an $EW_{\rm obs}$ near 100 eV can be attained
with $Z_{\rm Fe} \sim 1$. 

We also applied the new model ``MYTorus" \citep{Murphy09} to
our time-averaged data. This model is self-consistent and includes Compton reflection, Fe
line emission and line of sight absorption due to a dusty circumnuclear torus. It has the
benefit of being a physical model but also has the drawback of requiring
certain simplifications and assumptions made about the geometry of the torus
(namely that the half-opening angle of the torus is 60$\degr$, the torus cross section is circular, 
and the torus is uniform in density).  Note that ``MYTorus'' also uses the abundances 
of \citet{Anders89} which yield lower values of $N_\mathrm{H}$ than those of \citet{Wilms00}, 
which are used throughout the above analyses. However it gives us certain insights into possible
configurations of the material surrounding Cen~A.

The scenario tested by MYTorus was that of a torus that intercepts the line of sight,
accounting for the observed absorption.  In this case we obtained an
inclination angle for the torus of  $>$ 70$\degr$ to the line of sight, and a column density of 
$N_\mathrm{H}$  = 1.11$\pm 0.03 \times 10^{23}$ cm$^{-2}$. 
The torus was able to account for only half of the Fe line flux observed
and did not include any significant contribution from the Compton reflection hump.  
An additional toroidal component of material with column density 
$N_\mathrm{H}$ = 1.3$ \pm 0.5 \times 10^{23}$ cm$^{-2}$ outside of the direct line of sight to the nucleus
can account for the additional Fe line flux. 

\subsection{Inferences about a Clumpy Torus}

Temporal variations in column density are very common in Seyfert galaxies on various timescales \citep{Risaliti02,Lamer03}.
In Cen~A, the line of sight X-ray column density measured in 2003 March, 2004 January, 
and 2004 February  was (2.3, 2.6, 2.4)$\times$10$^{23}$ cm$^{-2}$,  respectively, whereas before 
and after it was about 1.6$\times$10$^{23}$ cm$^{-2}$.  For the following discussion, 
we assume that this represented the passage through our line of sight of a single clump of matter
in the circumnuclear torus representing a change in $N_\mathrm{H} \sim 1 \times 10^{23}$ cm$^{-2}$.

The clumpy torus model of Nenkova \citep{Nenkova02,Nenkova08a,Nenkova08b}, parameterizes 
the number of clumps along the line of sight at an angle $\Theta$ to the torus equatorial pole as 
$N_\mathrm{LOS}$(90$\degr - \Theta$). The model also characterizes the number of clumps along the line of sight as an
exponentially decreasing function of $\Theta$, i.e., the maximum number of clumps is seen in 
the equatorial plane of the torus, and the number decreases with angle from the plane. If one 
clump passing through the Cen~A line of sight has a column density that is a large fraction of 
the total column density seen, as is the case here, one can infer that the line of sight only contains 
a few clumps. This, in turn may be interpreted as viewing Cen A at an angle to the torus equatorial 
plane, such that our line of sight passes through the outer region of the torus. 

Furthermore,  using the angle of the torus to the line of sight as 62$\degr$.6 
\citep{Burtscher10} and $N_\mathrm{LOS}$(90$\degr - \Theta$)=N$_0$~exp($-$($\Theta/\sigma$)$^2$), 
where N$_0$ is the average number of clouds along the equatorial direction and $\sigma$ characterizes
angular distribution of clouds \citep{Almeida09}, we find for $\Theta$=62$\degr$.6, $\sigma$=60$\degr$
 \citep{Almeida09}, and $N_\mathrm{LOS}$(90$\degr - \Theta$)=3, that N$_0$=8. This is in accordance with the 
 statement in \citet{Nenkova08b} that ``N$_0$ is likely no larger than $\sim$10--15 at most'', since a 
 larger number would produce an infrared bump that is not seen. Therefore, viewing Cen~A through 
 the outer edges of the torus is consistent with the clumpy torus model.

If {\bf $N_\mathrm {H}$ for a representative clump is} 1$\times$10$^{23}$ cm$^{-2}$, then eight 
of them along the torus equatorial plane implies that the material in the equatorial plane
may be close to being Compton thick ($\sim 10^{24}$ cm$^{-2}$), but this value will drop 
as one goes farther from the plane of the torus, and the majority of the torus will be Compton-thin.  
Thus, the Compton thick aspects of the circumnuclear torus may be restricted to near the equatorial 
plane, and would explain the lack of a detected 
Compton reflection component. In addition, we have speculated previously that the high energy 
radiation from the nucleus could be anisotropic with very little if any of it illuminating the Compton 
thick material  \citep{Roth06}. A combination of a minimum of Compton thick material and an 
anisotropic radiation pattern may provide a more complete explanation for the lack of a Compton 
hump in Cen~A.

From Fig.~\ref{fig:power} (Middle) we can estimate how long the clump took to pass through our line 
of sight. Assuming a sharp transition in $N_\mathrm{H}$ as the clump passed into and out of the line 
of sight, we estimate a duration {\bf \textsl{T}} of a minimum of $\sim$1 year and a maximum of $\sim$4 years. Assuming 
a torus diameter of 0.6 pc  \citep{Meisenheimer07} and a black hole mass {\bf $M_{\rm BH}$} of 
{\bf 6}$\times$10$^7$ M$_\odot$ \citep{Cappellari09,Neumayer10}, 
the Keplerian velocity {\bf $V_\mathrm{K}$} of the clump can be used to estimate its linear size, if one knew the radial position 
of the clump within the torus. Using radial values  of the IR torus of {\bf $R_\mathrm{K}$}$\sim$0.1 to $\sim$0.3 pc, we find the velocity of 
the cloud {\bf $V_\mathrm{K} = (G M_{\mathrm{BH}}/R_\mathrm{K})^{0.5}$} to be on the order of 10$^3$ km s$^{-1}$ and has a linear dimension 
{\bf $L_\mathrm{c} = V_\mathrm{K} \times T$} on the order of 
3{\bf --12}$\times$10$^{13}$ m, 0.001{\bf--0.004} pc, or about one {\bf to four} light-days. Given the measured column density of the 
clump and the derived linear dimension, an inferred number density {\bf $n_\mathrm{c} = N_\mathrm{H}/L_\mathrm{c}$} is on the order of {\bf 1--}3$\times$10$^7$ cm$^{-3}$. 
This yields an approximate cloud mass {\bf $M_\mathrm{Cloud} = \frac{4}{3} \pi (\frac{1}{2} L_\mathrm{c})^3 m_\mathrm{p} n_\mathrm{c} \approx$ 1}$\times 10^{30}$ g, or about one Jupiter mass, {\bf where m$_\mathrm{p}$ is the proton mass}.
The linear dimension of the cloud is about a thousand times larger than those determined by \citet{Risilati09} for clouds in 
the broad line region, while the density is three to four orders of magnitude smaller.  The duration of the 
transiting event here is much longer than those seen in, e.g., NGC 1365 \citep{Risaliti09a,Risaliti09b} 
and NGC 3227 \citep{Lamer03}, which lasted for days to a few months, and were consistent with absorbing 
clouds in the optical broad line region of these objects. The lower density inferred from the Cen~A absorption event
could be due to the clouds in Cen~A being significantly further away from the supermassive black hole than 
those inferred for NGC~1365 or NGC~3227.

\subsection{X-ray Variability of Cen~A}

We can compare the break inferred in the observed PSD of Cen A with those breaks
measured in the PSDs of other Seyfert AGN by using the empirical relation 
between $T_{\rm b}$, $M_{\rm BH}$ and $\dot{m}$ quantified by \citet{McHardy06}.
This exercise relies on the assumption that accretion onto the supermassive black hole in Cen A proceeds
via the same mechanism and with the same efficiency of radiation as the Seyferts used in the sample of
\citet{McHardy06}. We assume a black hole mass of 
$6 \times 10^7 \mathrm{M}_\odot$, the average of the estimates discussed in Section 1.
The long-term, unabsorbed, 2--10 keV luminosity, {\bf L$_{2-10}$}, \citep{Rivers10} is $ 8.6 \times 10^{41}$ erg s$^{-1}$.
{\bf The bolometric luminosity {\bf $L_{\rm bol}$}, is estimated to be
$8 \times 10^{42}$ erg s$^{-1}$ (yielding an inferred accretion rate relative to Eddington of 0.1$\%$), 
assuming a conversion factor $L_{\rm bol}/L_{2-10}$ = 9 \citep{Marconi04}}. 
The best-fit empirical relation of \citet{McHardy06}, 
log($T_{\rm b}$(days)) = 2.10 log($M_{\rm BH}/10^6 \mathrm{M}_\odot$) -- 0.98 log($L_{\rm bol}/10^{44}$ erg s$^{-1}$) -- 2.32, 
yields a predicted break time scale $T_{\rm b,pred}$ of 317 days,
which corresponds to a break frequency of $3.7 \times 10^{-8}$ Hz.  
The value of $T_{\rm b,pred}$ is a factor of $\sim 17^{+36}_{-13}$ higher than
the observed PSD break time scale ($T_{\rm b} = 18.3^{+18.3}_{-6.7}$) measured above for Cen A.

The sample used by \citet{McHardy06} was derived from mainly radio-quiet
Seyferts accreting at $\dot{m}$ of a few percent or more, and it is thus
conceivable that radio-loud sources and/or sources accreting at
very low values of $\dot{m}$ do not strictly adhere to the relation derived by \citet{McHardy06}.
However, the measured PSDs for the radio-loud AGN 3C~120
and the low-luminosity AGN NGC 4258 seem to be at least roughly consistent with the relation.
3C 120 has $M_{\rm BH} \sim 5 \times 10^7\, \mathrm{M}_\odot$ and  $\dot{m} \sim 0.31$
\citep{Vasudevan09}, which yields  $T_{\rm b,pred} = 1.1$ days.
\citet{Marshall09} and \citet{Chatterjee09} have each
measured the X-ray PSD and found best-fit values of $T_{\rm b}$ = 6.5 days and 1.2 days, respectively,
consistent with $T_{\rm b,pred}$, considering the errors on $M_{\rm BH}$,  $\dot{m}$, and $T_{\rm b}$.
NGC 4258's black hole mass is well-determined via mega-masers, at $3.9 \times 10^7\, \mathrm{M}_\odot$
\citep{Herrnstein99}, 
and $\dot{m}$ = 10$^{-4}$ \citep{Lasota96}, so that $T_{\rm b,pred}$ is about 2000 days, consistent with the
lower limit of $T_{\rm b} = 4.5$ d measured by \citet{Markowitz05}.
We speculate that different physical mechanisms may dominate the observed X-ray variability in
Seyferts and in radio-loud and/or low-$\dot{m}$ AGN such as Cen A. That is, in such objects, the  X-ray 
emission may be dominated by emission from a jet component versus that coming from a corona
associated with the accretion disk (as in 3C 120 and the radio-quiet Seyferts).

\subsection{Lack of a Power Law Flux --- Power Law Index Correlation}

As presented in Fig.~\ref{fig:versus}, no correlation was seen in the plot of power law index versus unabsorbed 2--10 keV 
power law flux, when the flux varied over a factor of three (3--10$\times 10^{-10}$ ergs cm$^{-2}$ s$^{-1}$). This is in 
contrast to what is commonly reported for Seyfert 1s in the 2--10 keV band \citep[e.g., ][]{Shih02,Papadakis02}, where 
the phenomenon of increasing flux being positively correlated with increasing power law 
indices is virtually ubiquitous. All of the objects for 
which this correlation was seen are radio quiet Seyferts. While the radio quiet Seyferts have a positive correlation,  a relativistically broadened Fe K$\alpha$ emission component in about 2/3 of the cases \citep{Nandra07}, and 
significant Compton reflection components, Cen~A exhibits none of these,  and has a powerful jet. Perhaps the anisotropic 
radiation pattern at the base of the Cen~A jet only illuminates the accretion disk minimally, if at all, but does illuminate a 
portion of the clumpy torus. This could explain the lack of a measured Compton reflection 
component from the disk, if the disk is indeed Compton-thick. 
If the power law/flux correlation originates from processes associated with soft photon production and subsequent 
up-scattering of them by the hot electrons in the accretion disk corona \citep[e.g., ][]{Haardt93}, the anisotropic flux 
distribution mentioned above would minimize this process in Cen~A. This lack of a power law 
index/power law flux correlation may also be explained by  a lack of soft photons due to
the presence of a radiatively-inefficient accretion flow, \citep[e.g., ][]{Narayan98,Quataert01}, {\bf and thus a lack of disk-produced soft photons, which are the basis for Comptonization models of hard X-ray emission in Seyferts}.

\subsection{Continuum Emission from the Jet in Cen~A}

\citet{Chiaberge01} constructed the first spectral energy distribution (SED) for Cen~A 
showing that the SED consisted of two components: a synchrotron component in the 
sub-millimeter to near infrared and a second higher energy synchrotron self-Compton (SSC) 
component in the X-ray to gamma ray range, where the latter was defined by the \textsl{CGRO} results 
\citep{Kinzer95, Steinle98}. Later, \citet{Lubinski09} presented the \textsl{INTEGRAL}/IBIS 
spectrum of Cen~A extending to 400 keV or greater as a single power law. The lack of a 
significant deviation from a power law up to energies
as high as 2 MeV, as inferred from \textsl{RXTE}, can be 
interpreted in the misoriented BL Lac scenario \citep{Chiaberge01} as having the X-ray emission form the low energy side 
of the SSC hump, viewed at a large angle from the jet direction. Despite the 
overall SED being reminiscent of blazars, Cen~A's X-ray spectrum exhibits two signatures of 
Seyferts and quasars: a high column density of absorbing gas in the line of sight, and an iron 
emission line. In the Cen~A case the bright hard X-ray continuum emission may originate at the base of 
the jet, yet from behind the absorbing line of sight material in contrast to what is commonly 
observed from blazars. A jet origin to the continuum has also been put forth for FR I radio 
galaxy cores by \citet{Evans06}, but {\bf these authors} associate low absorption ($<$5$\times 10^{22}$ cm$^{-2}$) with 
sources having the hard X-ray emission associated with a jet. In this regard, Cen~A may 
be a counter example.

The nature of the accretion flow near the black hole
may contribute to defining the difference in nature between low-luminosity,
radio loud AGN, such as Cen~A, and radio-quiet Seyfert AGN, since
a radiatively inefficient flow may be responsible for
launching jet outflows \citep[e. g., ][]{Kording06}. In this case one would not have the soft 
photons for subsequent upscattering by 
the hot electrons in the accretion disk corona \citep{Haardt93}.
In Cen~A, the accretion disk may only supply matter
to the base of the jet and into the black hole, but may not 
make any contribution in terms of observational X-ray signatures.

\subsection{Conclusions}

Over more than a decade of observing, Cen~A has shown remarkably 
stable spectral behavior. Both the power law index and the flux of the iron line have remained essentially 
constant, in contrast to the range in power law flux and column density, which have varied by a factor of 3 
and 2 respectively. The stability of the power law index over the past 40 years (with the exception of the 
early 1970's when it flattened considerably \citet{Baity81}), implies 
a stability {\bf of the processes generating} the continuum at the base of the jet. The 
spectral aspects of these processes, i.e. the shape of the electron distribution and electron 
optical depth,  must therefore be relatively insensitive to the variations in the accretion rate experienced at the base of the jet, 
which is assumed to be directly related to the overall flux observed. From this 
we infer that the mass accretion rate in the inner portions of the accretion 
disk (\textsl{\.m} $\approx \mathrm{L_X}/ \mathrm{L_{Edd}} \sim$10$^{-4}$) has remained 
within the range where the {\bf radiation efficiency is} relatively unchanged 
even though the luminosity (a tracer of the mass accretion rate) has varied by a factor of 3.

Our observations with \textsl{RXTE} have led to the conclusions that we are viewing the supermassive 
black hole at the nucleus of this active radio galaxy through the outer portions of a clumpy molecular 
torus which is mainly Compton-thin, but may approach being Compton-thick only at or near its equator. 
The Compton thin torus {\bf may then} occupy 
a large solid angle as viewed from the Cen~A nucleus in order to be the site of the iron line emitting material.

The variability characteristics, when compared to Seyferts, imply that the accretion disk is not the location 
of the $>$3 keV X-ray flux, but that synchrotron self-Compton {\bf scattering} in the base of the 
jet may be responsible. In the case of Cen~A, the bright hard X-ray continuum emission may originate at the base of 
the jet, yet from behind the absorbing line of sight material in contrast to what is commonly 
observed from blazars. 
\acknowledgments

The authors acknowledge the excellent comments from the anonymous referee. We gratefully acknowledge NASA contract NAG5-30720 and NASA grants NNX08AY11G and NNX09AG79G for support of this analysis.

\clearpage

\clearpage

\clearpage

\clearpage

\clearpage

\clearpage

\clearpage

\clearpage

\clearpage

\clearpage

\clearpage

\clearpage

\setcounter{figure}{10}

\begin{figure}
\includegraphics[width=6.0in]{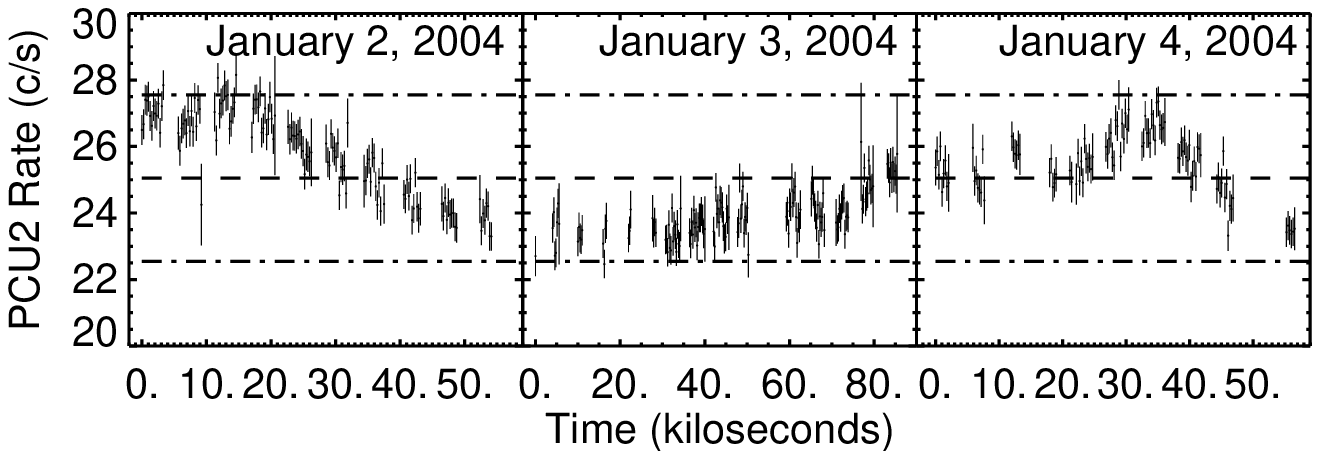}\\
\includegraphics[width=6.0in]{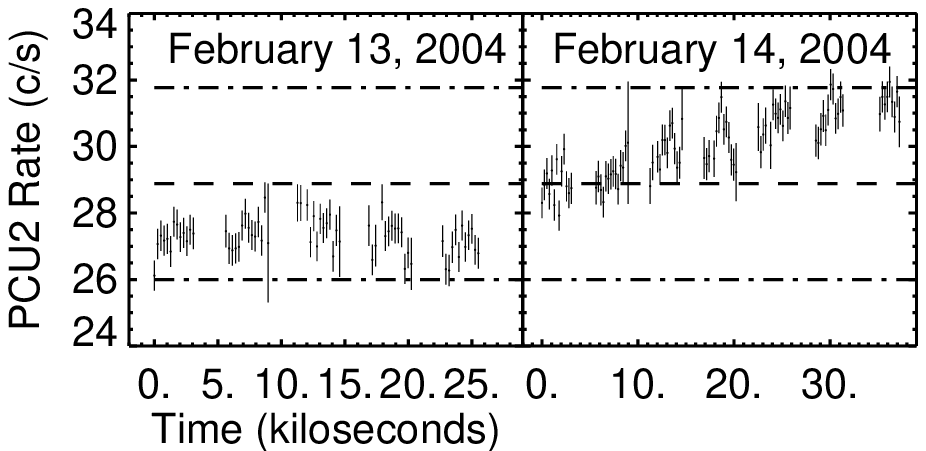}\\
\includegraphics[width=6.0in]{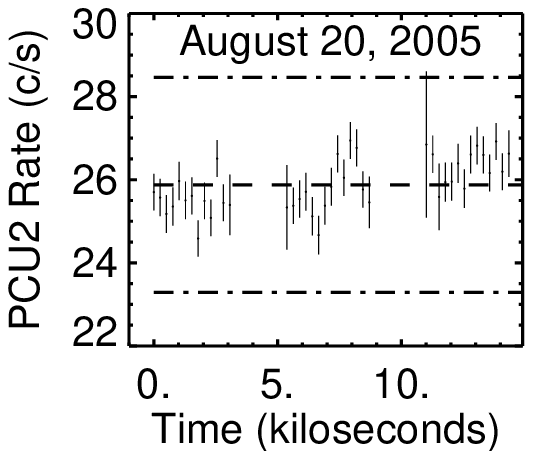}\\
\includegraphics[width=6.0in]{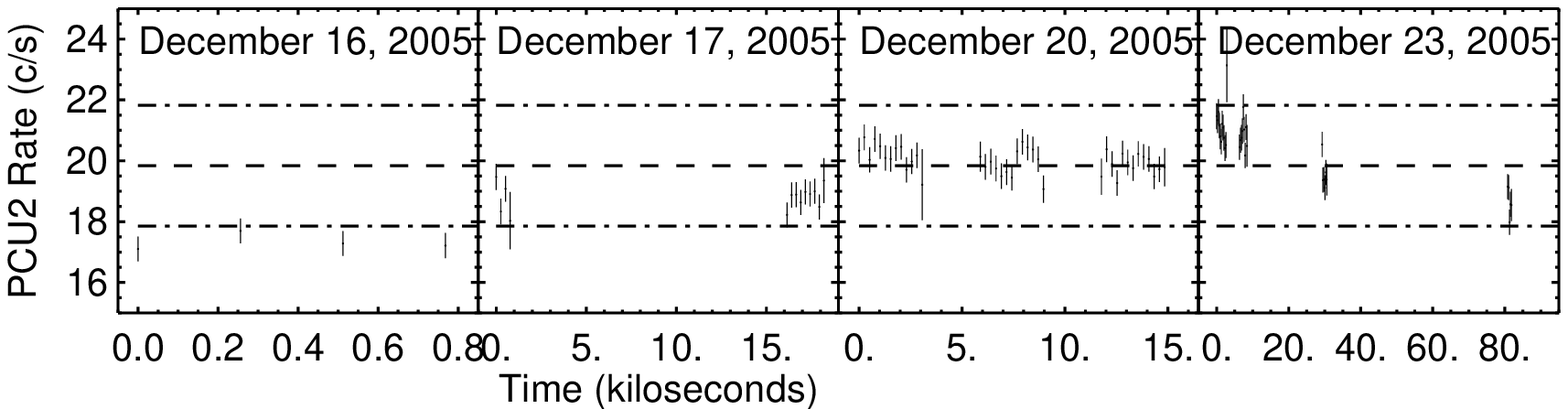}
\caption{Daily light curves of Cen~A with 256 s time bins over the full PCU2 energy range 
for observation intervals January 2004, February 2004, August 2005, and December 2005. 
Dashed line is the average rate for the observational interval and the dash-dot lines 
represent $\pm$10\% in the rate.}
\end{figure}

\clearpage

\setcounter{figure}{10}

\begin{figure}
\includegraphics[width=6.0in]{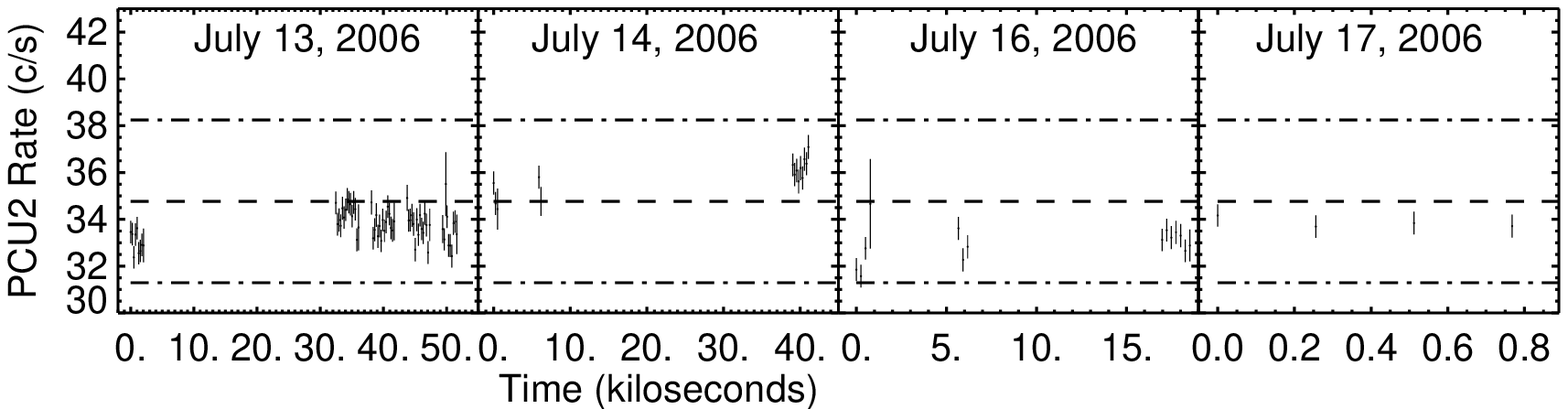}\\
\includegraphics[width=6.0in]{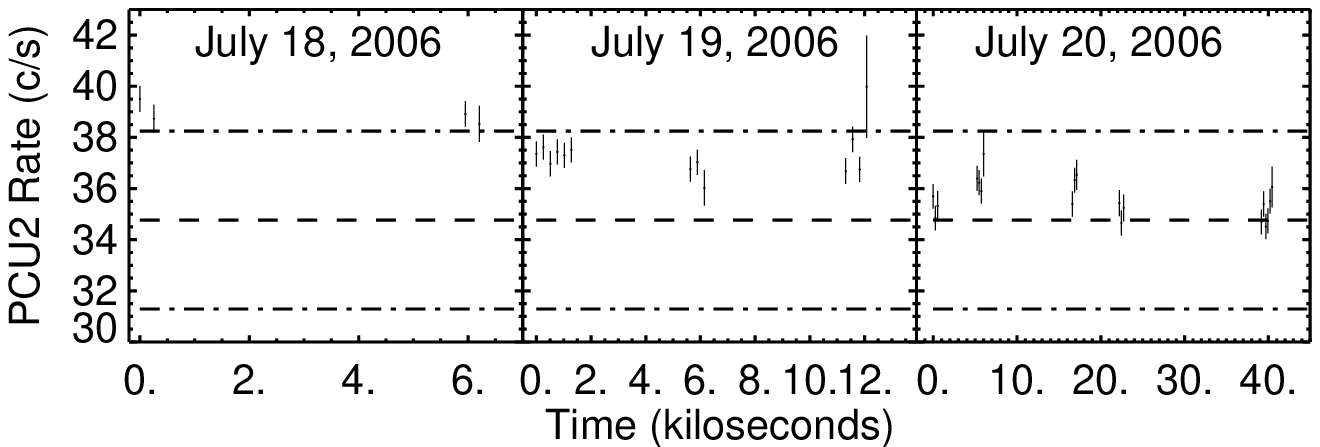}\\
\includegraphics[width=6.0in]{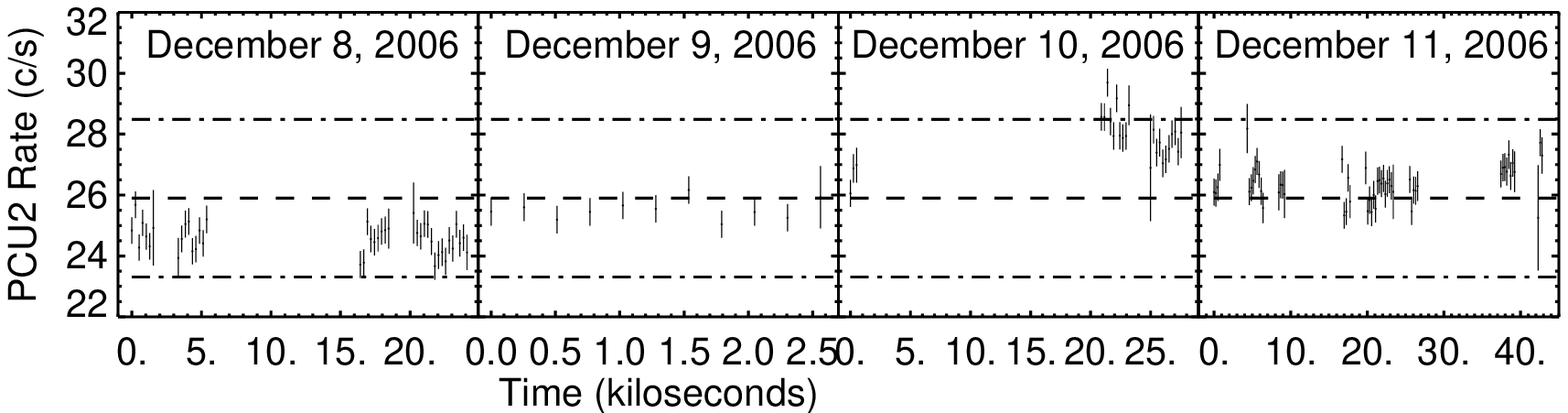}\\
\includegraphics[width=6.0in]{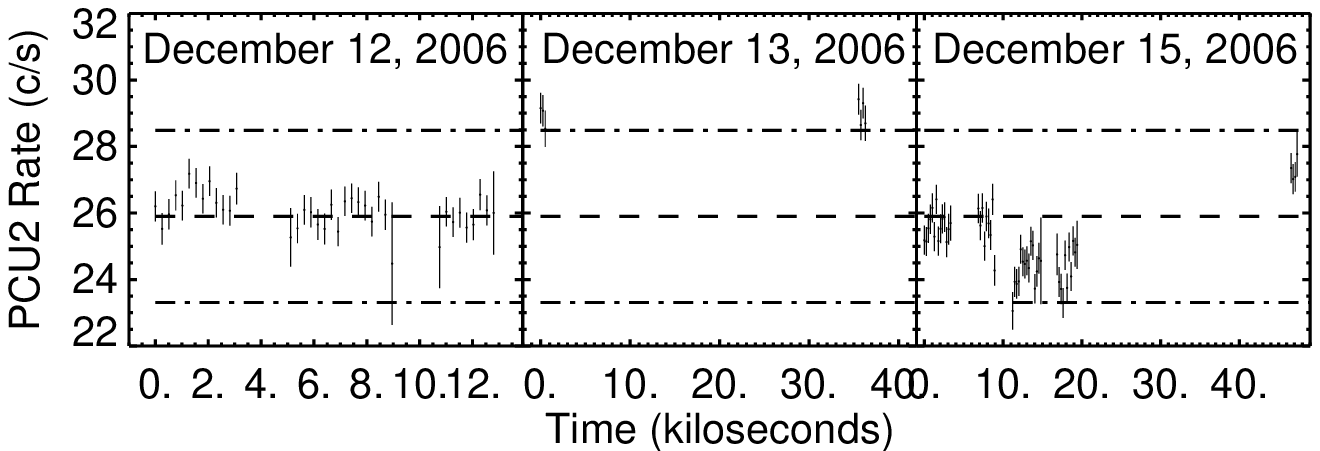}\\
\includegraphics[width=6.0in]{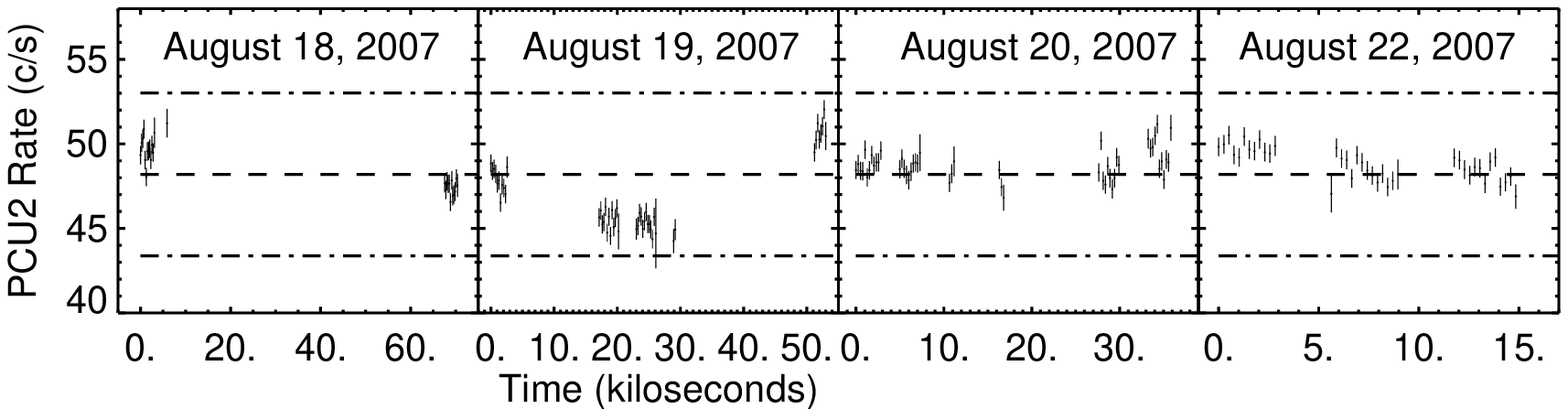}
\caption{Daily light curves of Cen~A with 256 s time bins over the full PCU2 energy range 
for observation intervals July 2006, December 2006, and August 2007. 
Dashed line is the average rate for the observational interval and the dash-dot lines 
represent $\pm$10\% in the rate.}
\end{figure}

\clearpage

\setcounter{figure}{10}

\begin{figure}
\includegraphics[width=6.0in]{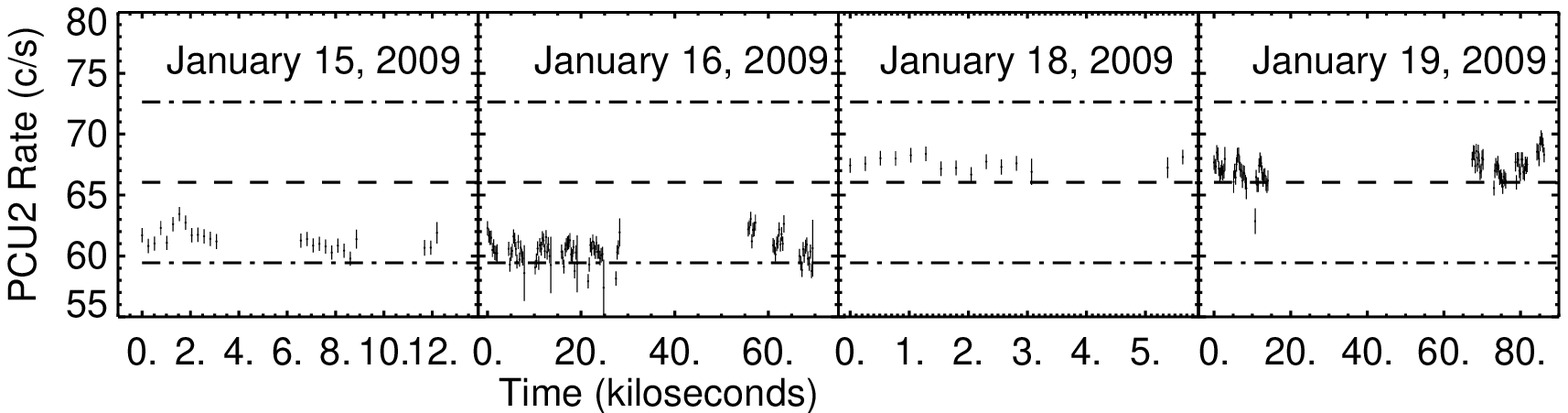}\\
\includegraphics[width=6.0in]{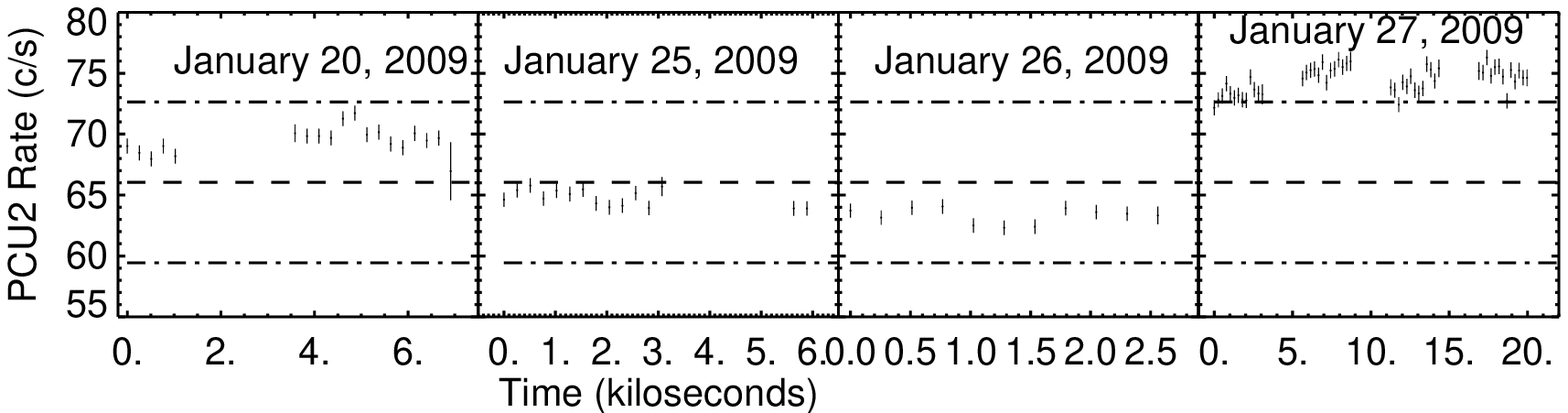}\\
\includegraphics[width=6.0in]{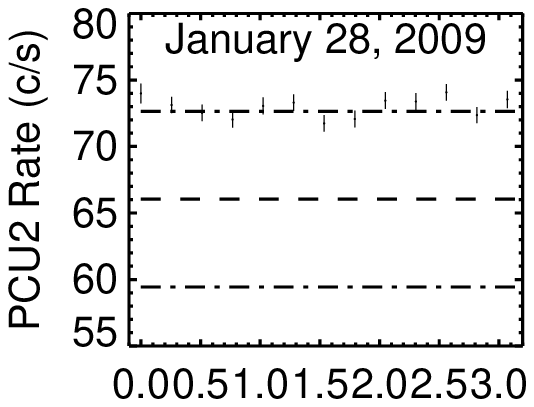}\\
\includegraphics[width=6.0in]{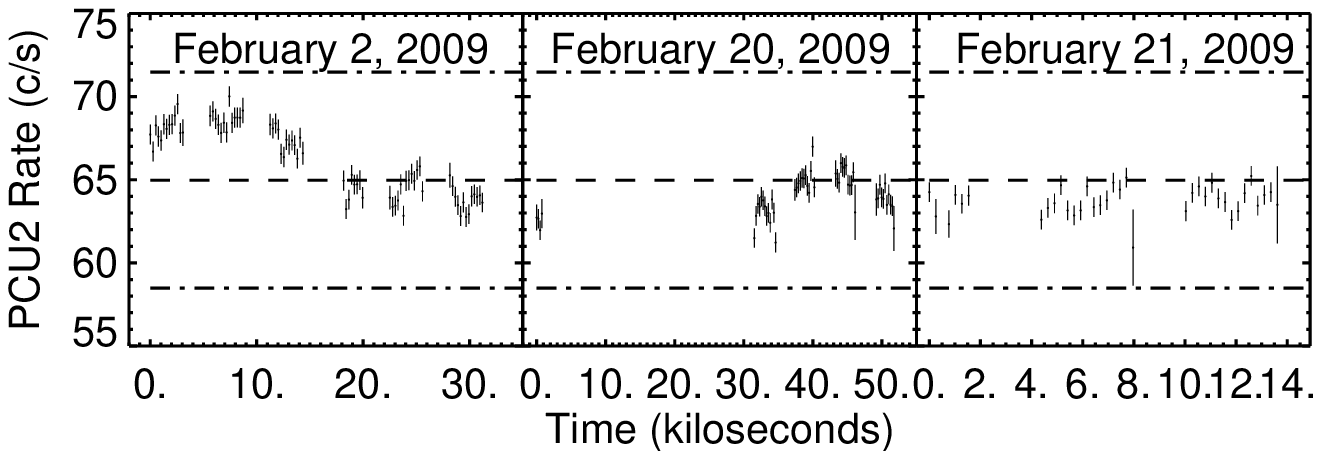}
\caption{Daily light curves of Cen~A with 256 s time bins over the full PCU2 energy range 
for observation intervals January 2009 and February 2009. 
Dashed line is the average rate for the observational interval and the dash-dot lines 
represent $\pm$10\% in the rate.}
\end{figure}

\clearpage

\begin{figure} 
\includegraphics[width=4.5in]{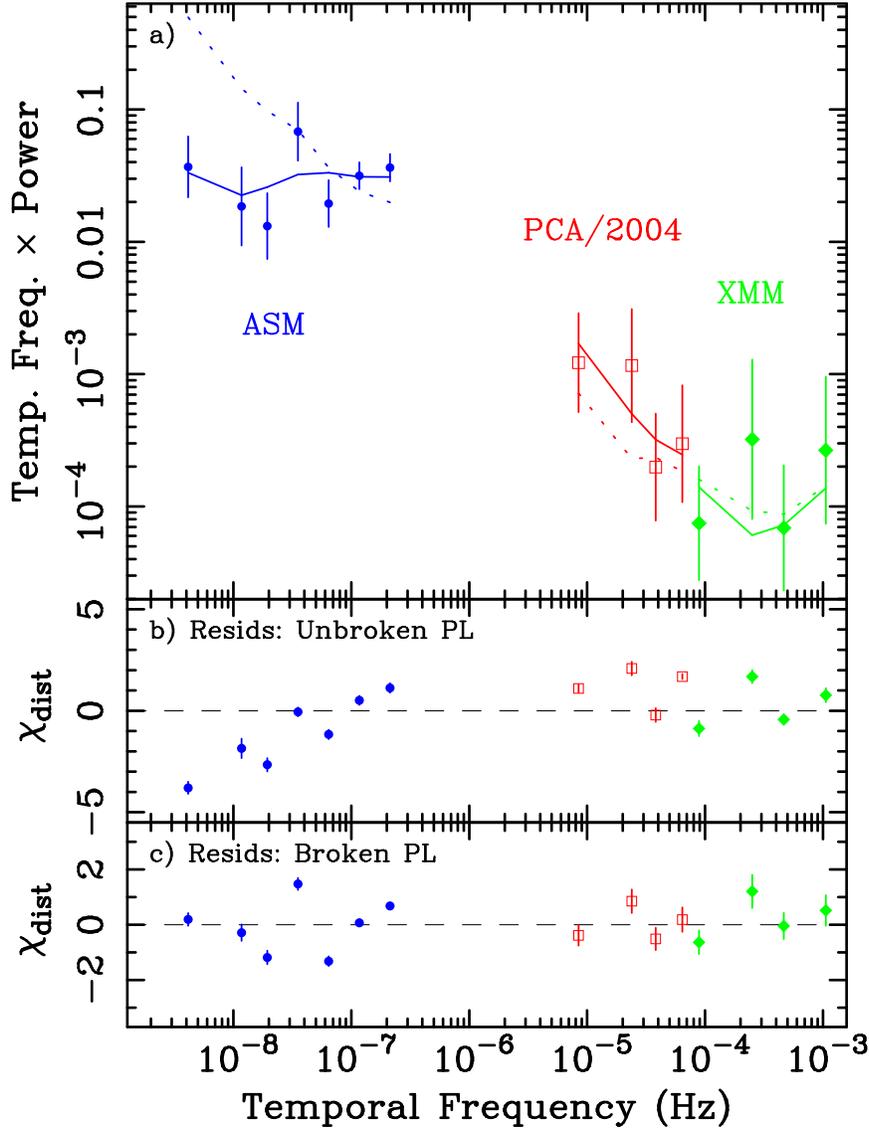}
\caption{The power spectral density function for Cen~A, derived from \textsl{RXTE} and \textsl{XMM} 
observations. {\bf In the top panel, the data points denote the observed binned PSD,
with uncertainties on each point derived from the Monte Carlo simulation procedure \citep{Utley02}.
The dotted and solid lines denote the best fit unbroken and broken power law models,
respectively; both are plotted in ``data'' space and contain the distortion effects 
inherent in PSD measurement.} The $\chi_{\rm dist}$ residuals to the unbroken and singly 
broken power law fits are given in b) and c), respectively. $\chi_{\rm dist}$ compares the model 
PSD (from the Monte Carlo simulation procedure) and observed PSD, using the distribution estimated from
the simulations \citep[see ][for details]{Uttley02}. }\label{fig:psd}
\end{figure}

\clearpage

\begin{deluxetable}{lcccccccc}
\tabletypesize{\scriptsize}
\tablecaption{Livetimes and Rates for Observations of Cen A with \textsl{RXTE}\label{tab:obstime}}
\tablewidth{0pt}
\tablehead{
\colhead{} & \colhead{PCU2 } & \colhead{PCU2} & \colhead{HEXTE-AB} & \colhead{HEXTE-AB} & \colhead{HEXTE-B} & \colhead{HEXTE-B} & \colhead{Num.} & \colhead{Num.}\\
\colhead{Date} & \colhead{Livetime\tablenotemark{a}} & \colhead{Rate\tablenotemark{b,c}} & \colhead{Livetime\tablenotemark{a}} & \colhead{Rate\tablenotemark{b,c}} & \colhead{Livetime\tablenotemark{a}} & \colhead{Rate\tablenotemark{b,c}} & \colhead{Obs.} & \colhead{Days} }
\startdata
Aug96 & 10,528 & 23.24$\pm$0.07 & 6,785   & 5.12$\pm$0.20    & 3,390   & 4.88$\pm$0.27    & 1   & 1\\
Aug98 & 67,872 & 21.23$\pm$0.03 & 42,567 & 4.41$\pm$0.08    & 18,521 & 3.77$\pm$0.12   & 7   & 4\\
Jan00 & 25,088 & 34.58$\pm$0.05  & 16,132 & 7.72$\pm$0.12    & 8,037   & 7.01$\pm$0.15   & 3   & 1\\
Mar03 & 75,280 & 42.46$\pm$0.03  & 58,574 & 11.52$\pm$0.04 & 29,302 & 10.40$\pm$0.06 & 13 & 4\\
Jan04 & 88,960 & 24.46$\pm$0.04  & 59,788 & 6.86$\pm$0.04    & 26,015 & 6.13$\pm$0.05   & 22 & 3\\
Feb04 & 36,112 & 28.14$\pm$0.04  & 23,680 & 7.94$\pm$0.06    & 14,405 & 7.11$\pm$0.08   & 4   & 2\\
Aug05 & 9,472   & 24.95$\pm$0.07  & 6,073   & 6.47$\pm$0.13    & 3,021    & 5.53$\pm$0.17   & 1   & 1\\
Dec05 & 20,994 & 19.15$\pm$0.07  & ---         & ---                           & 7,115    & 4.50$\pm$0.12   & 8   & 3\\
Jul06   & 29,968 & 33.46$\pm$0.04  & ---         & ---                           & 9,228    & 7.04$\pm$0.10   & 22 & 7\\
Dec06 & 53,088 & 25.01$\pm$0.03  & ---         & ---                           & 16,984  & 5.04$\pm$0.08   & 15 & 7\\
Aug07 & 38,784 & 46.50$\pm$0.04  & ---         & ---                           & 12,997  & 10.44$\pm$0.10 & 11 & 4\\
Jan09 & 81,230 & 64.31$\pm$0.03   & ---         & ---                           & 27,406  & 14.91$\pm$0.07 & 15 & 5\\
Feb09 & 38,800 & 63.90$\pm$0.05   & ---         & ---                           & 12,171  & 15.03$\pm$0.10 & 7   & 3\\
\enddata
\tablenotetext{a}{The Livetime in seconds}
\tablenotetext{b}{PCU2 3--60 keV counting rate in counts/s; HEXTE 15--200 keV counting rate in counts/s}
\tablenotetext{c}{\bf Uncertainties in rates are 68\% confidence}
\end{deluxetable}

\clearpage

\begin{deluxetable}{lcccccccc}
\tabletypesize{\scriptsize}
\rotate
\tablecaption{Best-fit Parameters for the 13 Observations of Cen A with \textsl{RXTE}\label{tab:13obs}}
\tablewidth{0pt}
\tablehead{
& \colhead{N$_H$\tablenotemark{a}} & \colhead{$\Gamma$\tablenotemark{b} } & \colhead{Norm(2-10)\tablenotemark{c}} & \colhead{E(Fe K-$\alpha$)\tablenotemark{d} } & \colhead{Flux(Fe K-$\alpha$)\tablenotemark{e}} & \colhead{EW(Fe K-$\alpha$)\tablenotemark{f}} & \colhead{$\chi ^2$/dof\tablenotemark{g}} & \colhead{$\chi^2_\nu$}
}
\startdata
Aug96 & 15.8$^{+0.8}_{-0.7}$  &  1.824$^{+0.028}_{-0.027}$ & 333.1$^{+8.7}_{-8.3}$  & 6.44$^{+0.06}_{-0.05}$  & 5.14$^{+0.63}_{-0.63}$ & 155$^{+40}_{-35}$    & 135.8/118 & 1.15 \\
Aug98 & 14.3$^{+0.3}_{-0.3}$  & 1.829$^{+0.011}_{-0.011}$  & 305.4$^{+3.1}_{-3.1}$  & 6.40$^{+0.03}_{-0.02}$  & 4.40$^{+0.24}_{-0.24}$ & 141$^{+14}_{-12}$    & 141.2/119  & 1.19 \\
Jan00 & 15.6$^{+0.4}_{-0.4}$   & 1.834$^{+0.014}_{-0.013}$ & 513.1$^{+6.8}_{-6.6}$  & 6.40$^{+0.05}_{-0.06}$  & 4.39$^{+0.50}_{-0.50}$ &  84$^{+17}_{-14}$      & 118.3/118  & 1.00 \\
Mar03\tablenotemark{h} & 22.7$^{+0.2}_{-0.2}$   & 1.799$^{+0.007}_{-0.007}$ & 711.8$^{+5.0}_{-4.2}$  & 6.32$^{+0.03}_{-0.03}$  & 4.72$^{+0.37}_{-0.37}$ & 63$^{+9}_{-8}$           & 118.3/117  & 1.01 \\
Jan04\tablenotemark{i} & 25.7$^{+0.3}_{-0.3}$   & 1.787$^{+0.010}_{-0.010}$ & 422.5$^{+3.9}_{-3.9}$  & 6.43$^{+0.03}_{-0.02}$  & 5.67$^{+0.30}_{_0.30}$ & 140$^{+11}_{-11}$    & 144.9/116  &  1.25 \\
Feb04\tablenotemark{h} & 24.2$^{+0.4}_{-0.4}$   & 1.796$^{+0.013}_{-0.014}$ & 478.6$^{+5.7}_{-6.3}$  & 6.39$^{+0.03}_{-0.04}$  & 5.15$^{+0.44}_{-0.45}$ & 106$^{+14}_{-14}$     & 125.2/117  & 1.07 \\
Aug05 & 19.4$^{+0.8}_{-0.8}$  & 1.848$^{+0.027}_{-0.027}$ & 408.5$^{+10.9}_{-10.5}$ & 6.41$^{+0.11}_{-0.10}$ & 3.17$^{+0.77}_{-0.77}$ & 79$^{+28}_{-29}$     &  96.4/118   &  0.82 \\
Dec05 & 18.2$^{+0.6}_{-0.6}$  & 1.823$^{+0.022}_{-0.023}$ & 301.5$^{+6.7}_{-6.4}$  & 6.32$^{+0.06}_{-0.06}$  & 3.51$^{+0.47}_{-0.48}$ & 112$^{+25}_{-23}$      & 105.9/118 & 0.90 \\
Jul06   & 18.8$^{+0.4}_{-0.4}$  & 1.844$^{+0.013}_{-0.013}$ & 544.5$^{+6.8}_{-6.9}$  & 6.34$^{+0.05}_{-0.05}$  & 4.21$^{+0.51}_{-0.50}$ & 77$^{+15}_{-14}$        & 107.3/118  & 0.91 \\
Dec06\tablenotemark{h} & 18.7$^{+0.3}_{-0.3}$  & 1.832$^{+0.012}_{-0.011}$ & 402.7$^{+4.5}_{-4.6}$  & 6.37$^{+0.03}_{-0.04}$  & 4.46$^{+0.33}_{-0.33}$ & 109$^{+13}_{-13}$      & 134.4/117  &  1.15 \\
Aug07\tablenotemark{h} & 17.3$^{+0.2}_{-0.3}$  & 1.818$^{+0.009}_{-0.009}$ & 736.4$^{+6.5}_{-6.4}$  & 6.30$^{+0.05}_{-0.05}$  & 4.27$^{+0.51}_{-0.50}$ & 56$^{+11}_{-11}$         & 129.2/117  & 1.10 \\
Jan09 &  17.0$^{+0.2}_{-0.1}$  & 1.826$^{+0.005}_{-0.006}$ & 985.8$^{+5.1}_{-5.0}$  & 6.29$^{+0.03}_{-0.04}$  & 5.01$^{+0.40}_{-0.40}$ & 49$^{+6}_{-7}$              & 137.1/118  & 1.16\tablenotemark{j} \\
Feb09 &  17.9$^{+0.2}_{-0.2}$  & 1.830$^{+0.008}_{-0.007}$ & 992.3$^{+7.5}_{-7.5}$  & 6.28$^{+0.04}_{-0.05}$  & 4.81$^{+0.59}_{-0.59}$ & 46$^{+10}_{-8}$           & 134.2/118   & 1.14 \\
\enddata
\tablenotetext{a}{Inferred Line of Sight Equivalent Hydrogen Column Density (10$^{22}$ cm$^{-2}$)}
\tablenotetext{b}{Power Law Photon Index}
\tablenotetext{c}{Power Law Normalization (2-10 keV; 10$^{-12}$ ergs cm$^{-2}$ s$^{-1}$)}
\tablenotetext{d}{The Line Energy (keV)}
\tablenotetext{e}{The Line Flux (10$^{-4}$ Photons cm$^{-2}$ s$^{-1}$)}
\tablenotetext{f}{The Line Equivalent Width (eV); {\bf Uncertainties are 99\% confidence}}
\tablenotetext{g}{Total Chi-squared/Degrees of Freedom}
\tablenotetext{h}{Additional systematics line at 29.5 keV included in fit}
\tablenotetext{i}{Additional systematics lines at 4.5 and 29.5 keV included in fit}
\tablenotetext{j}{A single PCU2 bin contributed $\chi^2$=21; without it, $\chi^2$=113.1/117=0.97}
\end{deluxetable}

\clearpage

\begin{deluxetable}{lccc|ccc|cc}
\tabletypesize{\scriptsize}
\rotate
\tablecaption{Fluxes and Limits on Reflection and Cutoffs for the 13 Observations of Cen A with \textsl{RXTE}\label{tab:13_other_obs}}
\tablewidth{0pt}
\tablehead{\colhead{Date} & \colhead{Flux(1 keV)\tablenotemark{a}}
& \colhead{Flux(2-10 keV)\tablenotemark{b}} & \colhead{Flux(20-100keV)\tablenotemark{b}} & \colhead{$\Gamma$} & \colhead{E$_{cut}$\tablenotemark{c}} & \colhead{$\chi^2_\nu$} & \colhead{R\tablenotemark{d}} & \colhead{$\chi^2_\nu$} 
}
\startdata
Aug96 & 0.099$^{+0.007}_{-0.009}$ & 1.857$^{+0.007}_{-0.007}$ & 4.99$^{+0.19}_{-0.18}$  & 1.759$^{+0.058}_{-0.092}$ & 154$^{+539}_{-86}$   & 1.12  & $\leq$8.7 &  1.16\\             
Aug98 & 0.092$^{+0.002}_{-0.003}$ & 1.773$^{+0.003}_{-0.002}$ & 4.54$^{+0.06}_{-0.06}$  & 1.804$^{+0.015}_{-0.027}$ & $\geq$204                    & 1.17  & $\leq$2.1 &  1.20\\  
Jan00 & 0.155$^{+0.005}_{-0.005}$ & 2.843$^{+0.005}_{-0.006}$ & 7.52$^{+0.08}_{-0.11}$   & 1.832$^{+0.014}_{-0.018}$& $\geq$689                     & 1.01 & $\leq$7.9 &  1.01 \\  
Mar03 & 0.203$^{+0.004}_{-0.003}$ & 3.351$^{+0.002}_{-0.003}$ & 11.23$^{+0.07}_{-0.07}$ &1.796$^{+0.007}_{-0.006}$ & $\geq$1399                  & 1.03 & $\leq$4.9 &  1.01 \\  
Jan04 & 0.120$^{+0.002}_{-0.003}$ & 1.909$^{+0.002}_{-0.002}$ & 7.01$^{+0.08}_{-0.05}$    & 1.765$^{+0.015}_{-0.015}$& 607$^{+747}_{-225}$ & 1.00 & $\leq$1.0 &  1.27\\  
Feb04 & 0.136$^{+0.005}_{-0.004}$ & 2.207$^{+0.004}_{-0.003}$ & 7.63$^{+0.07}_{-0.09}$   & 1.793$^{+0.013}_{-0.021}$ &$\geq$693                     & 1.07  & $\leq$4.4 &  1.08 \\  
Aug05 & 0.126$^{+0.005}_{-0.008}$ & 2.048$^{+0.007}_{-0.007}$ & 5.78$^{+0.14}_{-0.12}$   & 1.845$^{+0.031}_{-0.049}$ &$\geq$318                     & 0.82 & $\leq$18.2 &  0.82\\  
Dec05 & 0.090$^{+0.004}_{-0.005}$ & 1.579$^{+0.005}_{-0.004}$ & 4.55$^{+0.11}_{-0.11}$    & 1.816$^{+0.027}_{-0.049}$ &$\geq$209                    & 0.92 & $\leq$12.6 & 0.91\\  
Jul06   & 0.167$^{+0.005}_{-0.005}$ & 2.767$^{+0.003}_{-0.004}$ & 7.79$^{+0.11}_{-0.12}$   & 1.842$^{+0.012}_{-0.016}$ &$\geq$814                     & 0.92 & $\leq$8.2   &  0.92 \\  
Dec06 & 0.121$^{+0.004}_{-0.003}$ & 2.071$^{+0.002}_{-0.003}$ & 5.89$^{+0.10}_{-0.09}$   & 1.809$^{+0.026}_{-0.025}$ & $\geq$242                    & 1.12 & $\leq$7.1 &  1.14 \\  
Aug07 & 0.216$^{+0.005}_{-0.004}$ & 3.887$^{+0.004}_{-0.003}$ & 11.11$^{+0.10}_{-0.11}$ & 1.815$^{+0.009}_{-0.013}$ & $\geq$878                    & 1.11 & $\leq$5.7 &  1.10 \\  
Jan09 & 0.293$^{+0.004}_{-0.003}$ & 5.236$^{+0.003}_{-0.003}$  &  14.68$^{+0.10}_{-0.09}$ & 1.823$^{+0.006}_{-0.004}$ &$\geq$2796                 & 1.21 &  $\leq$4.6 & 1.16 \\  
Feb09 & 0.297$^{+0.006}_{-0.005}$ & 5.139$^{+0.003}_{-0.005}$  & 14.60$^{+0.12}_{-0.11}$  &  1.828$^{+0.008}_{-0.008}$ & $\geq$2499               & 1.17 & $\leq$3.9  & 1.15\\  
\enddata
\tablenotetext{a}{Power law normalization at 1 keV (photons cm$^{-2}$ s$^{-1}$ keV$^{-1}$)}
\tablenotetext{b}{Flux in units of 10$^{-10}$ ergs cm$^{-2}$ s$^{-1}$}
\tablenotetext{c}{Exponential Cut-off Energy (keV)}
\tablenotetext{d}{Compton Reflection Strength (\%)}

\end{deluxetable}

\clearpage

\begin{deluxetable}{lllllllll}
\tabletypesize{\scriptsize}
\rotate
\tablecaption{Start/Stop Times for the 51 PCU2 Light Curves of Cen A\label{tab:lctimes}}
\tablewidth{0pt}
\tablehead{
\colhead{Obs. Interval} &\colhead{Date} & \colhead{Time (UT)} & \colhead{Obs. Interval} &\colhead{Date} & \colhead{Time (UT)} & \colhead{Obs. Interval} &\colhead{Date} & \colhead{Time (UT)} 
}
\startdata
August 1996   & 8/14 & 2:43--8:03  & December 2005 & 12/16 & 0:08--0:21   & August 2007 & 8/18  & 3:54--23:27\\
August 1998   & 8/9   & 1:24--23:56   &                               & 12/17 & 17:06--22:09 &                         & 8/19 & 0:19--15:03\\
                          & 8/10 & 0:01--7:18   &                               & 12/20 & 20:00--24:07 &                         & 8/20  & 14:10--24:07\\
                          & 8/14 & 3:08--23:58   &                               & 12/23 & 0:08--22:52    &                        & 8/22  & 3:44--7:51\\
                          & 8/15 & 0:02--8:57   & July 2006            & 7/13    & 10:08--24:27 & January 2009 & 1/15 & 20:34--23:58\\
January 2000 & 1/23 & 7:38--23:24   &                               & 7/14    & 11:03--22:28 &                        & 1/16  & 0:02--19:20\\
March 2003    & 3/7    & 12:02--20:58 &                              & 7/16    & 17:56--23:04  &                         & 1/18  & 22:24--23:58\\  
                          & 3/8    & 1:04--17:28   &                               & 7/17   & 0:13-- 17:43    &                         & 1/19  & 0:02--23:58\\ 
                          & 3/9    & 3:33--18:48   &                               & 7/18   & 10:54--12:38  &                          & 1/20   & 0:02--1:57\\ 
                          & 3/10 & 8:00--16:02    &                               & 7/19   & 13:32--16:54  &                          & 1/25   & 22:19--23:58\\
                          & 3/11 & 7:43--8:17    &                               & 7/20   & 11:39--22:55  &                          & 1/26   & 0:02--0:44\\
January 2004 & 1/2    & 8:56--23:58   & December 2006 & 12/8   & 7:03--13:45     &                        & 1/27  & 18:16--23:49\\  
                          & 1/3    & 0:03--23:47    &                               & 12.9   & 1:39--2:21    &                           & 1/28  & 0:32--1:23\\
                          & 1/4    & 0:46--16:29    &                               & 12/10 & 1:47--9:26    & February 2009 & 2/2  & 13:58--22:39\\
February 2004& 2/13 & 9:41--16:46    &                               & 12/11  & 4:26--16:19   &                         & 2/20 & 4:59--19:19\\
                           & 2/14 & 9:19--19:39    &                               & 12/12  & 9:39--13:12   &                         & 2/21  & 4:06--7:53\\   
August 2005    & 8/20 & 3:59--7:58    &                               & 12/13  & 9:09--19:14\\                           
                           &          &             &                                          & 12/15  & 5:10--18:19\\
\enddata
\end{deluxetable}

\end{document}